\documentclass{aa}
\usepackage{etoolbox}
\usepackage{graphicx}
\usepackage{txfonts}
\usepackage{float}
\usepackage{hyperref}
\hypersetup{
     colorlinks = true,
     linkcolor = blue,
     anchorcolor = blue,
     citecolor = blue,
     filecolor = blue,
     urlcolor = blue
     }
\usepackage{pdflscape} 
\usepackage{afterpage} 

\usepackage{tabularx}
\usepackage{breqn}

\makeatletter
\renewcommand*\aa@pageof{, page \thepage{} of \pageref*{LastPage}}
\usepackage{caption}
\usepackage{multirow}
\usepackage{lscape}
\usepackage{orcidlink}
\defcitealias{diemer2019}{DJ19}

\begin{document} 

   \title{The galaxy-halo connection of disc galaxies over six orders of magnitude in stellar mass} 
   \titlerunning{The galaxy-halo connection over six orders of magnitude in stellar mass}
\authorrunning{Mancera Pi\~na et al.}

   \author{Pavel E. Mancera Pi\~na\,\orcidlink{0000-0001-5175-939X}
   \inst{1}\fnmsep\thanks{\email{pavel@strw.leidenuniv.nl}}, Justin I. Read\,\orcidlink{0000-0002-1164-9302}\inst{2}, 
   Stacy Kim\,\orcidlink{0000-0001-7052-6647}\inst{3}, 
Antonino Marasco\,\orcidlink{0000-0002-5655-6054}\inst{4},
   Jos\'e A. Benavides\,\orcidlink{0000-0003-1896-0424}\inst{5},
   Marcin Glowacki\,\orcidlink{0000-0002-5067-8894}\inst{6,7}, 
   Gabriele Pezzulli\,\orcidlink{0000-0003-0736-7879}\inst{8},
   and Claudia del P. Lagos\,\orcidlink{0000-0003-3021-8564}\inst{9} 
          }
   \institute{Leiden Observatory, Leiden University, P.O. Box 9513, 2300 RA, Leiden, The Netherlands 
         \and Department of Physics, University of Surrey, Guildford, Surrey GU2 7XH, United Kingdom
         \and Carnegie Theoretical Astrophysics Center, Carnegie Observatories, 813 Santa Barbara St, Pasadena, CA 91106, USA
         \and INAF – Padova Astronomical Observatory, Vicolo dell’Osservatorio 5, I-35122 Padova, Italy
         \and Department of Physics and Astronomy, University of California, Riverside, 900 University Avenue, Riverside, CA 92521, USA
         \and Institute for Astronomy, University of Edinburgh, Royal Observatory, Edinburgh, EH9 3HJ, United Kingdom
         \and Inter-University Institute for Data Intensive Astronomy, Department of Astronomy, University of Cape Town, Cape Town, South Africa
         \and Kapteyn Astronomical Institute, University of Groningen, Landleven 12, 9747 AD Groningen, The Netherlands
                  \and International Centre for Radio Astronomy Research, Curtin University, Bentley, WA 6102, Australia
          }
    
   \date{Accepted for publication in A\&A}

  \abstract{

The relations between stellar ($M_\ast$), gas ($M_{\rm gas}$), baryonic ($M_{\rm bar} = M_\ast + M_{\rm gas}$), and dark matter halo mass ($M_{200}$) provide unique constraints on galaxy formation and cosmology. The shape of the relations constrains how galaxies regulate their growth through gas accretion, star formation, and feedback, and their scatter probes the stochasticity of galaxy assembly, which depends on the underlying cosmological model.

In this paper, we assemble a sample of 49 nearby gas-rich dwarf and massive disc galaxies with unmatched ancillary data. We obtain their gas kinematics and derive their dark matter properties through rotation curve decomposition. Our sample is representative of the regularly rotating gas-rich galaxy population and allowed us to study the galaxy-halo connection across nearly six orders of magnitude in $M_\ast$. 
We find that the $M_{\rm gas}-M_{200}$ relation rises monotonically, with galaxies having around 4\% of the average cosmological baryon fraction in cold gas. Contrastingly, the $M_\ast-M_{200}$ relation shows a more complex behaviour. A particularly interesting finding is that of a population of `baryon-deficient' dwarfs (BDDs) with stellar masses $\sim 1-1.5$ orders of magnitude lower than expected from current models. Yet, baryon-rich galaxies also exist, and we find a large spread in the baryon retention fraction across our galaxies. 
We compare our findings with semi-analytic (DarkLight) and hydrodynamical (TNG50, Simba) galaxy formation simulations. While the simulations broadly reproduce most observed features, they struggle to match the BDDs and do not capture the diversity in baryon fractions. Understanding these differences will shed new light on how feedback regulates galaxy formation. 
Finally, we study the dark matter halo concentration-mass relation. We find that below $M_{200} \sim 10^{11}\,M_\odot$, the concentrations are systematically lower than expected from pure-dark matter simulations. We discuss whether these results stem from the influence of baryonic physics or the environment. Understanding this is crucial if gas-rich galaxies are to be used to test cosmological models.
  }


   \maketitle
%
\section{Introduction}
\label{sec:intro}

According to the current understanding of galaxy evolution, galaxies form in the centre of massive dark matter haloes. Cosmological accretion and mergers feed both dark matter and baryons, and together with stellar and active 
galactic nucleus feedback, they regulate galaxies through cosmic time (e.g. \citealt{press1974,binney1977,white1978,fall1980,blumenthal1984,dekel1986,frenk1988,duffy2010,lilly2013,somerville2015,bookFilippo,marasco2021}). 
Despite the expected complexity in the above processes and the apparent stochasticity of some of them, they all result in the high degree of universality and regularity imprinted in the strong self-similarity observed in present-day galaxies. Examples of this are the existence of different scaling relations between various fundamental parameters such as mass (of dark matter and baryons), rotational speeds, angular momentum, star formation histories (SFHs), gas content, and sizes (e.g. \citealt{faberjackson1976,tullyfisher,fall83,kennicutt1989,courteau2007,vanderkruit2011,cappellari_atlasXV,cortese2016,anastasia_SED,catinella2018,lelli_rar,posti_galaxyhalo,stone2021,paperIIBFR,nikki}).

One of the most powerful tools for studying the dynamics of the galaxy-halo connection is mass modelling through rotation curve decomposition of high-resolution kinematics (see also e.g. \citealt{yasin2023,yasin2024} for explorations with unresolved kinematics). With this technique, the observed kinematics of galaxies, typically traced with the neutral atomic hydrogen line (H\,{\sc i}), are explained through a combination of the gravitational potentials provided by the observed baryons and a putative dark matter halo (e.g. \citealt{freeman1970,roberts1973,bosma1978,vanalbada86,begeman,marc_phd,bullock2017,bookFilippo,salucci2019}). Since one of the outcomes of rotation curve decomposition is the halo mass, this approach opens the possibility to investigate the stellar-, gas-, and baryonic-to-dark-matter mass fractions and to constrain how efficient galaxies have been at retaining their baryons, accreting gas, and forming stars through their evolution (e.g. \citealt{vladimir_baryons,moster2010,aldo2015,trujillogomez2015,gonzalez2016,chauhan2020,girelli2020,guo2020,posti_galaxyhalo,korsaga2023,dev2024,lu2024}).

The literature on the mass models of massive galaxies is extensive, and in recent years researchers have exploited high-quality samples using modern statistical techniques (e.g. \citealt{frank2016,ren_SIDM_SPARC,li_massmodels,zenter2022,paper_massmodels}). Among the most recent results are the findings that many of the most massive spiral galaxies have converted nearly all of their available baryons into stars (i.e. they have little to no `missing baryons'; e.g. \citealt{postinomissing, enrico_massmodels_ss}) and that even sophisticated abundance-matching techniques and advanced cosmological hydrodynamical simulations struggle to reproduce galaxies with stellar masses (at fixed halo mass) as high as observed in nearby galaxies, arguably due to an overly efficient feedback implementation and/or wrong halo occupation fractions (e.g. \citealt{postinomissing,posti_galaxyhalo,marasco_DMinsim,enrico_massmodels_ss}, and references therein).

Studying the low-mass regime (e.g. \citealt{deblok2001,oh2015,readAD,kaplinghat2020,paper_massmodels,banares2023,roberts2024,sylos2025}) is not straightforward, partially due to the difficulty of having dwarf galaxies with extended and high-resolution kinematics, near-infrared (NIR) photometry to trace the stellar mass ($M_\ast$), and accurate distance measurements. Yet, dwarfs are crucial for the study of dynamical scaling relations and can set strong and insightful constraints on galaxy formation and evolution theories given that their dynamical properties are susceptible to the underlying cosmological model, stellar feedback, their accretion histories and SFHs, and even the time at which reionisation took place (e.g. \citealt{dicintio2014,benitezllambay2015,trujillogomez2015,oman_missingdarkmatter,readAD,read2017, robles2017,fitts2018,fitts2019,read2019,edge,coreEinasto,nadler2020,cuspcoreII,danieli2023,kim2024,koudmani2024,oman2024,muni2025}). This makes the galaxy-halo connection in dwarfs a particularly powerful probe of both galaxy formation models and cosmology.

With this in mind, in this work, we build a curated sample of 49 nearby gas-rich dwarf and massive galaxies with archival high-resolution H\,{\sc i} kinematics, NIR photometry, and accurate distance measurements.
After obtaining robust kinematic models for our galaxies, we derived their mass models. This allowed us to infer our sample's dark matter halo parameters and build a set of dynamical scaling laws to study the galaxy halo connection across six orders of magnitude in stellar mass with an unmatched quality sample.
This work is organised as follows. We present our galaxy sample in Sec.~\ref{sec:data}. In Sec.~\ref{sec:kinmodels}, we build kinematic models for 22 of our dwarf galaxies, which we complement with literature models for the rest of our sample. The kinematic measurements are used in Sec.~\ref{sec:massmodels} to perform rotation curve decomposition. 
We present our resulting mass models in Sec.~\ref{sec:massmodels_results}, together with the derived best-fitting dark matter halo parameters, mass-to-light ratios, and the gas scale heights of our sample. In Sec.~\ref{sec:scalinglaws}, we study different scaling laws between baryonic and dark matter properties across six orders of magnitude in stellar mass, and we discuss the implications of our results. Finally, we summarise our main results in Sec.~\ref{sec:conclusions}. Throughout this work, we adopt a $\Lambda$ cold dark matter (CDM) cosmology with $\Omega_{\rm m} = 0.3$, $\Omega_{\Lambda} = 0.7$ and $H_0 = 70~\rm{km\,s^{-1}\,Mpc^{-1}}$.

\section{Data and sample}
\label{sec:data}
To obtain rotation curve decomposition as reliably as possible, it is necessary to perform mass modelling on a galaxy sample with the following characteristics: \emph{i}) high-resolution H\,{\sc i} data to accurately trace the gas spatial distribution and kinematics; \emph{ii}) regular kinematic patterns of rotation to ease kinematic and dynamical modelling; \emph{iii}) kinematic models simultaneously obtaining the gas rotational velocity and velocity dispersion; \emph{iv}) inclination angles $\gtrsim40^\circ$ (to avoid large uncertainties when deprojecting line-of-sight velocities) but lower than $\sim80^\circ$ (for which current kinematic modelling is not well suited); \emph{v}) deep NIR imaging to trace their stellar component robustly (e.g. \citealt{mcgaugh_ML,anastasia_SED,marasco_mstar}); and \emph{vi}) precise redshift-independent distances such as those coming from the tip of the red giant branch (TRGB) or Cepheids. Meeting all of these requirements is not straightforward. 
In this work, we have compiled a sample of nearby field dwarf and massive galaxies with those characteristics based on available public data.
The following sections describe our galaxy sample and the gas and NIR imaging data we employed.

\subsection{Massive galaxies}
We select massive ($M_\ast \gtrsim 10^{9.5}\,M_\odot$) galaxies following \cite{paper_massmodels}. Specifically, we start with a base sample of spiral galaxies with high-resolution H\,{\sc i} data from \cite{enrico_radialmotions}. Of that initial sample, we keep only galaxies with regular kinematics, inclination angles $35^\circ \lesssim i \lesssim 80^\circ$, archival NIR photometry, available 2D bulge-disc decompositions (see below), available distances from standard candles (exclusively TRGB, Cepheids, or supernovae), and CO imaging (tracing molecular gas, H$_2$). Our selection cuts result in a sample of 16 massive spiral galaxies: NGC 0253, NGC 1313, NGC 2403, NGC 3198, NGC 3351, NGC 3621, NGC 3992, NGC 4535, NGC 4536, NGC 4559, NGC 4651, NGC 4725, NGC 4736, NGC 5005, and NGC 5055.

The H\,{\sc i} surface densities are taken directly from \cite{enrico_radialmotions}, while the H$_2$ surface densities come from \cite{paper_massmodels}. Both H\,{\sc i} and H$_2$ surface densities were derived from azimuthally averaged profiles of the total intensity maps and corrected by a factor of 1.36 to account for helium (the typical correction factor for our average $M_\ast$, see \citealt{mcgaugh2020}), i.e. $\Sigma_{\rm gas} = 1.36\,(\Sigma_{\rm HI} + \Sigma_{\rm H_2}$).
Obtaining robust constraints on the stellar surface densities ($\Sigma_\ast$) is particularly important for our massive galaxies since $M_\ast$ dominates their baryonic budget.
All our massive galaxies have (by selection) Spitzer \citep{spitzer} 3.6$\,\mu$m surface brightness profiles. Their morphology can be complex, and careful 2D bulge-disc decomposition is necessary. Because of this, we rely upon the 2D bulge-disc\footnote{For simplicity, for some of the galaxies (see \citealt{salo2015,paper_massmodels}), we ignore a small bar component (accounting for less than $5-10$\% of the flux). We expect their dynamical contribution to be small given their low mass, and we corroborate that this is the case in Appendix~\ref{app:bars} (see first Sec.~\ref{sec:massmodels_results}).} decomposition by \cite{salo2015}. The only three exceptions are NGC~2403, NGC~3198, and NGC~3621 (all discs without bulges, see Sec. 2.2.3 in \citealt{paper_massmodels}), whose 3.6$\,\mu$m surface brightness profiles are taken from \cite{paper_massmodels}. 

The distances (TRGB or Cepheids) to all these galaxies are taken from the NASA/IPAC Extragalactic Database (NED) and the original references are \cite{parodi2000,saha2006,rizzi2007,takanashi2008,dalcanton_angst,jacobs2009,radburn2011,foster2014,mcquinn2017,sabbi2018}. We note that these distances can differ from those reported in \cite{enrico_radialmotions} and \cite{paper_massmodels}.

\subsection{Dwarf galaxies}

To assemble our dwarf sample, we first start with the sample from \cite{iorio}, who obtained kinematic models from a subset of galaxies from LITTLE THINGS (Local Irregulars That Trace Luminosity Extremes, The H\,{\sc i} Nearby Galaxy Survey, \citealt{hunterLT}). After selecting galaxies with inclinations $35^\circ \lesssim i \lesssim 80^\circ$, distances from standard candles, NIR photometry, and regular kinematics (see also \citealt{read2017}), we end up with 11 dwarfs. The H\,{\sc i} surface density profiles are taken from \cite{iorio} and are azimuthally averaged radial profiles. Since molecular gas is expected to be dynamically negligible (e.g. \citealt{leroy}, but see also \citealt{hunter_codark}), the gas surface densities are computed as $\Sigma_{\rm gas} = 1.33\,\Sigma_{\rm HI}$. NIR (Spitzer 3.6$\,\mu$m) surface brightness profiles are taken from \cite{zhang} and \cite{galexs4g}.

To expand this sample, we inspected different public H\,{\sc i} surveys, namely: LVHIS (The Local Volume H\,{\sc i} survey, \citealt{lvhis}), THINGS (The H\,{\sc i} Nearby Galaxy Survey, \citealt{things}), LITTLE THINGS, WHISP (Westerbork observations of neutral Hydrogen in Irregular and SPiral galaxies, \citealt{whisp}), VLA-ANGST (Very Large Array - ACS Nearby Galaxy Survey Treasury, \citealt{vla_angst}) and HALOGAS (Hydrogen Accretion in LOcal GAlaxieS, \citealt{halogas}). 
To ensure a high-quality sample of dwarfs, we examine all the galaxies in the above surveys and select those with NIR images, TRGB/Cepheids distances, clear gradients in their velocity fields with rotational velocities $\lesssim100$ km/s, kinematic maps traced with at least five beams, and inclinations in the range $40^\circ-80^\circ$. Our selection results in 22 galaxies. For these dwarfs, we derive $\Sigma_{\rm gas} = 1.33\,\Sigma_{\rm HI}$ from azimuthally averaged profiles.

For 16 out of these 22 dwarfs, Spitzer imaging at 3.6$\,\mu$m is available. Our 3.6$\,\mu$m surface brightness data come either from (the Sérsic parameters of) \cite{salo2015}, from the S4G survey \citep{galexs4g}, or, if the profiles are not publicly available, we download the 3.6$\,\mu$m imaging from the NASA/IPAC Infrared Science Archive and derive azimuthally averaged surface brightness profiles following \cite{marasco_halogas}. For the remaining six galaxies (LVHIS 019, LVHIS 020, LVHIS 026, LVHIS 055, LVHIS 060, LVHIS 072), \cite{kirby08} and \cite{young} provide $H-$band (1.65$\,\mu$m) imaging; specifically, we use their Sérsic profiles fits. All our dwarf galaxies have distances derived from the TRGB or Cepheids methods and have been determined by \cite{tully2006,saha2006,dalcanton_angst,jacobs2009}, and \cite{tully2013}.

\subsection{The final sample}

From the above, we end up with a curated sample of 49 nearby, central, gas-rich galaxies. Since the galaxies come from different surveys and lie at different distances, they have a range of spectral and spatial resolutions. In the case of the former, it ranges between $\sim$ 1 and 5 km/s, and the latter between $\sim$ 0.1 and 2~kpc. These resolutions, coming from the most complete and dedicated publicly available H\,{\sc i} surveys, are sufficient to provide robust kinematic and dynamic constraints (e.g. \citealt{deblok_2000,marc_phd,barolo,oh2015,sparc,readAD,iorio,marasco_halogas}).
Despite coming from different surveys, we emphasise that our sample is homogenous: all galaxies have high-resolution H\,{\sc i} (and CO for the massive ones) interferometric data, gas surface densities derived in the same way, distances from standard candles, and NIR photometry. Moreover, their kinematics were derived following similar approaches with the same fitting software (Sec.~\ref{sec:kinmodels}). 

\begin{figure}
    \centering 
    \includegraphics[width=0.47\textwidth]{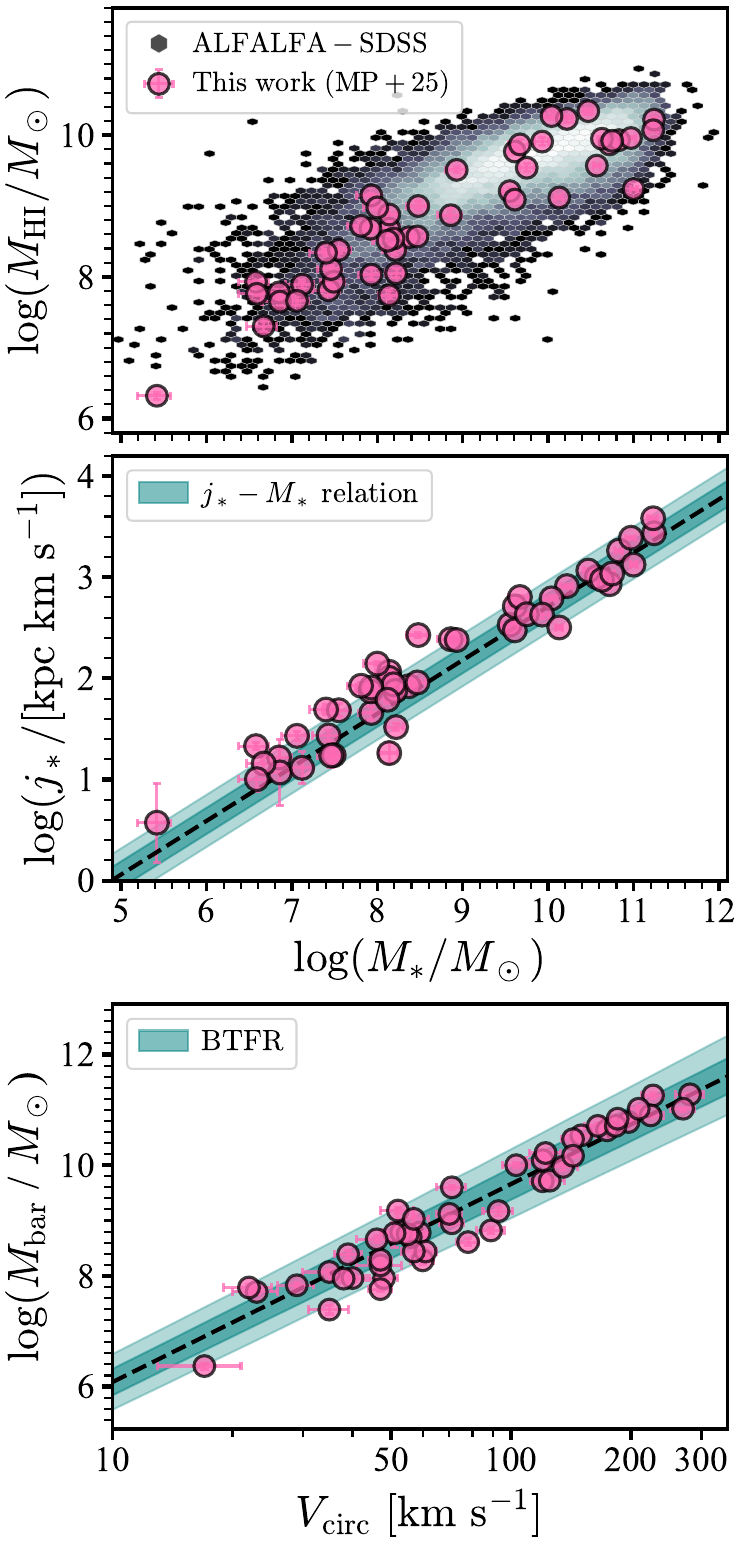}
    \caption{Overview of our galaxy sample. We place our galaxies in different scaling relations. 
    \emph{Top:} Stellar mass vs. H\,{\sc i} mass plane. The representative ALFALFA-SDSS sample \citep{alfalfa_sdss} is shown for comparison. \emph{Middle:} Baryonic Tully-Fisher relation (BTFR; from \citealt{enrico_massmodels_ss}). \emph{Bottom:} Stellar specific angular momentum vs. mass relation ($j_\ast-M_\ast$; from \citealt{paperIBFR}). In the top panel, the colour scale increases logarithmically with the density of the data counts. The shaded bands in the middle and top panels correspond to $1\,\sigma$ and $2\,\sigma$ from the best-fitting relations (black dashed lines).
    } 
    \label{fig:sample}
\end{figure}

While our sample is driven by the data quality and is therefore not volume complete and has a complex selection function (similar to samples at higher masses in the literature; e.g. \citealt{sparc,anastasia1,paperIIBFR}), it is representative of the nearby dwarf galaxy population with regular kinematics and large gas reservoirs. To illustrate this, Fig.~\ref{fig:sample} presents an overview of our sample's mass and kinematic properties. Specifically, the figure shows our sample in \emph{i}) the $M_\ast-M_{\rm HI}$ plane contrasted with the complete ALFALFA-SDSS sample of gas-rich galaxies \citep{alfalfa_sdss}, \emph{ii}) the baryonic Tully-Fisher relation of \cite{enrico_massmodels_ss}, and \emph{iii}) the stellar specific angular momentum vs mass ($j_\ast-M_\ast$) relation from \cite{paperIBFR}. The stellar, gas, and baryonic masses of our galaxies are computed in Sec.~\ref{sec:scalinglaws}, while $V_{\rm circ}$ is obtained in Sec.~\ref{sec:kinmodels}. The derivation of $j_\ast$ follows \cite{paperIBFR}. Fig.~\ref{fig:sample} highlights that our full sample has similar properties to the broad population of nearby star-forming galaxies across six orders of magnitude in $M_\ast$. Table~\ref{tab:sample} lists the galaxy names (and SIMBAD link to check cross-IDs), their distance, characteristic outer circular speed $V_{\rm circ,out}$, and average inclination. 

\section{Kinematic modelling}
\label{sec:kinmodels}

Gas kinematics can suffer from beam smearing if the size of the beam (PSF) is not negligible compared to the extent of the galaxies (e.g. \citealt{swatersPhD,barolo,iorio}). This effect should be weak in our sample of massive galaxies because of the high spatial resolution, but ideally, it should be taken into account. For the dwarfs, with a smaller physical extent, correcting for beam smearing is imperative \citep{barolo,iorio}. In addition, determining the rotation velocity of the gas ($V_{\rm rot}$) and its velocity dispersion ($\sigma_{_{\rm HI}}$) self-consistently is crucial to estimating the circular speed of the galaxies and their thickness (see below). Considering this, one of the primary requisites when building our sample was to have (or to be able to derive) kinematic measurements free of beam smearing and with simultaneous determination of $V_{\rm rot}$ and $\sigma_{_{\rm HI}}$.

In the case of our 16 massive galaxies, \cite{enrico_radialmotions} derived their kinematic parameters using the software $^{\rm 3D}$Barolo\footnote{v1.7, \url{https://editeodoro.github.io/Bbarolo/}} \citep{barolo}. $^{\rm 3D}$Barolo implements a forward modelling approach that mitigates beam smearing while fitting simultaneously $V_{\rm rot}$ and $\sigma_{_{\rm HI}}$ (together with radial motions and geometric parameters if needed and if the spatial resolution is high). In practice, $^{\rm 3D}$Barolo makes realisations of tilted-ring models \citep{rogstad1974} and, after convolution with the observational beam, compares them against the data using all the channel maps of the data cube. $^{\rm 3D}$Barolo has been well-tested for a variety of data with different resolutions, emission lines, and redshifts (e.g. \citealt{enrico_z1, iorio,huds2019,ceci_turbulence,enrico_radialmotions,alpaka2023,agc114905_deep,deg2024,rowland2024,liu2025}). 

From our sample of low-mass galaxies, 11 of them already have kinematic models derived using $^{\rm 3D}$Barolo by \cite{iorio} \footnote{We note that \cite{oh2015} derived kinematic models for some galaxies in our sample but using a less sophisticated method, upon which \cite{iorio} improved. See \cite{iorio} for details.}, which we adopt\footnote{\label{foot:ddo133}For DDO 133, \cite{iorio} adopted a distance of 6.1 Mpc following \cite{hunterLT}, which was derived from the galaxy's recessional velocity. We instead adopted a TRGB distance of 5.11 Mpc.}. We use $^{\rm 3D}$Barolo to obtain the kinematic models for the remaining 22 galaxies, namely DDO 181, DDO 183, DDO 190, IC 2574, LVHIS 009, LVHIS 011, LVHIS 012, LVHIS 017, LVHIS 019, LVHIS 020, LVHIS 026, LVHIS 055, LVHIS 060, LVHIS 072, LVHIS 077, LVHIS 078, LVHIS 080, NGC 0925, NGC 2541, NGC 4190, NGC 7793, and UGC 1501. In the remaining part of this section, we describe the methodology used to derive the kinematic measurements.

As the first step for our kinematic modelling, we obtain initial estimates for the centre, position angle, and inclination of the H\,{\sc i} discs. For this we use the software \texttt{cannubi}\footnote{\url{https://www.filippofraternali.com/cannubi}} (see \citealt{fraternali2017,huds2020,fernanda2023}), which uses a Markov Chain Monte Carlo (MCMC) forward modelling to find the centre, position angle, and inclination of a model H\,{\sc i} map that better matches the observed data. Reassuringly, the results found by \texttt{cannubi} agree with the values found from photometry \citep{kirby08,zhang,young,lvhis} within a few degrees. Once the initial guesses for the geometrical parameters are estimated, we proceed to use $^{\rm 3D}$Barolo to derive azimuthal models and obtain the best-fitting position angle, inclination, $V_{\rm rot}$, and $\sigma_{_{\rm HI}}$. We note that the position angle and inclination are only allowed to vary within 20 and 10 degrees from our input values (and generally, the resulting values move much less from the initial estimates), which is twice the typical uncertainty found by \texttt{cannubi} and our isophotal fitting. We performed our kinematic modelling following the next steps for each galaxy.

\begin{enumerate}
    \item We set the ring separation for the tilted ring modelling (\texttt{RADSEP} within the $^{\rm 3D}$Barolo environment). As a compromise between having a suitable rotation curve sampling but not oversampling the fit (having a strong correlation between the pixels), we chose \texttt{RADSEP}~=~$0.75\,(b_{\rm maj}\,b_{\rm min})^{1/2}$, with $b_{\rm maj}$ and $b_{\rm min}$ the major and minor axes of the beam, respectively. 
    
    \item We built a mask. In practice, we first generated a preliminary mask using the \texttt{SEARCH} option in $^{\rm 3D}$Barolo, which we then enlarged with a custom code by two or three pixels. This strategy avoids significant noise from being included, but it keeps faint H\,{\sc i} emission, which is vital to tracing $\sigma_{_{\rm HI}}$ and $V_{\rm rot}$ in the outer discs.
    
    \item We performed the first iteration of the kinematic modelling. We allowed $^{\rm 3D}$Barolo to estimate the systemic velocity ($V_{\rm sys}$) of the galaxies (based on the global H\,{\sc i} spectrum) and their centre (based on the total H\,{\sc i} map), which we found in good agreement with the results from \texttt{cannubi}. As free fitting parameters, we considered $V_{\rm rot}$, $\sigma_{_{\rm HI}}$, the position angle and the inclination. The last two parameters were regularised with smooth functions (typically straight lines or polynomials of degree one or two, see \citealt{barolo}).
    
    \item We inspected the first $^{\rm 3D}$Barolo fit. We corroborated that $V_{\rm sys}$ is correct based on inspection of the position-velocity (PV) diagrams of the data and the model, which would be offset if $V_{\rm sys}$ is wrong. If needed, we manually adjusted $V_{\rm sys}$, but this occurred only seldom. We fixed the value of $V_{\rm sys}$.
    
    \item We inspected whether $^{\rm 3D}$Barolo finds evidence for a change (i.e. a warp) in position angle or inclination. At the same time, using the PV diagrams, we investigated if there are hints of radial motions (visible as gradients in the minor-axis PV; see \citealt{enrico_radialmotions}). 
    
    \item We obtained a new kinematic model. If in the previous step we found evidence (from the visual inspection of the $^{\rm 3D}$Barolo outputs) for warps in the position angle or inclination (or radial motions), we considered them as free parameters again together with $V_{\rm rot}$ and $\sigma_{_{\rm HI}}$; otherwise (most of the time in practice), the only free parameters are $V_{\rm rot}$ and $\sigma_{_{\rm HI}}$. 
    
    \item We performed a final check on the kinematic models (moment maps, channel maps, PV diagrams). If they could be improved, we repeated steps 4 to 6.
\end{enumerate}

\begin{figure*}
    \centering 
    \includegraphics[width=0.905\textwidth]{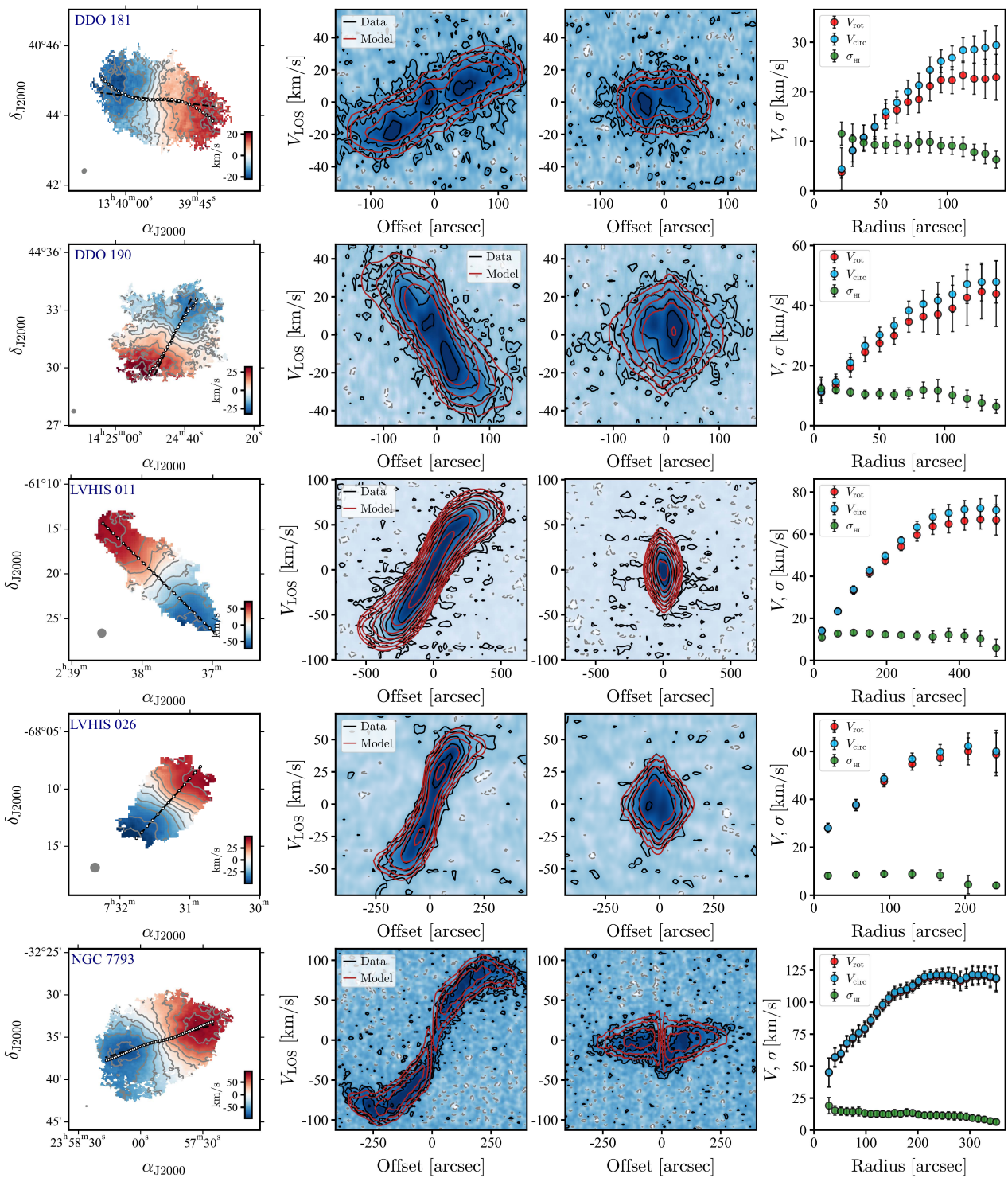}
    \caption{Kinematics of four representative galaxies in our sample (see also Fig.~\ref{fig:kinematics_rest}). \emph{a):} Velocity field (first moment map). We also show the kinematic (white circles) and average (black dashed lines) position angles as well as the beam (grey ellipse) of the observations. \emph{b) and c):} Position-velocity slices along the average major (\emph{b}) and minor (\emph{c}, perpendicular to \emph{b}) axes. The data are represented with a blue background and black contours (grey for negative values), while the best-fitting $^{\rm 3D}$Barolo model is shown with red contours. Contours are plotted at $2^n\times$ the noise, with $n=1,2,...,n$. \emph{d)} Rotation curve, circular speed, and gas velocity dispersion. We emphasise that the velocity fields are shown for illustration purposes, but $^{\rm 3D}$Barolo fits the entire data cube.}
    \label{fig:kinematics_examples}
\end{figure*}

In this way, we fit our data cubes with $^{\rm 3D}$Barolo. We obtain satisfactory models for all 22 galaxies, closely resembling the observations. Fig.~\ref{fig:kinematics_examples} shows the kinematics (velocity field, PV slices for data and best-fitting model, and kinematic radial profiles, see below) for four representative galaxies; similar plots for the remaining galaxies can be found in Fig.~\ref{fig:kinematics_rest}. 

As expected by our sample selection, our dwarfs show ordered kinematics. On top of the velocity fields in Figs.~\ref{fig:kinematics_examples} and \ref{fig:kinematics_rest}, we show the best-fitting (white circles) and average (black dashed line) kinematic position angle for our galaxies. As can be seen, most galaxies have constant position angles (so the white circles and the black dashed line overlap), with only a few showing compelling position angle warps (but still ordered rotation; e.g. DDO 181, LVHIS 009, LVHIS 077, NGC 2541, NGC 4190). The only galaxies with a warp in inclination are LVHIS 0012, LVHIS 078, NGC 0925, NGC 2541, and NGC 7793, albeit their effect is minor since the changes are of the order of $5^\circ-10^\circ$. Three galaxies (LVHIS 019, LVHIS 055, and LVHIS 078) also show some evidence of radial motions (as seen in their X-shaped minor-axis PVs). However, the amplitudes of the radial velocities are negligible compared to the circular motions. 

The final step before going into the mass modelling is to compute the circular speed ($V_{\rm circ}$) of the galaxies, i.e. correcting the observed $V_{\rm rot}$ for pressure-supported motions with the asymmetric drift correction (e.g. \citealt{binney,iorio,bookFilippo,paperIBFR}). For this, we follow the procedure detailed in \cite{iorio} and refer the reader to that work (see their Sec. 4.3) for details. To show the effect that the asymmetric drift correction has on our data, we show in the right panels of Figs.~\ref{fig:kinematics_examples} and \ref{fig:kinematics_rest} the $V_{\rm rot}$, $\sigma_{_{\rm HI}}$, and $V_{\rm circ}$ profiles for the 22 galaxies with kinematics derived in this work. We make our kinematic profiles publicly available under \href{https://www.dropbox.com/scl/fo/oc1tai7t4f4vp6f7iivls/AJOFSqUDfltJ9x3DNsSQSzE?rlkey=9iocjnkbzn4zfydy2p6effito&st=j2mcmbfi&dl=0}{this link}.  As shown below, these kinematic measurements can be used to test dynamical and gravity theories in the dwarf-galaxy regime. As mentioned before, we take the public kinematic measurements from the remaining galaxies from \cite{iorio} and \cite{enrico_radialmotions}. With this, we have all the kinematic information for our entire sample, which is used in the following section during our rotation curve decomposition.

\section{Building the mass models}
\label{sec:massmodels}

The technique of obtaining mass models through rotation curve decomposition consists of reproducing the observed circular speed of galaxies considering the gravitational potential provided by baryons and dark matter, i.e. 
\begin{equation}
\label{eq:massmodel}
    V_{\rm circ}^2 = \Upsilon_{\rm d} V_{\rm d}|V_{\rm d}| + \Upsilon_{\rm b} V_{\rm b}|V_{\rm b}| + V_{\rm HI}|V_{\rm HI}| + V_{\rm H_2}|V_{\rm {H_2}}| + V_{\rm DM}|V_{\rm DM}|~,
\end{equation}
where $V_{\rm d}$, $V_{\rm b}$, $V_{\rm{HI}}$, $V_{\rm{H_2}}$ and $V_{\rm DM}$ are the contributions provided by the stellar disc (up to its mass-to-light ratio), stellar bulge (up to its mass-to-light ratio), H\,{\sc i} disc, H$_2$ disc and dark matter halo, respectively. The parameters $\Upsilon_{\rm d}$ and $\Upsilon_{\rm b}$ are the disc and bulge stellar mass-to-light ratios, normalising $V_{\rm d}$, and $V_{\rm b}$, respectively. Mass modelling assumes dynamical equilibrium (e.g. \citealt{binney}) such that the rotational motions trace the gravitational potential (for possible caveats see \citealt{downing2023}). Our mass modelling technique follows closely \cite{paper_massmodels}. In the following sections, we provide an overview of our main steps. We note that for galaxies without bulges (molecular gas), $V_{\rm b}\ (V_{\rm{H_2}}) = 0$.

\subsection{The baryonic contribution}
\label{sec:massmodels_bar}
We determine $V_{\rm d}$, $V_{\rm b}$, and $V_{\rm HI}$, and $V_{\rm H_2}$ using the software \textsc{galpynamics}\footnote{\url{https://gitlab.com/iogiul/galpynamics/}}, which computes numerically the gravitational potential of a given mass density profile. In our case, we determine the gravitational potential of the stellar disc and bulge (modulo $\Upsilon$) and the gas components, together with their vertical distributions. 
\textsc{galpynamics} takes as input a continuous function for its numerical integration (see \citealt{cuddeford1993,iorio_phd,paper_massmodels}), so we fit the observed stellar and gas radial profiles with one of the following functional forms:

\begin{equation}
\label{eq:Frat_disc}
    \Sigma(R) = \Sigma_{0}~e^{-R/{\rm R_{1}}} (1+R/{\rm R_{2}})^\alpha~,
\end{equation}

\begin{equation}
\label{eq:sersic}
\Sigma(R) = \Sigma_{\rm e} \exp \left\{ -b(n) \left[ \left( \frac{R}{R_{\rm e}} \right)^{\frac{1}{n}} - 1 \right] \right\}~,
\end{equation}

\begin{equation}
\label{eq:polyexp}
    \Sigma(R) = \Sigma_{\rm 0,pex}~e^{-R/R_{\rm pex}}~(1 + c_1 R + c_2 R^2 + c_3 R^3 + ... + c_n R^n)~.
\end{equation}

\noindent
We used Eq.~\ref{eq:Frat_disc} (e.g. \citealt{Tom891}) to fit most of our H\,{\sc i} profiles, as it mimics the typical behaviour of H\,{\sc i} discs having a plateau or sink in their centre to then decay exponentially. Eq.~\ref{eq:sersic} is the well-known Sérsic profile \citep{sersic}, which we use to fit most of the stellar surface profiles (for the massive galaxies \cite{salo2015} provides directly the bulge\footnote{We note that we assume our bulges to be spherical and to have a projected 2D radial density that follows the given Sérsic profiles.} and disc Sérsic parameters). Finally, Eq.~\ref{eq:polyexp} is a `poly-exponential' disc, whose flexibility we use to fit the H$_2$ profiles as well as those H\,{\sc i} and stellar profiles with more complex behaviour. For the galaxies from \cite{enrico_radialmotions} and \cite{iorio} we take the coefficients ([$\Sigma_0$, $R_1$, $R_2$, $\alpha$], [$\Sigma_{\rm e}$, $n$, $R_{\rm e}$], and [$\Sigma_{\rm 0,pex}$, $R_{\rm pex}$, $c_1$, $c_2$, $c_3$, $c_4$]) from \cite{paper_massmodels}. For the remaining 22 dwarfs, we fit the different free parameters using the MCMC fitting software \texttt{emcee} \citep{emcee}, adopting uniform priors. 

Next, we specify the vertical distribution of the stellar and gas discs. For the former, we assume a sech$^2$ profile with a constant thickness $z_{\mathrm{d}}/\rm{kpc} = 0.196\, (\mathit{R}_{\rm d}/\rm{kpc})^{0.633}$ (with $R_{\rm d}$ the scale length of $\Sigma_\ast$), as typically found in star-forming galaxies \citep{bershady_thickness}. For the gas discs, which are known to flare with radius (e.g. \citealt{romeo1992,olling1995,karlberla2008,marasco2011,yim2014,patra_HIdwarfs,jeffreson2022}), we assume a Gaussian vertical profile and determine their flaring at the same time as deriving the mass models (see below) by exploiting the balance between the gravitational potential and the gas velocity dispersion in disc systems in vertical hydrostatic equilibrium (see \citealt{romeo1992,iorio_phd,ceciVSFL,paper_massmodels}).

The stellar mass-to-light ratios are the final ingredients to specify the baryonic contribution to the total $V_{\rm circ}$. We treat $\Upsilon_{\rm d}$ and $\Upsilon_{\rm b}$ as free parameters but take advantage of well-known priors on NIR mass-to-light ratios, as we discuss in Sec.~\ref{sec:themethod}.

\subsection{The dark matter contribution}
To determine the gravitational contribution from the dark matter haloes, we parametrise their density with functional forms. The Navarro-Frenk-White (NFW, \citealt{nfw}) profile is a commonly used halo profile. The NFW haloes have a density given by 
\begin{equation}
    \rho_{\rm NFW}(r) = \dfrac{4\,\rho_{\rm s}}{(r/r_{\rm s})\,(1 + r/r_{\rm s})^2}~,
\end{equation}
with $r~=~\sqrt{R^2+z^2}$ the spherical radius, $r_{\rm s}$ a scale radius (for an NFW, $r_{\rm s}=r_{-2}$, with $r_{-2}$ the radius at which the log-slope of the profile equals $-2$), and $\rho_{\rm s}$ the volume density at $r_{\rm s}$. 

The enclosed mass of the profile is given by
\begin{multline}
    M_{\rm NFW}(<r) = \dfrac{M_{200}}{\ln(1+c_{200}) - \dfrac{c_{200}}{1+c_{200}}} \\ 
    \times\, \left[\ln\left( 1 + \dfrac{r}{r_{\rm s}}\right) - \dfrac{r}{r_{\rm s}} \left( 1 + \dfrac{r}{r_{\rm s}}\right)^{-1} \right]~,
\end{multline}
where $M_{200}$ is the halo mass within the radius $R_{200}$ (where the average density is 200 times the critical density of the universe, $M_{200}/R_{200}^3=(4\,\pi/3)\,200\,\rho_{\rm crit(\mathit{z}=0)}$), and $c_{200} \equiv R_{200}/r_{\rm -2} = R_{200}/r_{\rm s}$ is the concentration parameter. 

However, different studies suggest that galaxies (especially dwarfs) can have dark matter profiles that are cored (e.g. \citealt{moore1994,burkert1995,deblok08,oman2015,bullock2017,salucci2019,sales_review_dwarfs,collins_feedback}, but see also \citealt{roeper2023}) for which the cuspy NFW is not appropriate, and different profiles with flatter inner dark matter densities have been proposed (e.g. \citealt{begeman,burkert1995,coreNFW,freundlich2020,coreEinasto}). In this work, we use the \textsc{coreNFW} profile \citep{coreNFW,readAD}, which gives extra freedom to the NFW profile to develop a core.

The \textsc{coreNFW} halo has the density profile
\begin{equation}
\label{eq:corenfw_1}
    \rho_{\rm\textsc{coreNFW}}(r) = f^n\, \rho_{\rm NFW}(r) + \dfrac{n\,f^{n-1}\,(1-f^2)}{4\,\pi\, r^2\, r_{\rm c}} M_{\rm NFW}(r)~.
\end{equation}
In this equation, $\rho_{\rm NFW}$ and $M_{\rm NFW}$ represent the density and mass for a NFW halo, $f=\tanh(r/r_{\rm c})$ is a function that generates a core of size $r_{\rm c}$, and $n$ is a parameter that regulates the cusp-core transition ($n=0$ reduces to the NFW halo, while $n=1$ produces a completely cored profile). 

It follows then that the \textsc{coreNFW} has originally four free parameters $M_{200}$, $c_{200}$, $n$, and $r_{\rm c}$. Based on simulations, \citet{coreNFW,readAD,read2017} proposed calibrations for $n$ and $r_{\rm c}$. For $n$, those authors propose $n = \tanh(\kappa\, t_{\rm SF} / t_{\rm dyn})$, with $\kappa = 0.04$, $t_{\rm SF}$ the time whilst the galaxy has been forming stars (assumed to be 14~Gyr), and $t_{\rm dyn}$ the NFW dynamical time at $r_{\rm s}$. For $r_{\rm c}$, they propose $r_{\rm c} = \eta\,R_{\rm e}$, with $\eta\approx1.75$ and $R_{\rm e}$ the stellar half-light radius. 
However, this work treats $n$ and $r_{\rm c}$ differently. To avoid relying entirely on the results from the simulations, we assume $r_{\rm c} = \eta\,R_{\rm e}$ but treat $\eta$ as a free parameter (with priors defined below).
Moreover, we fix $n=1$ to ensure that $r_{\rm c}$ is unambiguously defined and matches the inner radius where the \textsc{coreNFW} visually departs from the NFW. We note that fixing $n=1$ does not preclude haloes from being cuspy: If the data prefer the NFW profile, then the fit can do so with $r_{\rm c} \rightarrow 0$. Therefore, the free parameters of our \textsc{coreNFW} haloes are $M_{200}$, $c_{200}$, and $\eta$. With $\Upsilon_{\rm d}$, $\Upsilon_{\rm b}$, and the vertical scale height of the gas discs, these parameters were later determined with our rotation curve decomposition. The next section explains the fitting procedure \citep{paper_massmodels}.

\subsection{Self-consistent mass models with gas disc flaring}
\label{sec:themethod}

The free parameters in our mass models ($\log(\Upsilon_{\rm d})$, $\log(\Upsilon_{\rm b})$, $\log(M_{200})$, $\log(c_{200})$, and $\log(\eta)$) are found through a Bayesian Monte Carlo approach. Specifically, we use the software \texttt{dynesty} \citep{dynesty} to efficiently estimate their posterior distributions and evidence through nested sampling. This way, we can explore the full parameter space and build trial mass models to minimise Eq.~\ref{eq:massmodel}. For the minimisation we use a likelihood of the form $\exp(-0.5 \chi^2)$, with $\chi^2 = (V_{\rm circ}-V_{\rm circ,mod})^2 / \delta_{V_{\rm circ}}^2$. Here $V_{\rm circ}$ and $V_{\rm circ,mod}$ are the observed and model circular speed profiles, and $\delta_{V_{\rm circ}}$ the uncertainty in $V_{\rm circ}$.

The exploration of the parameter space follows a set of priors. For the sampling parameter $\log(\Upsilon_{\rm d})$ we use Gaussian priors on $\Upsilon_{\rm d}$ (so lognormal priors on $\log(\Upsilon_{\rm d})$). The priors are centred in the empirical luminosity--$\Upsilon_{\rm d}$ relations described in Appendix~\ref{app:m2l} and are motivated by model $\Upsilon_{\rm d}$ values obtained through SED fitting in some of our galaxies, following the procedure described in \cite{marasco_mstar}. Moreover, the priors also agree with stellar population models (SPMs) and dynamical estimates (e.g. \citealt{mcgaugh_ML, meidt2014, querejeta2015, marasco_mstar}), as discussed in Appendix~\ref{app:m2l}.
For $\Upsilon_{\rm b}$, SPMs in the NIR suggest $\Upsilon_{\rm b} \approx 1.2-1.6\,\Upsilon_{\rm d}$ \citep{schombert2022}. Therefore, we assumed the flat prior $\Upsilon_{\rm d} < \Upsilon_{\rm b} < 2\,\Upsilon_{\rm d}$.

For the halo mass, we explore a flat prior bounded within $6 < \log(M_{200}/M_\odot) < 14$. In turn, for $\log(c_{200})$ we use a Gaussian prior centred on the $c_{200}-M_{200}$ relation of \citet[][hereafter \citetalias{diemer2019}]{diemer2019}, for which we assume a $1\,\sigma$ standard deviation (in log space) of 0.16~dex \citep{diemer2015}. This prior is physically justified since the $c_{200}-M_{200}$ relation is inherent to structure formation in the $\Lambda$CDM framework (e.g. \citealt{bullock2001, ludlow2014,correa2015}). In particular, the relation from \citetalias{diemer2019} is adequate for galaxies with masses such as those in our sample (\citetalias{diemer2019}; \citealt{wang2020}).
As discussed in \cite{paper_massmodels}, for some galaxies the data has enough sampling and accuracy to constrain $\log(c_{200})$ even with a flat prior (typically recovering the $c_{200}-M_{200}$ relation within some scatter), but for some cases the concentration remains unconstrained, which motivates us to use our Gaussian prior (a common practice; e.g. \citealt{postinomissing,enrico_massmodels_ss}). 

Finally, for $\eta$, we adopt the flat prior $\log(0.1) \leq \log(\eta) \leq \log(3.75)$. The lower bound is small enough to allow for the recovery of cuspy profiles if preferred by the data; the upper bound comes from considerations on supernovae energy, which is not strong enough to create cores larger than $\sim 3.75\,R_{\rm e,\ast}$ (\citealt{coreNFW, read2017, benitezllambay19, coreEinasto}) for realistic supernovae coupling efficiencies ($\sim1$\%; e.g. \citealt{coreNFW, ceci_turbulence}). With this, all our priors and fitting parameters have been introduced.\\

\noindent
Next, we detail how to self-consistently determine the gas disc thickness and the best-fitting dark matter halo. As introduced in \cite{paper_massmodels}, we achieve this through an iterative process based on \textsc{galpynamics} and our nested sampling Monte Carlo routine. The steps are as follows.

\begin{enumerate}
    \item Assuming $\Upsilon_{\rm d}=1$ and $\Upsilon_{\rm b}=1$ (if a bulge is present), we use \textsc{galpynamics} (Sec.~\ref{sec:massmodels_bar}) to compute a preliminary gravitational potential for the stars ($\Phi_\ast = \Phi_{\rm d} + \Phi_{\rm b}$) and its circular speed $V_\ast$. We assume the H\,{\sc i} and H$_2$ (if a molecular gas disc is present) discs to be razor-thin and obtain a first estimate of $\Phi_{\rm HI}$, $\Phi_{\rm H_2}$, and their corresponding circular speeds. 
    
    \item With our initial circular speeds we fit Eq.~\ref{eq:massmodel} to obtain a preliminary ($\log(\Upsilon_{\rm d})$, $\log(\Upsilon_{\rm b})$, $\log(M_{200})$, $\log(c_{200})$, $\log(\eta)$) set. With this, we update $\Phi_\ast$ and obtain $\Phi_{\rm DM}$ and the corresponding $V_\ast$ and $V_{\rm DM}$. 
    
    \item With \textsc{galpynamics} and relying on the condition of vertical hydrostatic equilibrium, we computed the thickness of the gas disc taking into account $\Phi_{\rm DM}$, $\Phi_\ast$, $\Phi_{\rm H_2}$, and the H\,{\sc i} disc self-gravity\footnote{
In practice, solving numerically the equation 
\begin{equation}
\label{eq:rho_h}
    \rho_{\rm_{HI}}(R,z) = \rho_{\rm_{HI}}(R,0)\exp\left[ - \dfrac{\Phi_{\rm tot}(R,z) - \Phi_{\rm tot}(R,0)}{\sigma_{\rm_{HI}}^2(R)} \right]~,
\end{equation}
with $\rho_{\rm_{HI}}(R,z)$ and $\Phi_{\rm tot}(R,z)$ the gas volume density and total galactic potential evaluated at a given vertical distance $z$ from the midplane. For details see \cite{iorio_phd,ceciVSFL,paper_massmodels,ceci_q3d}.}. This allowed us to obtain a new estimate of $\Phi_{\rm HI}$ and $V_{\rm HI}$ for our flared H\,{\sc i} disc.

    \item \textsc{galpynamics} computes the H$_2$ taking into account $\Phi_{\rm DM}$, $\Phi_\ast$, and $\Phi_{\rm HI}$ (derived in the previous step), and the H$_2$ disc self-gravity.

    \item Steps 3 and 4 are repeated iteratively until the changes in both scale heights at all radii are at least smaller than 15\% (an appropriate threshold given the observational uncertainties, see \citealt{paper_massmodels}) compared to the previous iteration. Once convergence is achieved, we obtain the thickness radial profile for H\,{\sc i} and H$_2$. The thickness updates the gravitational potentials $\Phi_{\rm HI}$ and $\Phi_{\rm H_2}$ and the corresponding circular speeds.
    
    \item With the new total gravitational potential, our \texttt{dynesty} routine finds a new best-fitting ($\log(\Upsilon_{\rm d})$, $\log(\Upsilon_{\rm b})$, $\log(M_{200})$, $\log(c_{200})$, $\log(\eta)$) set, updating the mass model and generating a new set of potentials $\Phi_\ast$, $\Phi_{\rm gas}$, $\Phi_{\rm DM}$ (and their corresponding circular speeds).
    
    \item Steps 3 to 6 are repeated iteratively until all free parameters converge. In practice, we iterate until the changes in the parameters between the last and penultimate iterations are less than 3\%. For most galaxies, three iterations are enough to reach convergence. 
\end{enumerate}

\noindent
Following this iterative procedure, we simultaneously derive the scale height of the gas discs, the stellar mass-to-light ratios, and the dark matter halo parameters for our galaxy sample in a self-consistent way. In the following sections, we present the results of our mass models.

\section{Mass models}
\label{sec:massmodels_results}

\subsection{Mass models}
\begin{figure*}
    \centering 
    \includegraphics[width=1\textwidth]{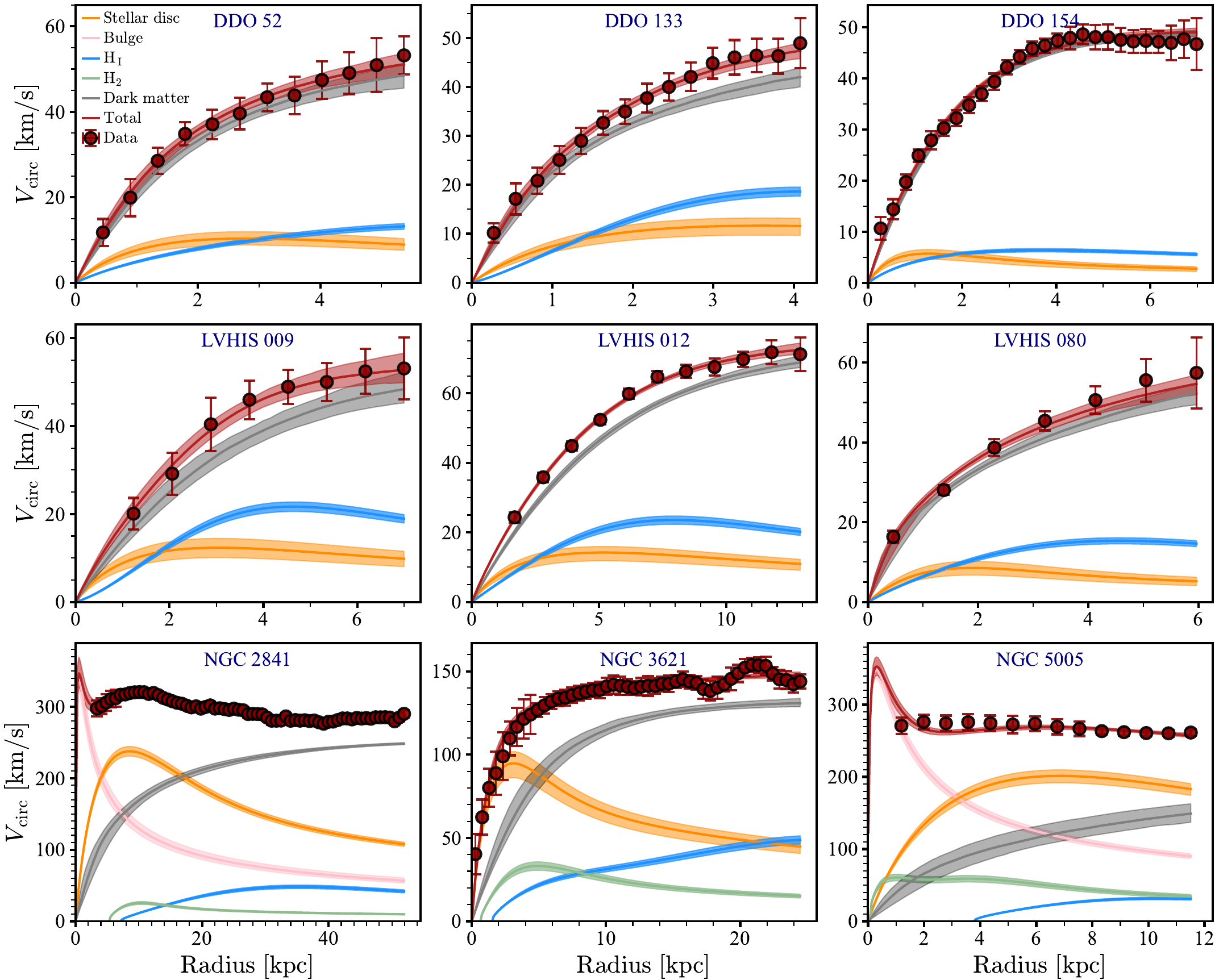}
    \caption{Mass models of nine representative galaxies in our sample (similar plots for our full sample and the corresponding posterior distributions are available in \href{https://www.dropbox.com/scl/fo/oc1tai7t4f4vp6f7iivls/AJOFSqUDfltJ9x3DNsSQSzE?rlkey=9iocjnkbzn4zfydy2p6effito&st=j2mcmbfi&dl=0}{this link}). In each panel, the observed circular speed profiles are shown with dark red circular markers. The colour lines and bands represent the gravitational contribution from the stellar disc (orange), bulge (pink, if present), H\,{\sc i} (blue), H$_2$ (green, if present) and dark matter (grey) to the total mass model (red) and their corresponding $1\,\sigma$ confidence bands. }
    \label{fig:massmodels_examples}
\end{figure*}

Examples of our resulting mass models for nine representative galaxies are shown in Fig.~\ref{fig:massmodels_examples}. The remaining mass models and all the corresponding corner plots of the posterior distributions are available at \href{https://www.dropbox.com/scl/fo/oc1tai7t4f4vp6f7iivls/AJOFSqUDfltJ9x3DNsSQSzE?rlkey=9iocjnkbzn4zfydy2p6effito&st=j2mcmbfi&dl=0}{this link}. In Appendix~\ref{sec:comparison}, we discuss similarities and discrepancies between our mass models and literature values. Additionally, in Appendix~\ref{app:bars}, we discuss the potential (negligible) effect of bars in our results.
 
We found satisfactory fits for all of our 49 galaxies. In particular, $\log(M_{200})$, $\log(c_{200})$, $\Upsilon_{\rm d}$ and $\log(\Upsilon_{\rm b})$ are always well constrained. We note that except for two galaxies (WLM and IC 2574) $r_{\rm c}<r_{\rm s}$.
The posterior of $\log(\eta)$ is well-constrained for most galaxies but not all. For one-fifth of our sample, at the low-mass regime, $\eta$ tends to go to the upper limit allowed by our priors, i.e. the data would prefer larger cores. At the high-mass end, some galaxies (but not all) prefer cuspy haloes, so $\eta$ tends to go to the lower bound of the prior. We will further discuss the core size distribution of our sample in an upcoming paper (Mancera Piña et al. in prep); we note, however, that within our prior, $\eta$ shows no strong degeneracies with $M_{200}$, $c_{200}$ or the mass-to-light ratios, so the results we present below are valid and robust. Similarly, thanks to our MC approach, the uncertainties in all the other parameters account for the behaviour of $\eta$ since we marginalise over it.

There is a final aspect regarding our mass models to be considered. One of the main advantages of our sample is that all the galaxies have accurate distance measurements ($D$), which reduces the associated impact that distance uncertainties ($\delta_D$) can have on the masses ($\propto D^2$) and radii ($\propto D$). One possibility to incorporate the effect of $\delta_D$ is to include $D$ as a (nuisance) free parameter in the fit (e.g. \citealt{li_massmodels,agc114905}). However, this approach is prohibitively time-consuming for us since, for each single trial $D$ within our MC routine, the galactic potentials would need to be updated together with the scale height (see Sec.~\ref{sec:themethod}), which is computationally expensive for the thousands of iterations taken to explore the parameter space of our mass models. Instead, we perform the following exercise to examine the impact of $\delta_D$. We obtain a new set of mass models considering $D+\delta_D$ and $D-\delta_D$ and look at their posterior distributions. Upon inspection, we find those posteriors are essentially indistinguishable from our fiducial posteriors (deriving assuming $D$), and the $50$th, $16$th, and $84$th percentiles remain the same. Therefore, we conclude that thanks to our selection of galaxies with accurate distance measurements, we can ignore the impact of $\delta_D$ in our mass models since the dominant source for uncertainty is the kinematic measurements.

In Table~\ref{tab:mm}, we report our best-fitting (50th percentiles) parameters ($\log(M_{200})$, $\log(c_{200})$, $\log(\eta)$, $\log(\Upsilon_{\rm d})$, and $\log(\Upsilon_{\rm b})$) and their corresponding uncertainties, which correspond to the absolute difference between the $50$th percentiles (median) and the $16$th, and $84$th percentiles of our posterior distributions. In Sec.~\ref{sec:scalinglaws}, we delve into our findings regarding the dark matter parameters for our galaxy sample. However, before that, in the following sections, we discuss two interesting by-products of our mass modelling, namely the resulting mass-to-light ratios and gas scale heights.

\subsection{Stellar mass-to-light ratios}
During our rotation curve decomposition, we fitted $\log(\Upsilon_{\rm d})$ with wavelength-dependent priors determined by empirical luminosity--$\Upsilon_{\rm d}$ relations informed by SED fitting and also supported by results from SPMs and dynamical models (see Appendix~\ref{app:m2l}). The prior for $\Upsilon_{\rm b}$ was flat but bounded within $\Upsilon_{\rm d} < \Upsilon_{\rm b} < 2\,\Upsilon_{\rm d}$, as suggested from SPMs.

\begin{figure}
    \centering
    \includegraphics[width=1\linewidth]{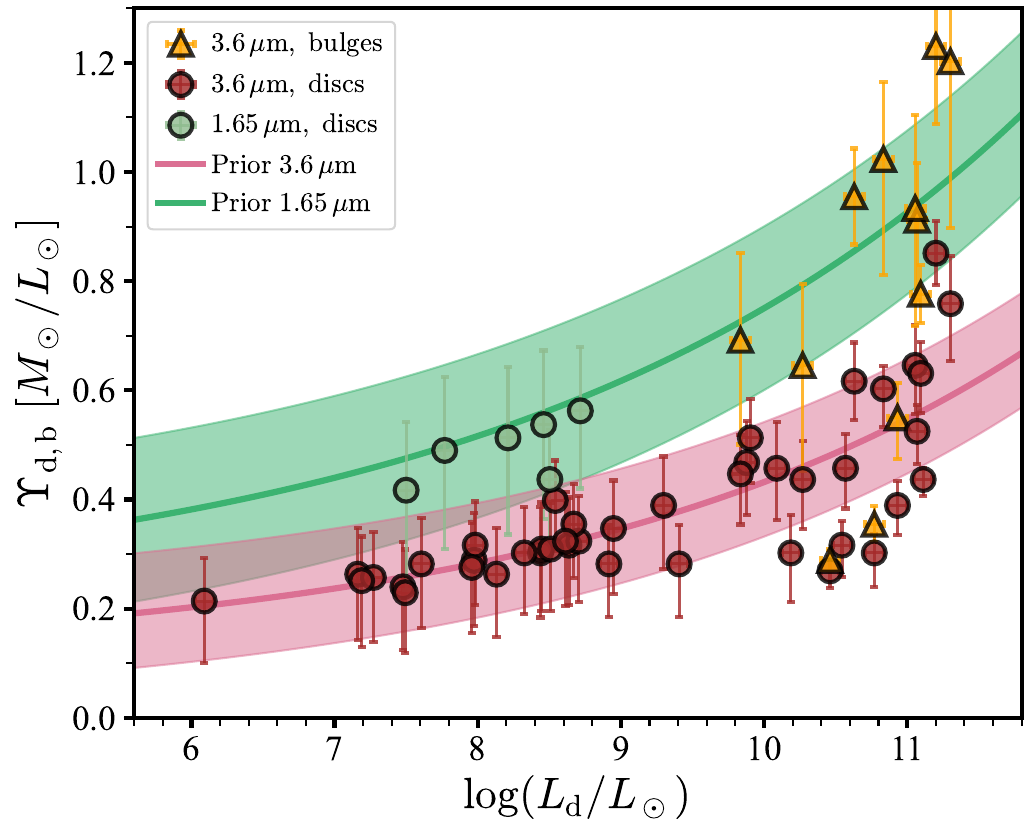}
    \caption{Our $\Upsilon_{\rm d}$ (circles) and $\Upsilon_{\rm b}$ (triangles) values as a function of disc luminosity ($L_{\rm d}$). We distinguish between galaxies with photometry at 1.65$\,\mu$m and 3.6$\,\mu$m. The lines and bands represent the Gaussian priors adopted for $\Upsilon_{\rm d}$ in our mass modelling (see Sec.~\ref{sec:massmodels} and Appendix~\ref{app:m2l}). The prior for $\Upsilon_{\rm b}$ was flat within $\Upsilon_{\rm d} < \Upsilon_{\rm b} < 2\,\Upsilon_{\rm d}$ (not shown).} 
    \label{fig:m2l}
\end{figure}

Fig.~\ref{fig:m2l} shows the $\Upsilon_{\rm d}$ and $\Upsilon_{\rm b}$ values obtained through our rotation curve decomposition as a function of their disc luminosity and distinguishing between galaxies with 3.6 $\mu$m and 1.65 $\mu$m data. The Gaussian priors imposed during our fit are also depicted. 
Focusing first on the galaxies with 3.6 $\mu$m photometry, we see that for $L^{3.6\,\mu m}_{\rm d} \gtrsim 10^{10}\,L_\odot$ the $\Upsilon_{\rm d}$ values can deviate from the centre of their prior and show a trend of less luminous galaxies having lower $\Upsilon_{\rm d}$, as reported in the literature (e.g. \citealt{marasco_mstar}). Instead, galaxies of lower $L^{3.6\,\mu m}$  move little from the central value of their prior. 

On the one hand, this tells us that the chosen priors do not disagree with the kinematics of the galaxies. On the other hand, the little scatter at low luminosities could also indicate little constraining power on $\Upsilon_{\rm d}$, a consequence of the dwarfs (unlike the massive spirals) being heavily dominated by the dark matter and gas components in their central parts (see Fig.~\ref{fig:massmodels_examples}). \cite{paper_massmodels} showed that this is also the case when imposing $\Upsilon_{\rm d}^{3.6}/[M_\odot/L_\odot] = 0.5\pm0.1$ as a prior, and argued that the uncertainties in their $\Upsilon_{\rm d}^{3.6}$ values for dwarfs were likely underestimated. This is true to some extent since the constraining power is not strong, but we emphasise that our $\Upsilon_{\rm d}^{3.6}$ and uncertainties are derived using a prior informed by SED fitting, ensuring they remain physically realistic. In Fig.~\ref{fig:m2l}, we also include the $\Upsilon_{\rm b}$ values for the 12 galaxies in our sample with bulges. We can see a trend of galaxies with increasing disc luminosity having larger $\Upsilon_{\rm b}$. The values of the $\Upsilon_{\rm b}/\Upsilon_{\rm d}$ ratio range between 1 and 1.75, with a mean (median) value of 1.45 (1.48).

For the six galaxies with 1.65$\,\mu$m photometry, their $\Upsilon_{\rm d}^{1.65}$ values are consistent within $1\,\sigma$ with their prior. \cite{kirby08} studied the $\Upsilon_{\rm d}^{1.65}$ in a sample of low-mass galaxies (including some of our galaxies) and based on the observed $B-H$ colour and the $\Upsilon_{\rm d}$--colour relation by \cite{bell2001_M2L} estimated $\Upsilon_{\rm d} = 0.9\pm0.6$. As also pointed out by \cite{kirby08}, \cite{bell2003} reports values in the range 0.7--1.3 for SDSS blue galaxies (somewhat more massive than ours) using $g-r$ colours. Our dynamical estimates broadly agree with the lower bounds of the above studies, but we see no conspicuous dynamical evidence for $\Upsilon_{\rm d}^{1.65} \gtrsim 0.8\,M_\odot/L_\odot$ for galaxies with $L^{1.65\,\mu m}_{\rm d} \lesssim 10^{9}\,L_\odot$. 

\subsection{Gas scale heights}

One of the most innovative aspects of our mass modelling, already introduced in \cite{paper_massmodels}, is that the gas scale height is derived simultaneously and consistently with the dark matter halo parameters. Studying the scale heights is vital to understanding the connection between the interstellar medium (ISM) kinematics and processes such as disc stability and star formation (see e.g. \citealt{romeo1992,romeo2013,iorio_phd,ceciVSFL,jeffreson2022,ceci_q3d}). Furthermore, the gravitational potential of a thick disc is weaker in the mid-plane than that of a thin disc with the same mass (e.g. \citealt{binney1977,paper_massmodels}), and so is their contribution to the circular speed, so incorporating gas flaring into the picture leads to a more robust determination of the halo parameters.

\subsubsection{Scale heights}

\begin{figure}
    \centering
    \includegraphics[width=1\linewidth]{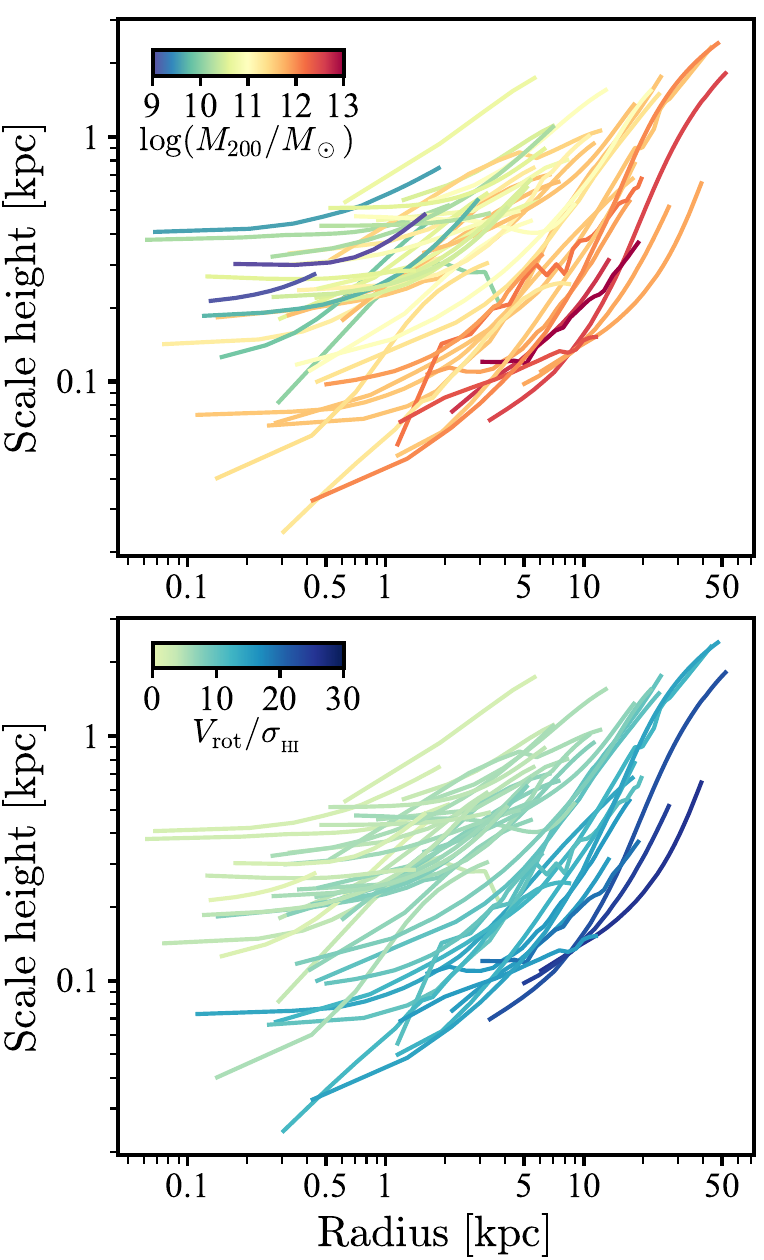}
    \caption{Scale heights of the H\,{\sc i} for our sample. The \emph{top} panel colour-codes each galaxy depending on its halo mass, while the \emph{bottom} panel uses the $V_{\rm rot}/\sigma_{\rm_{HI}}$ ratio.} \vspace{-0.4cm}
    \label{fig:scaleheights}
\end{figure}

In Fig.~\ref{fig:scaleheights}, we show the H\,{\sc i} scale heights for our sample (H$_2$ scale heights are less thick but otherwise exhibit a similar behaviour, see \citealt{paper_massmodels}). To better compare the different profiles, we plot them in log-log scale, but linear versions of each scale height can be found as in \href{https://www.dropbox.com/scl/fo/oc1tai7t4f4vp6f7iivls/AJOFSqUDfltJ9x3DNsSQSzE?rlkey=9iocjnkbzn4zfydy2p6effito&st=j2mcmbfi&dl=0}{this link}, together with ASCII tables containing the scale heights.

The interplay between the galactic gravitational potential and the gas pressure results in scale heights of increasing flare with increasing radii, as usually found in nearby galaxies (e.g. \citealt{kerr1957,renzo1979,olling1996a, romeo1992, nakanishi2003,yim2014,marasco2011,ceciVSFL,elmegreen2025}). This increase is typically from scales of $\sim0.1$ kpc in galaxy centres to $1-10$ kpc in the outskirts. The one exception in our galaxy sample is LVHIS 055, which has a scale height that rises from the centre to $R\sim2$ kpc to bend downwards. As discussed by \cite{ceci_q3d}, this happens for some gas-rich dwarfs (and some high-redshift galaxies) for which $\rho_{\rm_{HI}}(R,0)$ in Eq.~\ref{eq:rho_h} bends down due to $\sigma_{\rm_{HI}}^2$ decreasing faster with radius than the difference between $\Phi_{\rm tot}$ evaluated in the midplane and above the midplane (i.e. gas pressure cannot counteract the gravity from the potential).

It has been attempted to find a universal profile for the gas flaring (e.g. \citealt{patra_HIspirals,patra_HIdwarfs,paper_massmodels}), we conclude that the shapes remain too diverse to reconcile them with a universal profile (despite trying different normalisations, see Sec. 4.2 in \citealt{paper_massmodels}). Nevertheless, some general trends can be appreciated. The top and bottom panels in Fig.~\ref{fig:scaleheights} colour-code the scale heights depending on the galaxies' halo mass and the rotation-to-dispersion ratio $V_{\rm rot}/\sigma_{\rm_{HI}}$, respectively. Here, $V_{\rm rot}/\sigma_{\rm_{HI}}$ is computed as the ratio between the characteristic outer rotational speed (as in Table~\ref{tab:sample} but for $V_{\rm rot}$) and the median gas velocity dispersion.

Focusing on the top panel, we can see that, in absolute terms, dwarf galaxies usually have thicker H\,{\sc i} discs than massive spiral galaxies; however, the scatter at a given halo mass remains significant, driven by different $\sigma_{\rm_{HI}}$, mass-to-light ratios, and the presence of bulges in some of the massive spirals. The scatter is largely reduced when looking at the scale heights at fixed $V_{\rm rot}/\sigma_{\rm_{HI}}$ ratio (bottom panel in Fig.~\ref{fig:scaleheights}). Evidently, galaxies with lower $V_{\rm rot}/\sigma_{\rm_{HI}}$ values have thicker H\,{\sc i} scale heights. This trend is consistent with the expectations (e.g. Appendix A in \citealt{ceciVSFL}) since $V_{\rm rot}$ correlates with the galactic potential (pushing the scale heights towards the midplane) and $\sigma_{\rm_{HI}}$ with the gas pressure (counteracting the effect of gravity). These results are also in qualitative agreement with \cite{randriamampandry2021}, who reported that gas-rich galaxies typically have thicker H\,{\sc i} layers within their optical discs.

The scale heights here derived (and which we make publicly available) are relevant due to their impact on mass models (see the following section). Moreover, they can be used to study the processes driving star formation and turbulence in the ISM (e.g. \citealt{ceciVSFL,ceci_turbulence}). Additionally, they can also set constraints to discriminate between baryonic physics implementations in hydrodynamical simulations (e.g. \citealt{gensior2024}).

\subsubsection{Dynamical impact on the mass models}
\label{sec:impact}
We finish this section by touching briefly upon the gas flaring's effect on our mass models. As discussed in \cite{paper_massmodels}, the flaring has a negligible dynamical effect (i.e. the mass models with the flaring are almost identical to those using razor-thin discs) for \emph{massive} galaxies. There is a tendency for the flaring to allow higher halo masses (the flaring lowers the gravitational potential of the gas in the mid-plane, leaving more room for the dark matter), but usually the change ($\sim0.1$ dex) is within the uncertainties. 

The above holds truth for our sample. However, as in \cite{paper_massmodels}, we note that the dynamical effect of the flaring can severely impact the smallest and most gas-rich galaxies. Specifically, there is a trend of increasing differences in the recovery of halo mass for decreasing halo mass, stellar mass, and $V_{\rm rot}/\sigma_{_{\rm HI}}$. The impact becomes noticeable for $M_{200}\lesssim 10^{10}\,M_\odot$, $M_\ast \lesssim10^7\,M_\odot$, and $V_{\rm rot}/\sigma_{_{\rm HI}} \lesssim 3$. We refrain from doing a more in-depth quantification since a larger sample at those regimes is needed, but we raise caution on using the razor-thin approximation for such low-mass galaxies. 

For example, we highlight our sample's four most extreme cases. For the galaxies DDO 181, DDO 183, DDO 210, and CVn I dwA, the halo masses considering the gas flaring compared to the razor-thin case increase by a factor of 1.4, 1.4, 1.6, and 6.8, respectively. These four galaxies are also those with the lowest $V_{\rm circ}$ in our entire sample ($V_{\rm circ} \lesssim 30\ \rm{km/s}$) and are among the six galaxies with the lowest stellar mass and rotation-to-dispersion ratios ($M_\ast < 10^7\,M_\odot$ and $V_{\rm rot}/\sigma_{_{\rm HI}} \lesssim 3$).
These results highlight the importance of considering the effects of gas disc thickness and flaring when deriving the mass models of the smallest gas-rich dwarf galaxies. Upcoming surveys should consider such effects to provide robust measurements of dark matter halo properties at the above scales.

\section{Galaxy-halo connection}
\label{sec:scalinglaws}

Here we use the results of our rotation curve decomposition to study different dynamical scaling relations. These scaling relations rely on the best-fitting dark matter halo parameters and the baryonic content of the galaxies. We estimate $M_\ast$ and $M_{\rm gas}$ by integrating the functional forms fitted to the observed mass surface density profiles in Sec.~\ref{sec:massmodels_bar} up to $R_{200}$. We do this to be fully consistent with our $M_{200}$ estimate, but we emphasise that the integration limit has negligible impact on the integrated masses since $\Sigma \rightarrow 0$ shortly after the extent of our deep data. Finally, the baryonic mass is simply $M_{\rm bar} = M_\ast + M_{\rm gas}$. The associated uncertainties in $M_\ast$ and $M_{\rm gas}$ come from the 16th and 84th percentiles of Monte Carlo realisations that propagate the error from the distance, mass surface density profiles (typically of the order of 10\%), and, in the case of $M_\ast$, from $\Upsilon_{\rm d}$ and $\Upsilon_{\rm b}$. The uncertainty in $M_{\rm bar}$ also comes from Monte Carlo sampling propagating the errors in $M_\ast$ and $M_{\rm gas}$.

We looked into the (cold) baryon retention fraction relative to the cosmological baryon $f_{\rm bar, cosmic}$\footnote{$f_{\rm {bar,cosmic}} \equiv \Omega_{\rm b} / \Omega_{\rm m} \approx 0.16$ is the average cosmological baryon fraction; e.g. \citet{komatsu2011,planck2020_fbar}.}, i.e.:
\begin{equation}
    \tilde{f}_{i} = \dfrac{M_i}{f_{\rm bar, cosmic}\,M_{200}}~,
\end{equation}
with $i$ as the stellar, gas, or baryonic component of the galaxies. These quantities are of significant astrophysical relevance, as they allow some quantification of the fraction of baryons retained during galaxy formation and the amount converted into stars or ejected beyond the ISM (e.g. \citealt{vladimir_baryons,gonzalez2014, bookFilippo,pezzulli2019,posti_galaxyhalo}). As before, the uncertainties in $\tilde{f}_{i}$ are obtained through Monte Carlo realisations considering the distribution of the parameters involved in each calculation. Table~\ref{tab:mbar} lists $M_\ast$, $M_{\rm gas}$, $M_{\rm bar}$, $f_{\rm gas} = M_{\rm gas}/M_{\rm bar}$, $\tilde{f}_{\rm \ast}$, $\tilde{f}_{\rm gas}$, and $\tilde{f}_{\rm bar}$ for our sample.

We contrasted our results with theoretical expectations from models and cosmological hydrodynamical simulations.
Our goal is not to perform an in-depth comparison or to discuss the various physics implementations among the different models and simulations (see \citealt{wright2024} for a detailed study) but rather to see general trends and obtain an overview of the similarities and discrepancies between our data and state-of-the-art implementations of galaxy formation models.
We looked at two cosmological hydrodynamical simulations, TNG50 \citep{illustris_tng50} and Simba \citep{simba}. TNG50 generates galaxies within a 51.7~Mpc$^3$ volume with a baryon and dark matter particle resolution of $8.5\times10^4\, M_\odot$ and $4.5\times10^5\, M_\odot$, respectively. In the case of Simba, we focused on the high-resolution full-physics run with a box length of 25 Mpc, which has a dark matter and baryon particle resolution of $1.2\times10^7\,M_\odot$ and $2\times10^6\,M_\odot$, respectively. We made the following selection cuts to select simulated TNG50 and Simba galaxies with properties similar to those of our sample. First, we chose central (i.e. not satellites) systems. Next, we required $f_{\rm gas} \geq 0.01$ to avoid gas-depleted galaxies. Further, we imposed $\kappa_{\rm rot} > 0.45$ (with $\kappa_{\rm rot}$ being the fraction of kinetic energy invested in ordered rotation in the stars, as introduced by \citealt{sales2012}). While $\kappa_{\rm rot} \gtrsim 0.7$ is often adopted to select disc galaxies, it primarily works at the high-mass regime: simulated dwarfs (even gas-rich and rotating) have only very seldom $\kappa_{\rm rot} \gtrsim 0.7$. We find instead that $\kappa_{\rm rot} > 0.45$ results in simulated galaxies that match well our observed data in the $M_\ast-f_{\rm gas}$ plane, so we adopted this as our threshold. The selection cuts result in 327 galaxies from Simba and 2292 from TNG50.

Regarding models, we first considered as a reference the one by \cite{moster2010}, which uses abundance-matching. Although the \cite{moster2010} relation was initially derived for galaxies with $\log(M_\ast/M_\odot)>9$ and $\log(M_{200}/M_\odot)>11$, we compared it against our entire stellar and halo mass range, so this caveat should be kept in mind.  
In addition, we considered DarkLight \citep{kim2024}, a semi-empirical dwarf galaxy formation model calibrated such that the halo assembly histories match the relation between peak halo velocity and star formation rate in real galaxies \citep{read2017} and in the high-resolution zoom-in EDGE cosmological simulation \citep{edge}. DarkLight is particularly important for our purposes since it focuses on the low-mass regime, which is inaccessible to simulations due to their resolution limits. Furthermore, DarkLight galaxies are simulated in a void, which resembles the environment of our field galaxies more than a more crowded environment.
From the full DarkLight model, we chose those systems classified as blue and star-forming (see \citealt{kim2024}), resulting in 461 dwarfs.

The stellar, gas, and dark matter masses in TNG50 and Simba come directly by summing the particle masses associated with each halo (see \citealt{illustris_tng50,simba}). This approach is qualitatively similar to our approach of integrating the fitted functional forms describing $\Sigma_\ast$ and $\Sigma_{\rm gas}$ out to $R_{200}$; in fact, as we show in the upcoming sections, there is an overall good agreement in the stellar and gas masses of data and simulations. 
In the case of DarkLight, the stellar masses come from mapping the growth history of the haloes into SFHs so that the empirical relation between SFR and halo circular velocity is reproduced (see \citealt{kim2024} for details). As for $c_{200}$ (available for TNG50 and DarkLight), the estimates come from NFW fits to the density profile of the haloes and computing $c_{200}= R_{200}/r_{\rm s}$.

\subsection{Stellar-to-halo mass relation}

\begin{figure*}
     \centering
     \includegraphics[width=1\linewidth]{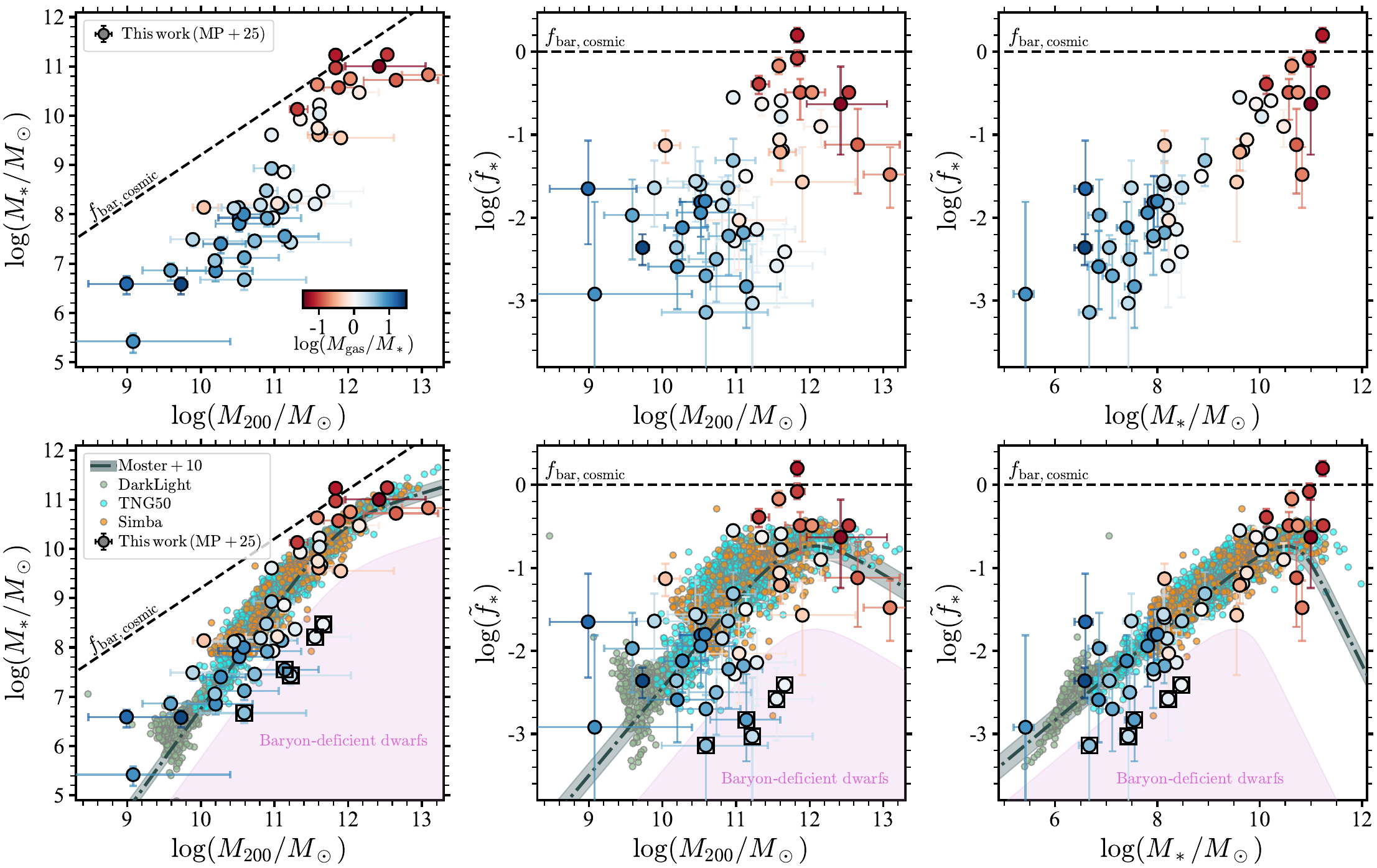}
     \caption{Relations between stellar and halo mass in our galaxy sample. The \emph{left} panels show $M_\ast$ vs. $M_{200}$, while the \emph{middle} and \emph{right} panels show $\tilde{f}_{\ast}$ as a function of $M_{200}$ and $M_{\ast}$, respectively. The observations are colour-coded by the $\log(M_{\rm gas}/M_\ast)$ ratio. All panels highlight the cosmological baryon fraction $f_{\rm bar, cosmic}$. 
     In the bottom panels, we contrast our observations against the abundance-matching model of \cite{moster2010}, the semi-empirical DarkLight model (light green, \citealt{kim2024}) and the TNG50 (cyan, \citealt{illustris_tng50}) and Simba (orange, \citealt{simba}) cosmological hydrodynamical simulations. The bottom panels highlight a population of observed baryon-deficient dwarfs (BDDs, squares enclosing circles), characterised by having stellar masses more than one order of magnitude below abundance-matching expectations (plum region).}
     \label{fig:shmr}
 \end{figure*}

The first dynamical scaling law we inspect is the stellar-to-halo mass relation (SHMR), which we complement with the $\log(\tilde{f}_{\rm \ast})-\log(M_{200})$ and $\log(\tilde{f}_{\rm \ast})-\log(M_\ast)$ planes. These three planes are shown in the top panels Fig.~\ref{fig:shmr}, where galaxies are colour-coded by their logarithmic gas-to-stellar mass ratios $\log(M_{\rm gas}/M_\ast)$. The colour scheme reflects that, unlike spirals, dwarf galaxies are gas-dominated. More importantly, we find that at fixed $M_{200}$ gas-poor galaxies tend to show higher $M_\ast$ (and therefore $\tilde{f_\ast}$), at least down to $M_{200}\approx 10^{11}\,M_\odot$, where the trend becomes less apparent (see also \citealt{geha2006,romeo2020_instabilities}). Interestingly, \cite{scholz2024} has reported a trend at fixed halo mass (for $M_{200}\gtrsim 3\times10^{10}\,M_\odot$) of galaxies with older ages having higher $M_\ast$. Since stellar population ages correlate with stellar mass (e.g. \citealt{gallazzi2005}) in a similar fashion as $M_{\rm gas}/M_\ast$ anticorrelates (i.e. gas-poor galaxies usually have older stellar ages), both trends are likely connected and evidence of the coupling between baryons and dark matter.

Regarding the shape of the SHMR, the data generally show the expected behaviour of more massive haloes hosting galaxies of higher $M_\ast$ (and a bend at high halo masses), although the scatter is significant at any given stellar or halo mass, as shown in the top middle and right panels of Fig.~\ref{fig:shmr}. From the $\log(\tilde{f}_{\rm \ast})-\log(M_\ast)$ plane, we can see that $\tilde{f_\ast}$ appears to grow nearly monotonically with $M_\ast$, with those spiral galaxies with the highest $M_\ast$ having baryon fractions that approach and even surpass the line of the cosmological baryon fraction (i.e. they have virtually no `missing baryons'), as also reported by \cite{postinomissing}, \cite{posti2021}, and \cite{enrico_massmodels_ss}.

The $\log(\tilde{f}_{\rm \ast})-\log(M_{200})$ plane instead shows a more complex distribution. From $\log(M_{200}/M_\odot) \sim 13$ to $\log(M_{200}/M_\odot) \sim 12$, there appears to be an anticorrelation until star-forming galaxies reach their peak of maximum $\tilde{f_\ast}$. At lower mass scales, galaxies show significant scatter (between $\sim0.001-60$\% of $f_{\rm bar,cosmic}$), perhaps with hints of an anticorrelation with $M_{200}$ in the range $9 \lesssim \log(M_{200}/M_\odot)\lesssim 12$. The scatter is large enough that below $M_{200}\sim 10^{12}\,M_\odot$ haloes of lower $M_{200}$ can host stellar components of larger $M_\ast$ than more massive haloes.

The $\log(\tilde{f}_{\rm \ast})-\log(M_{200})$ and $\log(\tilde{f}_{\rm \ast})-\log(M_\ast)$ plane also show a handful of `baryon-deficient' dwarf galaxies (BDDs) with halo masses $\log(M_{200}/M_\odot)\sim10.5-12$ but stellar masses as low as $\log(M_\ast/M_\odot)\sim7-9$. Below, we discuss this finding further, and show that this population also has low $\tilde{f}_{\rm gas}$ and $\tilde{f}_{\rm bar}$. In general, our results suggest that the galaxy-halo connection is more intricate than expected, likely driven by the stochasticity of star formation (particularly important in low-mass galaxies).

We contrast our galaxies against the models and simulations described above. The comparison is shown in the bottom panels of Fig.~\ref{fig:shmr}. First, we can see that the SHMR from abundance matching broadly follows the observations. However, it is clear that its scatter ($\sigma_{\log(M_\ast)} \approx 0.15\,\rm{dex}$, \citealt{moster2010}) in $\log(M_\ast)$ at fixed halo mass is way too narrow compared to the data, and it cannot match the $M_\ast\sim10^{10.5}\,M_\odot$ galaxies with no missing baryons (see also \citealt{postinomissing,enrico_massmodels_ss}), or the $M_\ast\sim10^{6.5}-10^{8.5}\,M_\odot$ galaxies with low and high $\tilde{f_\ast}$, which appear to follow a trend almost perpendicular to the one from abundance matching. In fact, we identify a population of BDDs (squares in the bottom panels of Fig.~\ref{fig:shmr}) with $M_\ast$ more than one order of magnitude below (plum-coloured regions) the predictions from abundance-matching. We delve into this galaxy population a few paragraphs below.

Next, we turn our attention to more sophisticated simulations. Fig.~\ref{fig:shmr} shows that, generally, TNG50 produces simulated galaxies with higher $\tilde{f}_{\rm \ast}$ than Simba. As discussed in \cite{wright2024}, this is likely due to the different feedback implementations. In TNG50, outflows driven by supernovae do not leave $R_{200}$, allowing for \emph{i)} gas accretion to continue and \emph{ii)} a more enriched CGM that is easier to cool and can be recycled into the ISM (see e.g. \citealt{pezzulli2017,bookFilippo}). In contrast, the outflows in Simba are stronger at $R_{200}$ scales, favouring a more depleted CGM and generating significantly lower inflow rates at ISM scales. 
Our analysis reveals that both TNG50 and Simba manage to match the overall behaviour of the data in the three observational planes, but they fall short of reproducing the observed scatter, despite their different feedback implementations and their larger volumes compared to the observational data set. Neither TNG50 nor Simba appears to produce systems with $\tilde{f}_{\rm \ast}\approx 1$, suggesting that these simulated galaxies might be too dark matter dominated. \cite{marasco_DMinsim} has suggested that advanced cosmological hydrodynamical simulations produce massive galaxies that are too dark matter dominated, in agreement with our observations (see also \citealt{glowacki2020} for indications of similar behaviour in the low-mass regime). As we discuss below, the simulations also appear to struggle to reproduce the BDDs.

Below $M_{200}\approx 2\times10^{10}\,M_\odot$ and down to $M_{200}\approx 3\times10^{9}\,M_\odot$ DarkLight provides the best comparison. As real galaxies, DarkLight blue dwarfs preferentially lie above the relation from \cite{moster2010} and show significant scatter, although arguably not enough within $2\times10^{10} \lesssim M_{200}/M_\odot \lesssim 2\times10^{11}$. This, however, can also be related to selection effects since the overall low-mass population traced by abundance matching could be quiescent (but see \citealt{boylan2012} for discrepancies also with quenched dwarfs), as red DarkLight dwarfs (not shown, see \citealt{kim2024}). 
While the low-mass end of the DarkLight model should be treated with caution, it is interesting to note the hinted tail towards higher $\tilde{f}_{\rm \ast}$ values in the direction of our observations. Overall, the different simulations/models seem to be able to reproduce some of the observed features in the three planes of Fig.~\ref{fig:shmr}, but none of them can match them all.\\

\noindent
We now focus on the population of BDDs (which are dwarfs in terms of their $M_\ast$) revealed by our analysis, which have stellar (halo) masses a factor $\sim20-60$ lower (higher) than expected.  
As discussed in detail in Appendix~\ref{app:vmax}, we find it unlikely that these BDDs are spurious (arising from problems with our observations or models). However, we remain cautious since the uncertainties in $M_{200}$ for the BDDs are relatively large, driven mainly by the fact that in some of the galaxies the flattening of the rotation curves becomes evident only at the edge of our data. To have a better assessment of the confidence level at which the BDDs are atypical, we look into the maximum circular speed of the halo $V_{\rm DM,max}$, which is easier to measure than $M_{200}$ (for details see Appendix~\ref{app:vmax}). We find that the BDDs also have high $V_{\rm DM,max}$ for their $M_\ast$. Out of the five galaxies highlighted in Fig.~\ref{fig:shmr}, three (DDO 190, LVHIS 080, and UGC 8508) have a $V_{\rm DM,max}$ lower than simulated TNG50 and DarkLight galaxies at a $2\,\sigma$ level, and the remaining two (LVHIS 017 and LVHIS 072, both at $M_{200}\sim10^{11.5}\,M_\odot$ and $V_{\rm DM,max} \sim 120\ \rm{km/s}$) at a level $\gtrsim3\,\sigma$. Therefore, although the uncertainties of the BDDs in the SHMR are relatively large (see Fig.~\ref{fig:shmr_lensing} below), we find statistical differences compared to simulated galaxies.

Clearly, enlarging our galaxy sample will be crucial to definitively establishing the existence of this baryon-deficient population\footnote{See also \cite{forbes2024} for hints of a gas-poor population with overly massive haloes (based on globular cluster counts) given their stellar mass.}. Nevertheless, we highlight an additional argument that lends support to our findings: determinations from weak lensing show a similar feature in the SHMR, as we show in Fig.~\ref{fig:shmr_lensing}. The figure shows our SHMR compared against a set of weak lensing measurements. Specifically, we show at high masses the relation from the KiDS+GAMMA surveys by \cite{dvornik2020}, and at intermediate masses those from central galaxies in the Dark Energy Survey by \cite{thornton2024} and the VOICE survey by \cite{luo2022}.
There is a reasonable agreement between those estimates and ours. In particular, one of the measurements from the Dark Energy Survey and one from VOICE lie close to our BDDs. Although observational uncertainties cannot be neglected, the agreement between these independent estimations suggests that the baryon-deficient population is authentic. The fact that among the VOICE measurements, the one consistent with the dip is that from blue galaxies (see \citealt{luo2022} for details) may point to a morphology-dependent effect in which the dip is driven by blue, gas-rich galaxies such as those in our sample. This is hard to establish with the current data, but potentially, it could explain the deviations from abundance matching if the model galaxies are more gas-poor and red than our observations. Still, this would require a biased selection function not only in the VOICE data but also in DES, which did not note any bias in preferentially selecting blue galaxies.

Finally, we come back to the models and simulations. 
The \cite{moster2010} abundance matching relation (or similar empirical models; e.g. \citealt{behroozi2019}) does not produce this baryon-poor population. If selection effects drive the observed feature (but see the previous discussion), it could be that the mismatch with abundance matching is due to the model being biased towards red and quenched galaxies. However, this would indicate a problem in the pairing between haloes and galaxies in abundance matching since most observed central galaxies within our $M_\ast$ range are not quenched systems \citep{geha2012}. 

\begin{figure}
    \centering
    \includegraphics[width=1\linewidth]{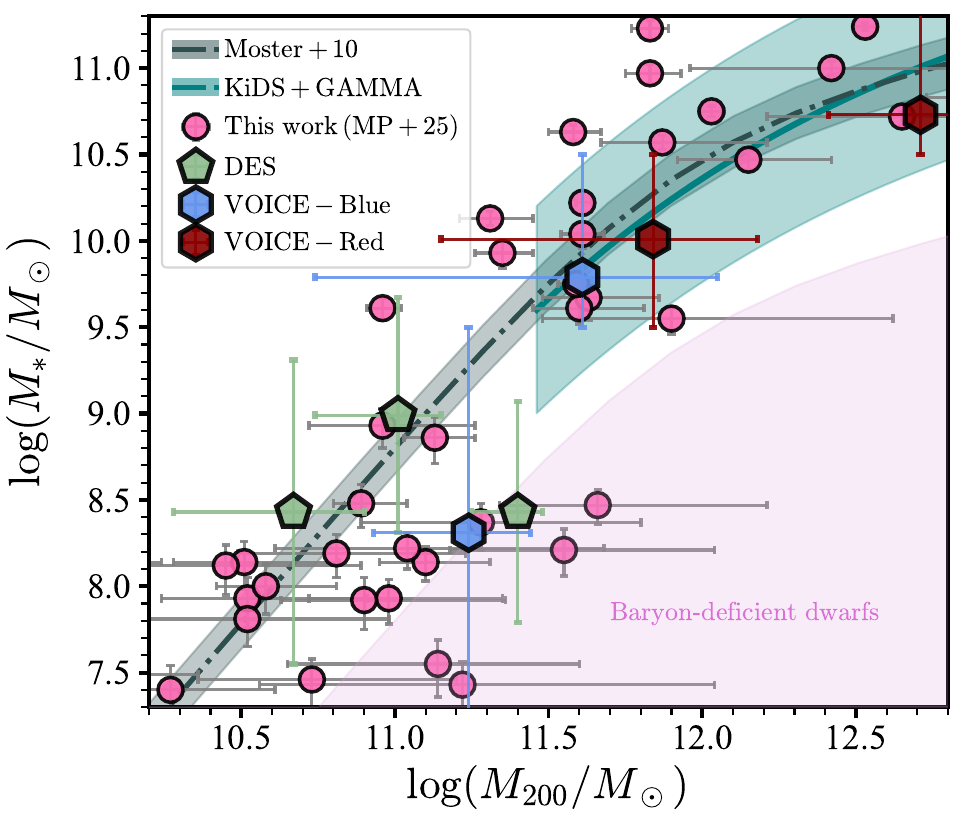}
    \caption{Zoom-in SHMR for our galaxy sample contrasted against estimates from weak lensing \citep{dvornik2020,luo2022,thornton2024}. For the lensing data, vertical error bars represent the width of the $M_\ast$ bin of each measurement. The independent results from weak lensing and rotation curve decomposition suggest the existence of a population of BDDs. Note the smaller mass span compared to Fig.~\ref{fig:shmr}. }
    \label{fig:shmr_lensing}
  \vspace{-0.3cm}
\end{figure}

The BDDs do not have clear counterparts in TNG50 or Simba, despite their relatively large cosmological volumes and having selected simulated galaxies with rotational support and matching our observed gas fractions. As shown in Fig.~\ref{fig:shmr} Simba produces a few galaxies in a position consistent with our BDDs in the SHMR (we also note the relative scarcity of Simba galaxies with $M_{200}\sim2\times10^{11}\,M_\odot$ and $M_\ast\sim10^{9}\,M_\odot$, as in our data). However, while encouraging, there is an essential difference between these Simba galaxies and the observations. The gas fractions of the simulated galaxies are about two times larger than for real galaxies. This makes the simulated galaxies stand out in terms of their $\tilde{f_\ast}$, but be `normal' when looking at their $\tilde{f}_{\rm gas}$ and $\tilde{f}_{\rm bar}$; instead, as we show below (see Figs.~\ref{fig:ghmr} and \ref{fig:bhmr}), our BDDs have low $\tilde{f_\ast}$, $\tilde{f}_{\rm gas}$ and $\tilde{f}_{\rm bar}$. A caveat in this comparison is the Simba resolution (systems with $M_\ast\approx10^8\,M_\odot$ are resolved with only $\sim50$ particles), which could complicate the quantification of gas fractions at low masses. Understanding the different features in our SHMR and the origin of BDDs will be key to further deciphering how feedback regulates galaxy formation at such mass scales.

\subsection{Gas-to-halo mass relation}
 \begin{figure*}
     \centering
     \includegraphics[width=1\linewidth]{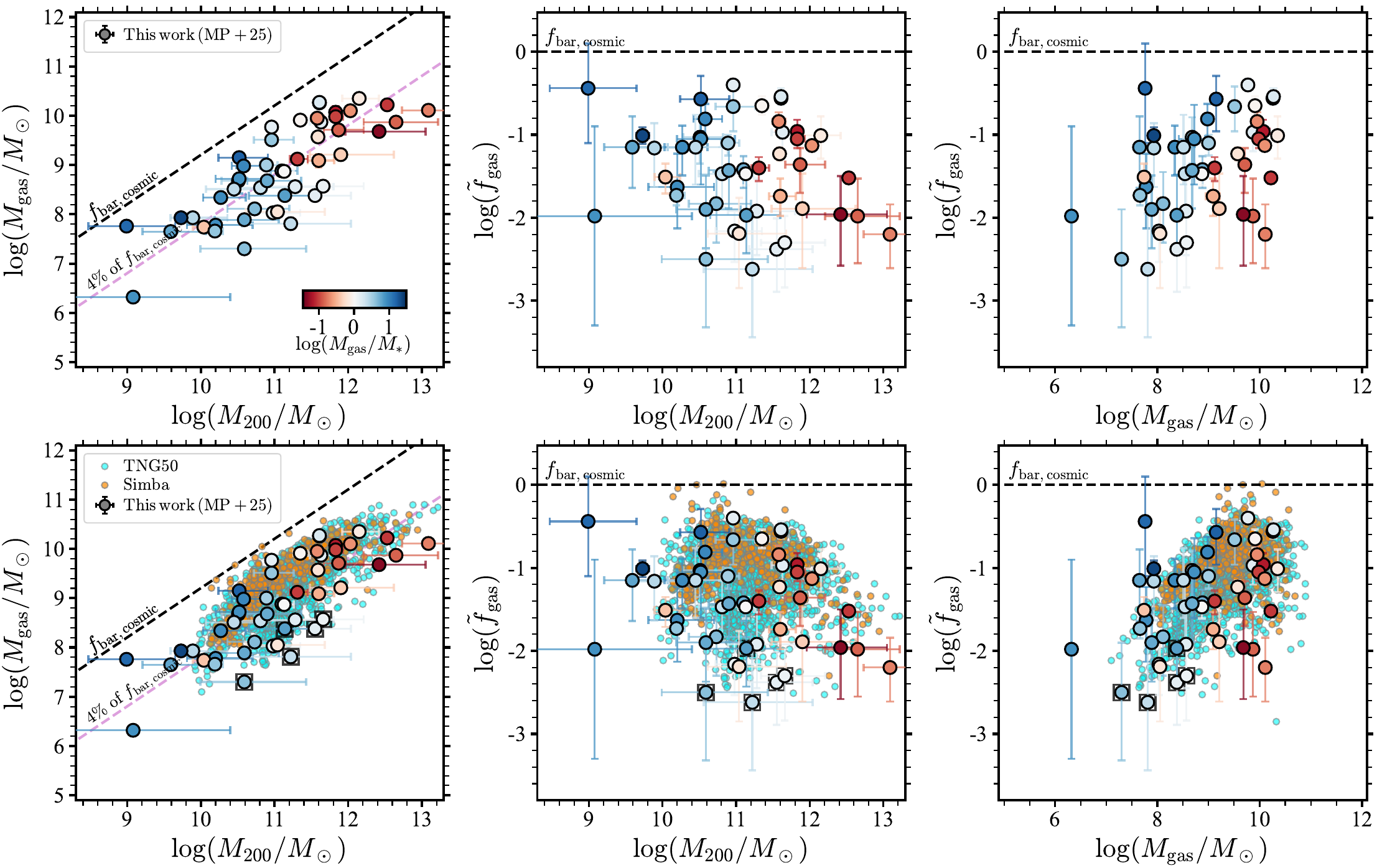}
     \caption{Relations between the cold gas and halo mass in our sample. The \emph{left} panel shows $M_{\rm gas}$ vs. $M_{200}$, while the \emph{middle} and \emph{right} panels show $\tilde{f}_{\rm{gas}}$ as a function of $M_{200}$ and $M_{\rm gas}$, respectively. The observations are colour-coded by the $M_{\rm gas}/M_\ast$ ratio. The plum dashed curve represents a gas mass equal to 4\% of $f_{\rm bar, cosmic}$, a median trend for our sample. All panels highlight the cosmological baryon fraction $f_{\rm bar, cosmic}$. TNG50 (cyan) and Simba (orange) galaxies are also included. The bottom panels highlight with squares those galaxies with $M_\ast$ lower than abundance-matching expectations by more than an order of magnitude.}
     \label{fig:ghmr}
 \end{figure*}

Next, we focus on the connection between haloes and the cold ISM. For simplicity, we refer to the cold ISM (i.e. H\,{\sc i} and H$_2$) as the total gas since the mass contribution from hotter gas phases to the ISM is thought to be small within our halo mass regime (e.g. \citealt{papastergis2012,bookFilippo,dev2024}). We remind the reader that for our dwarf sample, $M_{\rm gas}$ is given by the H\,{\sc i} mass corrected for helium and neglecting any potential contribution from H$_2$, expected to be minor. In contrast, the $M_{\rm gas}$ measurements of the massive spirals include both H\,{\sc i} and H$_2$. The comparison with theoretical models is limited to TNG50 and Simba since the abundance matching relation was valid only for stars (but see also e.g. \citealt{calette2018,calette2021_halo}), and the addition to the gas component in DarkLight is work in progress. For both TNG50 and Simba, we include H\,{\sc i} and H$_2$. This should not introduce strong biases when comparing against our dwarfs (assumed to have no molecular gas) since the H$_2$ content in the simulated dwarfs is subdominant compared to H\,{\sc i}.

Fig.~\ref{fig:ghmr} shows the relations between $M_{200}$, $M_{\rm gas}$, and $\tilde{f}_{\rm gas}$. As in the previous figure, the top panels display our measurements, while the bottom panels include the simulated galaxies.
First, we notice that the $\log(M_{\rm gas})-\log(M_{200})$ relation is much shallower than the SHMR (see also \citealt{chauhan2020}). Moreover, while the low-mass end shows scatter, the global relation appears to be parallel to the dashed line, indicating $f_{\rm bar, cosmic}$; that is, the gas content of star-forming galaxies is consistent with being a constant fraction of $f_{\rm bar, cosmic}$ across our observed halo mass range\footnote{At $M_{200}\gtrsim10^{12}\,M_\odot$, where AGN feedback is expected to become strong, some models and semi-empirical studies suggest a flattening on the $\log(M_{\rm gas})-\log(M_{200})$, which we perhaps start to see evidence of in our data. For a dedicated study and discussion on this high-mass end, see \cite{chauhan2020,spinelli2020,manasvee2024} and references therein.}. We can also see that the BDDs discussed in the previous section have a behaviour similar to the other galaxies, but they have an overly massive halo for their $M_{\rm gas}$ (or vice-versa).

As expected from the above, the $\log(\tilde{f}_{\rm gas})-\log(M_{200})$ plane shows no correlation, as also noticed by \cite{korsaga2023}. Our $\log(\tilde{f}_{\rm gas})$ has a median value of $-1.4$, indicating that across $9 \lesssim \log(M_{200}/M_\odot) \lesssim 13$, $M_{\rm gas}$ is about 4\% of $f_{\rm bar, cosmic}$ (or 0.005 of $M_{200}$), within a factor of around 4 (0.55 dex). To illustrate this, we add to the $\log(M_{\rm gas})-\log(M_{200})$ plots the curve (plum colour) of 4\% of $f_{\rm bar, cosmic}$.
Lastly, in the right panels of Fig.~\ref{fig:ghmr}, we see no compelling correlation between $\tilde{f}_{\rm gas}$ and $M_{\rm gas}$, a very different picture from the stellar content. At fixed halo and gas mass, the scatter in $\tilde{f}_{\rm gas}$ has been suggested to be driven by the specific angular momentum of the galaxies and their offset from the star-formation main sequence (see e.g. \citealt{stevens2019,chauhan2020,romeo2020_instabilities, paperIIBFR, manasvee2024}). 

Regarding the TNG50 and Simba galaxies, they tend to scatter towards higher $\tilde{f}_{\rm gas}$ values but also appear to lie parallel to lines of fixed baryon fraction. There are some indications of curvature for Simba and TNG50 in the $\log(\tilde{f}_{\rm gas})-\log(M_{200})$ and $\log(\tilde{f}_{\rm gas})-\log(M_{\rm gas})$ plane, respectively, but both their median distributions are broadly consistent with the observations down to the halo mass range in common. We note that while Simba and TNG50 overlap in the high $\tilde{f}_{\rm gas}$ galaxies, TNG50 produces more galaxies with lower $\tilde{f}_{\rm gas}$ values, and it has a scatter that resembles more closely the observed one (which likely goes beyond the fact that the simulated volume for TNG50 is larger that for Simba) although does not fully match our lowest $\tilde{f}_{\rm gas}$ values. Interestingly, there are a few Simba galaxies with $\tilde{f}_{\rm gas} \approx f_{\rm bar,cosmic}$ as a few of our dwarfs (see also Sec.~\ref{sec:bhmr}). 

Generally, the results shown in this subsection highlight the high degree of self-regulation and self-similarity in the cold ISM of star-forming galaxies across four orders of magnitude in halo mass (see also \citealt{romeo2020_instabilities,korsaga2023}). Similar to the SHMR, the gas-to-halo mass relation encodes important physical information that can provide insightful constraints to models and simulations. 

\subsection{Baryonic-to-halo mass relation}
\label{sec:bhmr}

 \begin{figure*}
     \centering
     \includegraphics[width=1\linewidth]{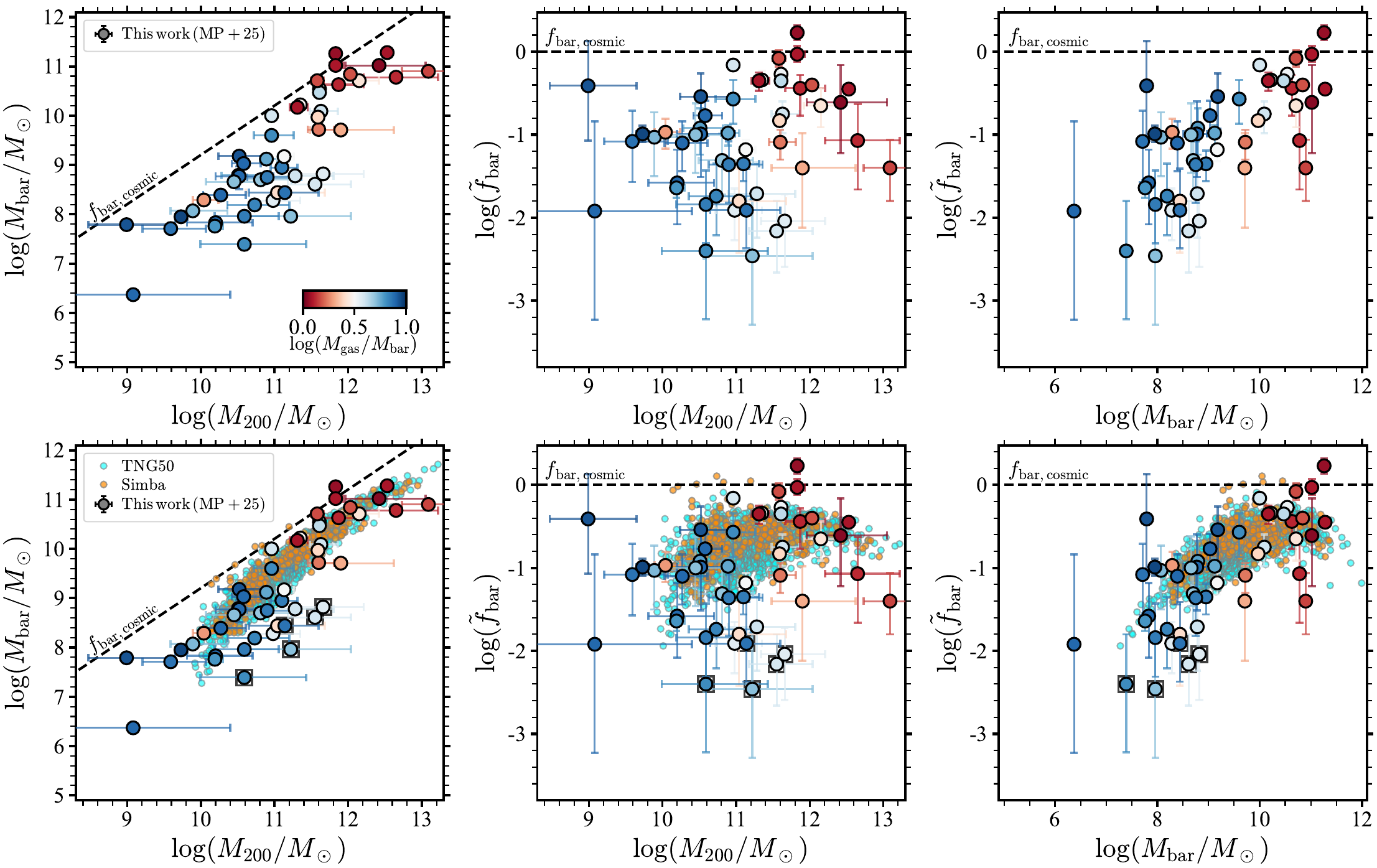}
     \caption{Relations between baryonic and halo mass in our sample. The \emph{left} panels show $M_{\rm bar}$ vs. $M_{200}$, while the \emph{middle} and \emph{right} panels show $\tilde{f}_{\rm{bar}}$ as a function of $M_{200}$ and $M_{\rm bar}$, respectively. The observations are colour-coded by the $M_{\rm gas}/M_{\rm bar}$ ratio. All panels highlight the cosmological baryon fraction $f_{\rm bar, cosmic}$. The bottom panels also include comparisons with TNG50 (cyan) and Simba (orange) galaxies. The bottom panels highlight with squares those galaxies with $M_\ast$ lower than abundance-matching expectations by more than an order of magnitude.}
     \label{fig:bhmr}
 \end{figure*}

We then combine the information on our galaxies' stellar and gas content to study their baryon-to-halo mass relation. From a physical viewpoint, this relation should be more fundamental and complete than the SHMR and the gas-to-halo mass relation, and it will give a fairer comparison between massive galaxies (dominated by the stars) and dwarfs (dominated by the gas). 

In Fig.~\ref{fig:bhmr} we present the scaling relations involving $M_{\rm bar}$, $\tilde{f}_{\rm bar}$, and $M_{200}$. Unlike in Figs.~\ref{fig:shmr} and \ref{fig:ghmr}, galaxies in Fig.~\ref{fig:bhmr} are colour coded by their gas fraction $f_{\rm gas}=M_{\rm gas}/M_{\rm bar}$. Above $M_{200}\sim10^{10}\,M_\odot$, our data suggests a monotonic relation with $M_{\rm bar}$ increasing for larger $M_{200}$. However, the scatter at $M_{\rm bar}\sim10^{8}-10^9\,M_\odot$ is large, in part driven by the BDDs discussed above. At lower halo masses, our few data points scatter around, suggesting a wide baryon retention fraction in dwarfs with $M_{200}\leq10^{10}\,M_\odot$. There is no clear correlation between $M_{200}$ and $\tilde{f}_{\rm bar}$ (middle panels in Fig.~\ref{fig:bhmr}). This changes when looking at $\tilde{f}_{\rm bar}$ as a function of $M_{\rm bar}$, where a positive correlation can be seen. As expected from the previous sections, the trend is weaker than for the $\tilde{f}_{\rm \ast}-M_\ast$ relation but stronger than the gas counterpart. The $\tilde{f}_{\rm bar}-M_{\rm bar}$ plane also suggests that, at fixed $M_{\rm bar}$, gas-rich galaxies tend to have retained more baryons than gas-poor ones.

The bottom panels of Fig.~\ref{fig:bhmr} include the TNG50 and Simba data. Given their resolution, we can only compare down to $M_{200}>10^{10}\,M_\odot$. While Simba produces more galaxies with $\tilde{f}_{\rm bar} \approx f_{\rm bar,cosmic}$, the physics implementation of TNG50 shows a broader spread in $\tilde{f}_{\rm bar}$. Both TNG50 and Simba align with our main trends on average. Nevertheless, they do not match the massive spirals with $\tilde{f}_{\rm bar} \approx f_{\rm bar,cosmic}$ neither the baryon-poor galaxies with $\log(\tilde{f}_{\rm bar}) < -1.2$. Once large-volume cosmological hydrodynamical simulations manage to reach our lowest $M_{200}$ and $M_{\rm bar}$ values with enough resolution (e.g. \citealt{muncshi2021,gutcke2022}), it will be interesting to test if the trends and diversity in $\tilde{f}_{\rm bar}$ can be reproduced.

Our analysis shows that dwarf and massive disc galaxies can have a large spread (up to a significant percentage of the cosmological average) in their baryon fractions. Our results represent a substantial addition to the growing evidence of a population of gas-rich low-mass galaxies with large scatter in their baryon fractions, from baryon-poor to baryon-rich cases (e.g. \citealt{enrico_smc,huds2019,demao,maccagni2024,agc114905_deep,barbara_udgs}). The improvement in simulations mentioned above should be paired with enlarging the samples of dwarfs with high-quality data and reducing the observational uncertainties. This will clarify whether broader feedback implementations (e.g. \citealt{silk2017,bradford2018,mina2021,azartash2024}) in simulations are necessary or if a deeper understanding of stochasticity in galaxy evolution is needed—ideas hinted at in our data (see also \citealt{kaplinghat2020}), as shown in Figs.~\ref{fig:shmr}--\ref{fig:bhmr}.

\subsection{Concentration--mass relation}
\label{sec:concentration}

An anticorrelation between halo mass and concentration arises naturally in N-body simulations in the CDM framework \citep{vladimir2001,bullock2001,wechsler2002,duttonmaccio2014,ludlow2014,correa2015,diemer2015,wang2020}. The scatter of the relation is, in principle, dictated by the different assembly histories of the haloes \citep{nfw,wechsler2002}, such that haloes that assemble earlier tend to have higher concentrations. Theoretically, the $c_{200}-M_{200}$ relation (not only its scatter, but also its median value) can be influenced by baryonic physics (e.g. \citealt{dicintio2014,dutton2016,coreEinasto,anbajagane2022}). In principle, the environment could also play a role, although there is still no consensus on systematic variations of the concentrations across most galaxy environments \citep{lemson1999,maccio2007,lee2017,hellwing2021}. All this makes the concentration-mass relation another insightful probe of the galaxy-halo connection.

\begin{figure}
    \centering
    \includegraphics[width=0.95\linewidth]{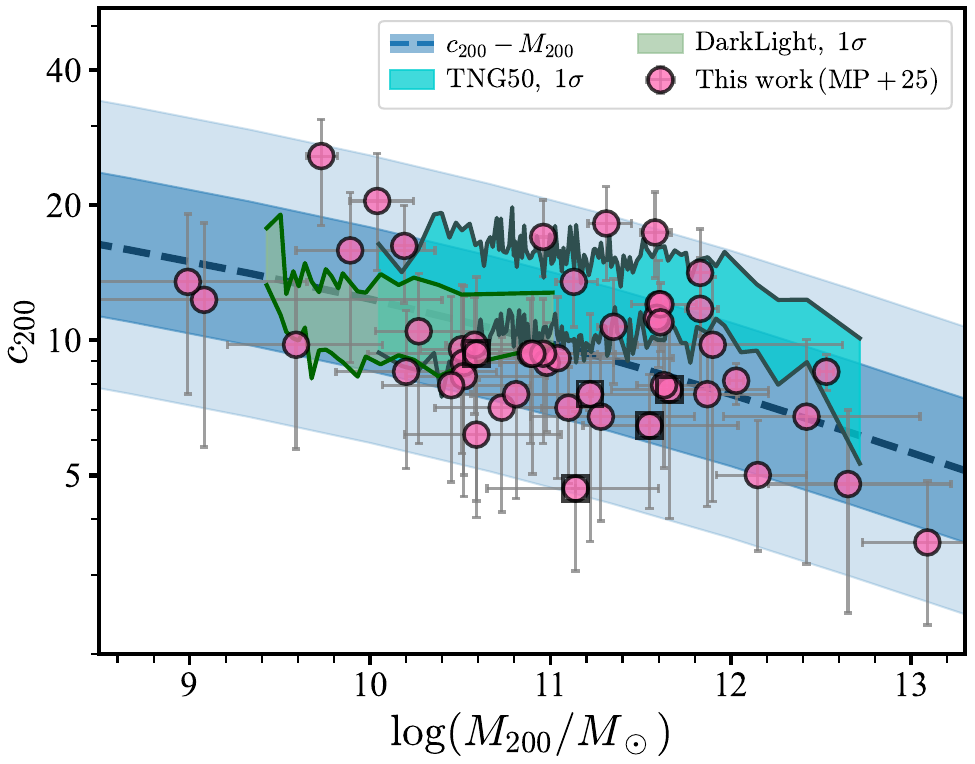}
    \caption{Concentration--mass relation for our galaxy sample (pink markers, squares highlight BDDs). The blue dashed line shows the theoretical relation by \citetalias{diemer2019} (imposed as a prior), and the bands show their $1\,\sigma$ and $2\,\sigma$ confidence bands. The $1\,\sigma$ (84th percentiles) distributions of DarkLight (green) and TNG50 (turquoise) simulated galaxies are also included for comparison.}
    \label{fig:c200M200}
\end{figure}

Despite having imposed the $\log(c_{200})-\log(M{_{200})}$ relation from \citetalias{diemer2019} as a Gaussian prior during our mass modelling, its generous standard deviation ($\sigma=0.16$ dex, see \citealt{diemer2015}) gives room for the data to deviate from the relation if that provides a better fit. We analyse this in Fig.~\ref{fig:c200M200}, where we show the $c_{200}-M{_{200}}$ plane for our galaxy sample compared to the \citetalias{diemer2019} relation and its $1\,\sigma$ and $2\,\sigma$ confidence bands. 
While all our concentration parameters are consistent within $2\,\sigma$ with the prior, we still see indications of substructure or `wiggles' in the relation. Below $M_{200}\sim10^{10}\,M_\odot$, the haloes appear to scatter around the mean relation. Between $10^{10} \lesssim M_{200}/M_\odot\lesssim10^{11}$, the galaxies instead systematically lie below the \citetalias{diemer2019} relation. Systems with halo masses within $10^{11} \lesssim M_{200}/M_\odot\lesssim10^{12}$ show large scatter, with some haloes under- and over-concentrated at a $2\,\sigma$ level. Finally, galaxies with  $M_{200}\gtrsim10^{12}\,M_\odot$ tend to be below the relation. We note that the first and last halo mass bins remain more uncertain, given the low number of galaxies within them. In this section, we speculate whether halo assembly, feedback, or environment are the drivers for the scatter and substructure in our $c_{200}-M{_{200}}$ relation.

First, we consider the possibility that the observed diversity in the $c_{200}-M{_{200}}$ relation is a consequence of different halo assembly histories. While there is no direct way to test this scenario, we deem it unlikely that all the features arise from halo growth histories. This is because \cite{sorini2025} has found that simulated TNG50 galaxies (which we further discuss below) follow typical concentration-mass relations in the dark-matter-only run (but see also \citealt{demao}) and only present deviations when hydrodynamical effects are considered.

Indeed, as discussed by \cite{lovell2018,anbajagane2022,paper_massmodels,sorini2025}, and references therein, the oscillations or wiggles are thought to result from the interplay between baryonic physics and the dark matter halo. In this context, the concentration parameters in our second and fourth mass bins could be low due to halo expansion driven by stellar and black hole feedback, respectively. At $10^{11} \lesssim M_{200}/M_\odot\lesssim10^{12}$, a competition between feedback and adiabatic contraction (for the latter, see \citealt{gnedin2004,schulz2010,dicintio2014,dutton2016,coreEinasto}) could drive the large scatter (although dark matter physics may also play a role, see \citealt{kaplinghat2020,kong2025}). At the low-mass end, feedback might not be strong enough to systematically lower the concentration parameters\footnote{Note that the dwarf with high $c_{200}$ at $\log(M_{200}/M_\odot)\approx9.8$ is WLM, whose concentration remains somewhat ambiguous since it has $r_{\rm s} < r_{\rm c}$, see also \cite{readAD}.}, but it is hard to establish this with our sample size.

To further explore the above ideas, we add to Fig.~\ref{fig:c200M200} the TNG50 and DarkLight samples (values for $c_{200}$ are not yet available for Simba, poor fits from DarkLight have been discarded). Specifically, we bin the simulated galaxies from TNG50 and DarkLight in 100 and 20 bins of equal galaxy counts, respectively, and show the $1\,\sigma$ distributions, assumed to be the $16\rm{th}-84\rm{th}$ percentile range. We highlight a few relevant implications found from this comparison.

The first regards the distribution of TNG50 in Fig.~\ref{fig:c200M200}. While it shows some overlap with the data and DarkLight, the TNG50 $c_{200}$ distribution for $\log(M_{200}/M_\odot)\gtrsim10.5$ is systematically shifted towards high values (see also \citealt{sorini2025}) and does not reproduce the entire observed distribution. Therefore, as previous studies \citep{lovell2018,heinze2024,li_manga2024}, our analysis indicates an enhanced adiabatic contraction in TNG50, which our data suggest is overly efficient. TNG50 has also been reported to be inefficient at lowering the central dark matter densities \citep{lovell2018}, and indeed, our data support this idea given the lower $c_{200}$ parameters seen in real galaxies.

The comparison with DarkLight is also insightful.
Below halo masses $\log(M_{200}/M_\odot)\sim10.6$, 
DarkLight galaxies fall systematically below the \citetalias{diemer2019} relation. Two considerations should be taken into account for these simulated galaxies. The first one is that DarkLight systems were simulated in a void, and the second one is that the current version of DarkLight does not include feedback-driven halo expansion. This is interesting for different reasons. The lower $c_{200}$ parameters with respect to \citetalias{diemer2019} are likely reflecting the low-density environment in which the galaxies formed. In fact, theoretical studies have found tentative evidence for haloes in low-density environments having lower $c_{200}$ than their high-density environment counterparts \citep{maccio2007,lee2017,hellwing2021}.

If environmental effects are at play, it could have important implications for our understanding of galaxy formation\footnote{We also note that this can have repercussions for the formation mechanisms of isolated gas-rich ultra-diffuse galaxies, which appear to have concentration parameters much lower than expected in the CDM model \citep{huds2019,sengupta2019,agc114905,shi2021,demao,agc114905_deep,afruni2025}.}, and we will further explore this possibility in future work by creating DarkLight realisations in different environments. In this context, some of the low concentrations in our galaxy sample could be due to their environment since, by selection, they tend to inhabit environments of relatively low density (they are central, relatively unperturbed, and with gas reservoirs). Still, a handful of our galaxies have $c_{200}$ parameters lower than those in DarkLight, specially for $\log(M_{200}/M_\odot)\gtrsim10.5$. As discussed above, the culprit behind these differences could be feedback-driven halo expansion, which is not yet implemented in DarkLight but might be at play in our galaxies -- most of which exhibit cored dark matter distributions.

\section{Conclusions}
\label{sec:conclusions}

In this work, we have studied the galaxy-halo connection for nearby disc galaxies through several dynamic scaling relations that connect baryons and dark matter. To do so, we assembled a sample of 49 galaxies (Fig.~\ref{fig:sample}) with unmatched data quality spanning six orders of magnitude in stellar mass ($M_\ast$). Of our total sample,
27 galaxies have kinematic models derived with the software $^{\rm 3D}$Barolo \citep{barolo} by \cite{enrico_radialmotions} and \cite{iorio}. For the remaining 22 (dwarf) galaxies, we built robust kinematic models using $^{\rm 3D}$Barolo (Fig.~\ref{fig:kinematics_examples}). Our work significantly enlarges the sample of dwarf galaxies with high-quality kinematics (which we make publicly available). These observations can be used as strong constraints to test galaxy formation models and their implementation in simulations.

Using a novel and sophisticated technique, we obtained mass models for our galaxies through rotation curve decomposition (Fig.~\ref{fig:massmodels_examples}). The free parameters of our mass model are mass-to-light ratios for the stellar disc ($\Upsilon_{\rm d}$) and bulge ($\Upsilon_{\rm b}$) and those describing the halo (assumed to be a \textsc{coreNFW}, see \citealt{coreNFW}), namely halo mass ($M_{200}$), concentration ($c_{200}$), and the parameter $\eta$ which regulates the size of the dark matter core ($r_{\rm c}~=~\eta\,R_{\rm e}$, with $R_{\rm e}$ the half-light radius of the stellar component).

Our rotation curve decomposition exploits the condition of vertical hydrostatic equilibrium to self-consistently derive the mass models and the vertical scale heights of the gas discs (Fig.~\ref{fig:scaleheights}). The scale heights are flared and show a moderate correlation with halo mass (the scale heights of galaxies with low-mass haloes are usually thicker) and a strong correlation with $V_{\rm rot}/\sigma_{\rm_{HI}}$ (galaxies with higher $V_{\rm rot}/\sigma_{\rm_{HI}}$ have thinner gas discs). For galaxies with $V_{\rm circ} < 30\,\rm{km/s}$, $M_\ast < 10^7\,M_\odot$, and $V_{\rm rot}/\sigma_{_{\rm HI}} \lesssim 3$ the flaring has a major dynamical effect on the rotation curve decomposition (Sec.~\ref{sec:impact}.)

Using our homogeneous sample, we built different scaling laws connecting baryons and dark matter. We also compared the location of our galaxies with simulated systems from abundance matching calibrations \citep{moster2010}, the DarkLight model \citep{kim2024}, and the TNG50 \citep{illustris_tng50} and Simba \citep{simba} cosmological hydrodynamical simulations. Our main findings are summarised as follows.

\begin{itemize}

    \item The relation between $M_\ast$, $M_{200}$, and their ratio normalised to the average cosmological baryon fraction (i.e. $\tilde{f}_{\rm \ast} = M_\ast/(f_{\rm bar,cosmic}\times M_{200})$) is complex, deviates from abundance matching, and implies a wide variety of $M_\ast/M_{200}$ ratios at a given $M_\ast$ or $M_{200}$ (Fig.~\ref{fig:shmr}). 
    
    \item We find tentative evidence for the existence of a population of baryon-deficient dwarf galaxies (BDDs) with stellar and baryonic masses $\sim1.5$ orders of magnitude below expectations. Observational uncertainties remain important, but our findings appear to be supported by recent weak lensing measurements (Fig.~\ref{fig:shmr_lensing}). Galaxies from the Simba simulation show hints of a population of BDDs similar to what we observe, but the simulated systems are too gas-rich to be consistent with our observations. Overall, our results imply a broad range in the baryon retention efficiency, which is not fully reproduced by the simulated galaxies.
    
    \item The cold gas mass ($M_{\rm gas}$) correlates more simply with $M_{200}$ (Fig.~\ref{fig:ghmr}). The $M_{\rm gas}-M_{200}$ relation is well described by a curve parallel to the cosmological baryon fraction but with 4\% (within a factor of three to four) of its normalisation. Consequently, $\tilde{f}_{\rm gas}$ shows no clear correlations with $M_{200}$ or $M_{\rm gas}$.
    
    \item The baryonic mass ($M_{\rm bar}=M_\ast + M_{\rm gas}$) correlates with $M_{200}$ (Fig.~\ref{fig:bhmr}). The fraction $\tilde{f}_{\rm bar}$ shows a large scatter and correlates weakly with $M_{\rm bar}$ and does not clearly vary as a function of $M_{200}$. Remarkably, galaxies with $\tilde{f}_{\rm bar}$ being a significant fraction of the average cosmological baryon fraction are present at low and high halo masses. Again, theoretical models reproduce the main trends but fall short of explaining the wide diversity in the baryon fractions. 
    
    \item We imposed the $c_{200}-M_{200}$ relation from dark-matter only cosmological simulations \citep{diemer2019} as a Gaussian prior in our mass models. Yet, the resulting $c_{200}-M_{200}$ relation shows substructure (Fig.~\ref{fig:c200M200}). In particular, below $\log(M_{200}/M_\odot) \approx 11$, galaxies have a significantly lower $c_{200}$ than expected. We discussed possible explanations for this and argued that the culprits might be the low-density environment and stellar feedback. We also find that TNG50 galaxies have concentration parameters that are too high compared to our data, which is a symptom of over-efficient adiabatic contraction.

\end{itemize}

\noindent
Overall, our results show strong correlations in the galaxy-halo connection and a wide range in the baryon retention fraction and halo structure of nearby disc galaxies. While observational uncertainties are discussed and should be considered, our results hint at the need for more diverse implementations of feedback and galaxy evolution stochasticity in galaxy formation models.

\begin{acknowledgements}

We want to thank the anonymous referee for their careful and constructive review, which helped strengthen our study. We thank Filippo Fraternali for stimulating comments regarding our results and Vladimir Avila-Reese and Haibo Yu for their valuable comments on our manuscript. We also thank Marijn Franx, Andrés Bañares-Hernández, Jorge Martin Camalich, and Anton Rudakovskyi for helpful discussions, and Giuliano Iorio for his assistance regarding \textsc{galpynamics}.

PEMP acknowledges the support from the Dutch Research Council (NWO) through the Veni grant VI.Veni.222.364.

JAB is grateful for partial financial support from NSF-CAREER-1945310 and NSF-AST-2107993 grants and data storage resources of the HPCC, which were funded by grants from NSF (MRI-2215705, MRI-1429826) and NIH (1S10OD016290-01A1).

MG is supported by the UK STFC Grant ST/Y001117/1. MG acknowledges support from the Inter-University Institute for Data Intensive Astronomy (IDIA). IDIA is a partnership of the University of Cape Town, the University of Pretoria and the University of the Western Cape. For the purpose of open access, the author has applied a Creative Commons Attribution (CC BY) licence to any Author Accepted Manuscript version arising from this submission.

We acknowledge the use of the ilifu cloud computing facility - www.ilifu.ac.za, a partnership between the University of Cape Town, the University of the Western Cape, Stellenbosch University, Sol Plaatje University, the Cape Peninsula University of Technology and the South African Radio Astronomy Observatory. The ilifu facility is supported by contributions from the Inter-University Institute for Data Intensive Astronomy (IDIA - a partnership between the University of Cape Town, the University of Pretoria and the University of the Western Cape), the Computational Biology division at UCT and the Data Intensive Research Initiative of South Africa (DIRISA).

We have used the services from SIMBAD, NED, and ADS extensively, as well the tool TOPCAT \citep{topcat} and the Python packages NumPy \citep{numpy}, Matplotlib \citep{matplotlib}, SciPy \citep{scipy}, spectral$\_$cube \citep{spectral_cube}, pandas \citep{pandas}, Astropy \citep{astropy}, and COLOSSUS \citep{colossus}, for which we are thankful.

\end{acknowledgements}

   \bibliographystyle{aa.bst} 
   \bibliography{references} 

\begin{appendix}
\onecolumn
\section{Galaxy sample}
\label{app:sample}
Table~\ref{tab:sample} lists the galaxies in our sample.

\begin{table}[H]
\caption{Galaxy sample used in this work.}
\vspace{-0.1cm}
\label{tab:sample}
\centering
\begin{tabular}{lccc}
	\hline \noalign{\vskip 1.1pt} 
 Name & Distance   & $V_{\rm circ,out}$ & $i$   \\ \noalign{\vskip 1.5pt}
      & [Mpc] &    [km/s] & [deg] \\ \noalign{\vskip 1.1pt}
   \hline \noalign{\vskip 1.5pt}  
CVn I dwA$^{\href{https://simbad.u-strasbg.fr/simbad/sim-basic?Ident=CVn+I+dwA}{\rm{SIMBAD}}}$ & 3.60 $\pm$ 0.17 & 22 $\pm$ 3 & 50\\ \noalign{\vskip 1.1pt}
DDO 52$^{\href{https://simbad.u-strasbg.fr/simbad/sim-basic?Ident=DDO+52}{\rm{SIMBAD}}}$ & 10.30 $\pm$ 0.47 & 52 $\pm$ 5 & 55 \\ \noalign{\vskip 1.1pt}
DDO 87$^{\href{https://simbad.u-strasbg.fr/simbad/sim-basic?Ident=DDO+87}{\rm{SIMBAD}}}$ & 7.40 $\pm$ 2.18 & 51 $\pm$ 4 & 43  \\ \noalign{\vskip 1.1pt}
DDO 126$^{\href{https://simbad.u-strasbg.fr/simbad/sim-basic?Ident=DDO+126}{\rm{SIMBAD}}}$ & 4.90 $\pm$ 0.38 & 39 $\pm$ 3 & 62 \\ \noalign{\vskip 1.1pt}
DDO 133$^{\href{https://simbad.u-strasbg.fr/simbad/sim-basic?Ident=DDO+133}{\rm{SIMBAD}}}$ & 5.11 $\pm$ 0.22 & 46 $\pm$ 5 & 39 \\ \noalign{\vskip 1.1pt}
DDO 154$^{\href{https://simbad.u-strasbg.fr/simbad/sim-basic?Ident=DDO+154}{\rm{SIMBAD}}}$ & 3.70 $\pm$ 0.07 & 47 $\pm$ 3 & 68 \\ \noalign{\vskip 1.1pt}
DDO 168$^{\href{https://simbad.u-strasbg.fr/simbad/sim-basic?Ident=DDO+168}{\rm{SIMBAD}}}$ & 4.30 $\pm$ 0.26 & 55 $\pm$ 6 & 62 \\ \noalign{\vskip 1.1pt}
DDO 181$^{\href{https://simbad.u-strasbg.fr/simbad/sim-basic?Ident=DDO+181}{\rm{SIMBAD}}}$ & 3.12 $\pm$ 0.11 & 29 $\pm$ 3 & 59 \\ \noalign{\vskip 1.1pt}
DDO 183$^{\href{https://simbad.u-strasbg.fr/simbad/sim-basic?Ident=DDO+183}{\rm{SIMBAD}}}$ & 3.22 $\pm$ 0.12 & 23 $\pm$ 3 & 68 \\ \noalign{\vskip 1.1pt}
DDO 190$^{\href{https://simbad.u-strasbg.fr/simbad/sim-basic?Ident=DDO+190}{\rm{SIMBAD}}}$ & 2.82 $\pm$ 0.10 & 48 $\pm$ 4 & 41 \\ \noalign{\vskip 1.1pt}
DDO 210$^{\href{https://simbad.u-strasbg.fr/simbad/sim-basic?Ident=DDO+210}{\rm{SIMBAD}}}$ & 0.90 $\pm$ 0.04 & 17 $\pm$ 4 & 63 \\ \noalign{\vskip 1.1pt}
IC 2574$^{\href{https://simbad.u-strasbg.fr/simbad/sim-basic?Ident=IC+2574}{\rm{SIMBAD}}}$ & 3.91 $\pm$ 0.07 & 71 $\pm$ 6 & 53 \\ \noalign{\vskip 1.1pt}
LVHIS 009$^{\href{https://simbad.u-strasbg.fr/simbad/sim-basic?Ident=LVHIS+009}{\rm{SIMBAD}}}$ & 4.57 $\pm$ 0.21 & 53 $\pm$ 5 & 44\\ \noalign{\vskip 1.1pt}
LVHIS 011$^{\href{https://simbad.u-strasbg.fr/simbad/sim-basic?Ident=LVHIS+011}{\rm{SIMBAD}}}$ & 4.99 $\pm$ 0.07 & 71 $\pm$ 5 & 77\\ \noalign{\vskip 1.1pt}
LVHIS 012$^{\href{https://simbad.u-strasbg.fr/simbad/sim-basic?Ident=LVHIS+012}{\rm{SIMBAD}}}$ & 5.76 $\pm$ 0.08 & 70 $\pm$ 5 & 71\\ \noalign{\vskip 1.1pt}
LVHIS 017$^{\href{https://simbad.u-strasbg.fr/simbad/sim-basic?Ident=LVHIS+017}{\rm{SIMBAD}}}$ & 6.06 $\pm$ 0.14 & 78 $\pm$ 5 & 74 \\ \noalign{\vskip 1.1pt}
LVHIS 019$^{\href{https://simbad.u-strasbg.fr/simbad/sim-basic?Ident=LVHIS+019}{\rm{SIMBAD}}}$ & 6.95 $\pm$ 0.32 & 47 $\pm$ 6 & 57 \\ \noalign{\vskip 1.1pt}
LVHIS 020$^{\href{https://simbad.u-strasbg.fr/simbad/sim-basic?Ident=LVHIS+020}{\rm{SIMBAD}}}$ & 9.02 $\pm$ 0.42 & 56 $\pm$ 5 & 48\\ \noalign{\vskip 1.1pt}
LVHIS 026$^{\href{https://simbad.u-strasbg.fr/simbad/sim-basic?Ident=LVHIS+026}{\rm{SIMBAD}}}$ & 4.63 $\pm$ 0.21 & 60 $\pm$ 4 & 49 \\ \noalign{\vskip 1.1pt}
LVHIS 055$^{\href{https://simbad.u-strasbg.fr/simbad/sim-basic?Ident=LVHIS+055}{\rm{SIMBAD}}}$ & 3.33 $\pm$ 0.15 & 47 $\pm$ 3 & 52 \\ \noalign{\vskip 1.1pt}
LVHIS 060$^{\href{https://simbad.u-strasbg.fr/simbad/sim-basic?Ident=LVHIS+060}{\rm{SIMBAD}}}$ & 3.40 $\pm$ 0.16 & 40 $\pm$ 3 & 58 \\ \noalign{\vskip 1.1pt}
LVHIS 072$^{\href{https://simbad.u-strasbg.fr/simbad/sim-basic?Ident=LVHIS+072}{\rm{SIMBAD}}}$ & 3.15 $\pm$ 0.13 & 89 $\pm$ 7 & 79 \\ \noalign{\vskip 1.1pt}
LVHIS 077$^{\href{https://simbad.u-strasbg.fr/simbad/sim-basic?Ident=LVHIS+077}{\rm{SIMBAD}}}$ & 5.50 $\pm$ 0.25 & 93 $\pm$ 8 & 68 \\ \noalign{\vskip 1.1pt}
LVHIS 078$^{\href{https://simbad.u-strasbg.fr/simbad/sim-basic?Ident=LVHIS+078}{\rm{SIMBAD}}}$ & 1.96 $\pm$ 0.09 & 61 $\pm$ 4 & 48 \\ \noalign{\vskip 1.1pt}
LVHIS 080$^{\href{https://simbad.u-strasbg.fr/simbad/sim-basic?Ident=LVHIS+080}{\rm{SIMBAD}}}$ & 4.37 $\pm$ 0.20 & 57 $\pm$ 4 & 77 \\ \noalign{\vskip 1.1pt}
NGC 0253$^{\href{https://simbad.u-strasbg.fr/simbad/sim-basic?Ident=NGC+0253}{\rm{SIMBAD}}}$ & 3.66 $\pm$ 0.05 & 197 $\pm$ 8 & 80  \\ \noalign{\vskip 1.1pt}
NGC 0925$^{\href{https://simbad.u-strasbg.fr/simbad/sim-basic?Ident=NGC+0925}{\rm{SIMBAD}}}$ & 9.29 $\pm$ 0.16 & 120 $\pm$ 11 & 53  \\ \noalign{\vskip 1.1pt}
NGC 1313$^{\href{https://simbad.u-strasbg.fr/simbad/sim-basic?Ident=NGC+1313}{\rm{SIMBAD}}}$ & 4.19 $\pm$ 0.62 & 125 $\pm$ 11 & 44  \\ \noalign{\vskip 1.1pt}
NGC 2366$^{\href{https://simbad.u-strasbg.fr/simbad/sim-basic?Ident=NGC+2366}{\rm{SIMBAD}}}$ & 3.40 $\pm$ 0.27 & 57 $\pm$ 5 & 65 \\ \noalign{\vskip 1.1pt}
NGC 2403$^{\href{https://simbad.u-strasbg.fr/simbad/sim-basic?Ident=NGC+2403}{\rm{SIMBAD}}}$ & 3.09 $\pm$ 0.06 & 135 $\pm$ 12 & 60  \\ \noalign{\vskip 1.1pt}
NGC 2541$^{\href{https://simbad.u-strasbg.fr/simbad/sim-basic?Ident=NGC+2541}{\rm{SIMBAD}}}$ & 12.60 $\pm$ 0.32 & 103 $\pm$ 8 & 60  \\ \noalign{\vskip 1.1pt}
NGC 2841$^{\href{https://simbad.u-strasbg.fr/simbad/sim-basic?Ident=NGC+2841}{\rm{SIMBAD}}}$ & 14.10 $\pm$ 0.39 & 281 $\pm$ 22 & 71  \\ \noalign{\vskip 1.1pt}
NGC 3198$^{\href{https://simbad.u-strasbg.fr/simbad/sim-basic?Ident=NGC+3198}{\rm{SIMBAD}}}$ & 14.50 $\pm$ 0.53 & 150 $\pm$ 13 & 73  \\ \noalign{\vskip 1.1pt}
NGC 3351$^{\href{https://simbad.u-strasbg.fr/simbad/sim-basic?Ident=NGC+3351}{\rm{SIMBAD}}}$ & 9.64 $\pm$ 0.22 & 174 $\pm$ 15 & 45  \\ \noalign{\vskip 1.1pt}
NGC 3621$^{\href{https://simbad.u-strasbg.fr/simbad/sim-basic?Ident=NGC+3621}{\rm{SIMBAD}}}$ & 7.11 $\pm$ 0.39 & 143 $\pm$ 9 & 65  \\ \noalign{\vskip 1.1pt}
NGC 3992$^{\href{https://simbad.u-strasbg.fr/simbad/sim-basic?Ident=NGC+3992}{\rm{SIMBAD}}}$ & 22.7 $\pm$ 3.14 & 227 $\pm$ 14 & 62 \\ \noalign{\vskip 1.1pt}
NGC 4190$^{\href{https://simbad.u-strasbg.fr/simbad/sim-basic?Ident=NGC+4190}{\rm{SIMBAD}}}$ & 2.83 $\pm$ 0.13 & 35 $\pm$ 5 & 46\\ \noalign{\vskip 1.1pt}
NGC 4535$^{\href{https://simbad.u-strasbg.fr/simbad/sim-basic?Ident=NGC+4535}{\rm{SIMBAD}}}$ & 17.80 $\pm$ 0.33 & 224 $\pm$ 17 & 36  \\ \noalign{\vskip 1.1pt}
NGC 4536$^{\href{https://simbad.u-strasbg.fr/simbad/sim-basic?Ident=NGC+4536}{\rm{SIMBAD}}}$ & 17.70 $\pm$ 0.73 & 165 $\pm$ 12 & 66  \\ \noalign{\vskip 1.1pt}
NGC 4559$^{\href{https://simbad.u-strasbg.fr/simbad/sim-basic?Ident=NGC+4559}{\rm{SIMBAD}}}$ & 8.91 $\pm$ 0.21 & 122 $\pm$ 11 & 66  \\ \noalign{\vskip 1.1pt}
NGC 4651$^{\href{https://simbad.u-strasbg.fr/simbad/sim-basic?Ident=NGC+4651}{\rm{SIMBAD}}}$ & 18.80 $\pm$ 1.13 & 183 $\pm$ 11 & 51  \\ \noalign{\vskip 1.1pt}
NGC 4725$^{\href{https://simbad.u-strasbg.fr/simbad/sim-basic?Ident=NGC+4725}{\rm{SIMBAD}}}$ & 13.50 $\pm$ 0.37 & 209 $\pm$ 11 & 56  \\ \noalign{\vskip 1.1pt}
NGC 4736$^{\href{https://simbad.u-strasbg.fr/simbad/sim-basic?Ident=NGC+4736}{\rm{SIMBAD}}}$ & 4.20 $\pm$ 0.25 & 143 $\pm$ 7 & 40  \\ \noalign{\vskip 1.1pt}
NGC 7793$^{\href{https://simbad.u-strasbg.fr/simbad/sim-basic?Ident=NGC+7793}{\rm{SIMBAD}}}$ & 3.61 $\pm$ 0.13 & 120 $\pm$ 8 & 46 \\ \noalign{\vskip 1.1pt}
NGC 5005$^{\href{https://simbad.u-strasbg.fr/simbad/sim-basic?Ident=NGC+5005}{\rm{SIMBAD}}}$ & 16.50 $\pm$ 1.29 & 270 $\pm$ 19 & 69  \\ \noalign{\vskip 1.1pt}
NGC 5055$^{\href{https://simbad.u-strasbg.fr/simbad/sim-basic?Ident=NGC+5055}{\rm{SIMBAD}}}$ & 8.87 $\pm$ 0.08 & 185 $\pm$ 16 & 55  \\ \noalign{\vskip 1.1pt}
UGC 1501$^{\href{https://simbad.u-strasbg.fr/simbad/sim-basic?Ident=UGC+1501}{\rm{SIMBAD}}}$ & 5.19 $\pm$ 0.23 & 59 $\pm$ 3 & 72 \\ \noalign{\vskip 1.1pt}
UGC 8508$^{\href{https://simbad.u-strasbg.fr/simbad/sim-basic?Ident=UGC+8508}{\rm{SIMBAD}}}$ & 2.60 $\pm$ 0.24 & 35 $\pm$ 4 & 68 \\ \noalign{\vskip 1.1pt}
WLM$^{\href{https://simbad.u-strasbg.fr/simbad/sim-basic?Ident=WLM+galaxy}{\rm{SIMBAD}}}$ & 1.00 $\pm$ 0.04 & 38 $\pm$ 3 & 74\\ \noalign{\vskip 1.pt}
   \hline
    \end{tabular} 
    \tablefoot{Name (SIMBAD link for cross-ID), distance, outer characteristic circular speed ($V_{\rm circ,out}$), and average inclination. $V_{\rm circ,out}$ is the mean value of $V_{\rm circ}(R)$ consistent with being flat within 5\%. The uncertainties correspond to the sum in quadrature of the standard deviation in the values used to estimate $V_{\rm circ,out}$ and the median error in $V_{\rm circ}(R)$.}
\end{table}

\section{Kinematics of our remaining galaxy sample}
\label{app:kinematics}
In this work, we derived kinematic models for 22 dwarf galaxies. Figs.~\ref{fig:kinematics_examples} and \ref{fig:kinematics_rest} present the main kinematic information (velocity fields, position velocity diagrams, and kinematic radial profiles). 

\begin{figure}[H]
    \centering 
    \includegraphics[width=0.92\textwidth]{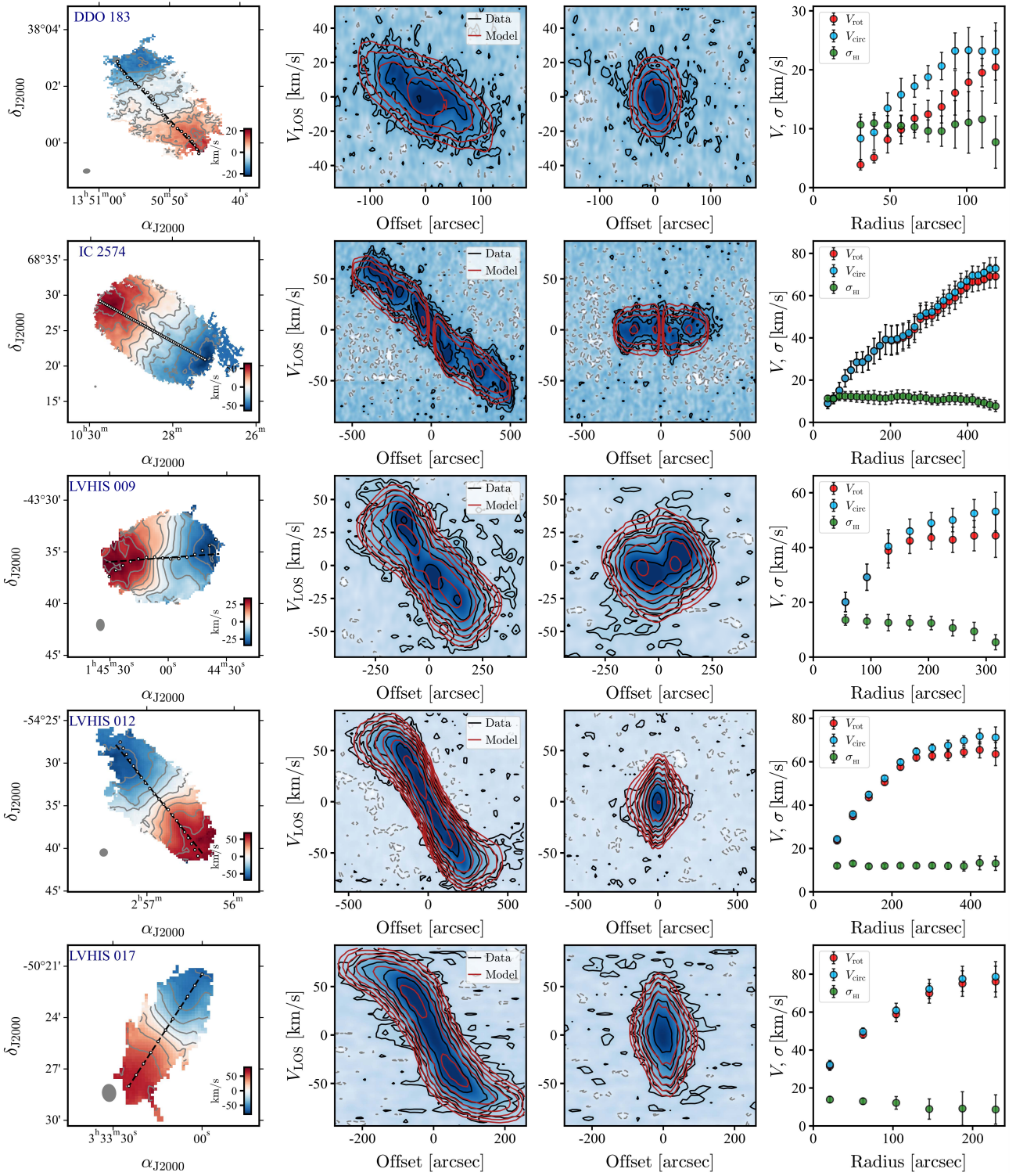}
    \caption{Kinematic models for our galaxy sample. This figure continues on the next page and is complemented by Fig.~\ref{fig:kinematics_examples}. Panels, markers, and symbols are as in Fig.~\ref{fig:kinematics_examples}}
    \label{fig:kinematics_rest}
\end{figure}

\begin{figure*}
    \ContinuedFloat
    \centering 
    \includegraphics[width=0.92\textwidth]{fig_kinmodels_extra2_final.pdf}
    \caption{Continuation.}
\end{figure*}

\begin{figure*}
    \ContinuedFloat
    \centering 
    \includegraphics[width=0.92\textwidth]{fig_kinmodels_extra3_final.pdf}
    \caption{Continuation.}
\end{figure*}

\clearpage

\section{Priors on luminosity$-\Upsilon_{\rm d}$ relations}
\label{app:m2l}
During our mass modelling, we have assumed a set of priors for the disc mass-to-light ratios ($\Upsilon_{\rm d}$). The priors are particularly important for dwarf galaxies, where the gravitational contribution from the stars is small compared to the gas and, especially, the dark matter. Our dwarfs have photometry in the $3.6\,\mu \rm{m}$ and $1.65\,\mu \rm{m}$ bands.

In practice, we fit $\log(\Upsilon_{\rm d})$ by assuming Gaussian priors on $\Upsilon_{\rm d}$. We determine the centre and standard deviation of the Gaussians from an empirical relation between NIR luminosities and $\Upsilon$, motivated by the work by \cite{marasco_mstar}. Those authors performed careful SED fitting (using up to 10 photometric bands, ranging from far-UV to mid-IR) with the software Bagpipes \citep{bagpipes} on galaxies from the SPARC database \citep{sparc}; they adopted the 2016 version of the \cite{bc03} SPs models, a \cite{kroupa2002} IMF, and the dust attenuation model from \cite{charlot2000}. \cite{marasco_mstar} found that the $M_\ast$ (and $\Upsilon^{3.6\,\mu \rm{m}}$) values derived with their method are in excellent agreement with those estimated from dynamics (rotation curve decomposition) by \cite{postinomissing}. Moreover their $\Upsilon^{3.6\,\mu \rm{m}}$ estimates match well the expectations from SPMs (e.g. \citealt{mcgaugh_ML, meidt2014, querejeta2015}).

\begin{figure}[h]
    \centering
    \begin{minipage}{0.6\linewidth}
        \includegraphics[width=\linewidth]{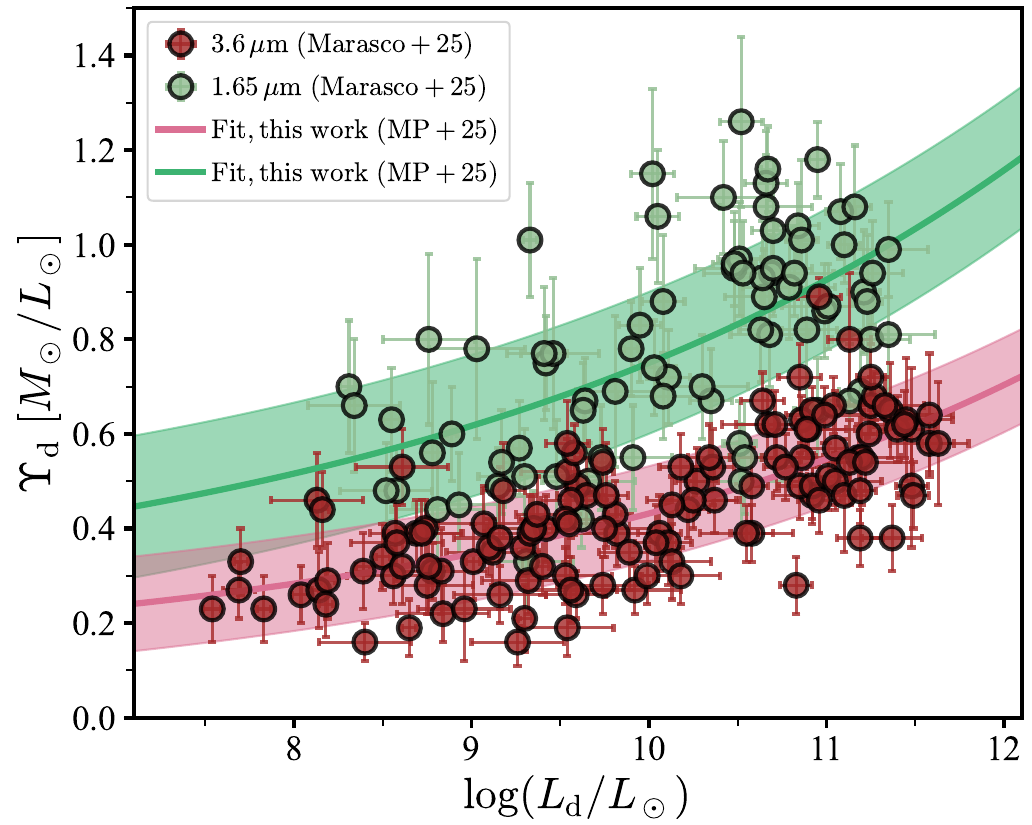}
    \end{minipage}%
    \hspace{0.05\linewidth}%
    \begin{minipage}{0.3\linewidth}
        \captionof{figure}{Relations between disc luminosity and disc mass-to-light ratio exploited to set our mass models priors. Data (from \citealt{marasco_mstar}) are shown with red (3.6\,$\mu$m) and green (1.65\,$\mu$m) markers. Our best-fitting exponential functions and the adopted uncertainties are shown with colour lines and bands. These exponential fits are imposed as priors during our mass models in Sec.~\ref{sec:massmodels}.}
        \label{fig:m2l_priors}
    \end{minipage}
\end{figure}

As shown in Fig.~\ref{fig:m2l_priors}, there is a clear trend in the data from \cite{marasco_mstar} of increasing $\Upsilon^{3.6\,\mu \rm{m}}$ with increasing $3.6\,\mu \rm{m}$ disc luminosity (red markers, from Spitzer data). In addition to this, a trend for $H-$band (1.662\,$\mu$m) can also be seen (green markers, from 2MASS data), albeit with a somewhat larger scatter, likely driven by the lower 2MASS data quality compared to Spitzer (see \citealt{marasco_mstar} for details). We exploit these correlations (assuming that our 1.65\,$\mu$m data follows the 1.662\,$\mu$m from \citealt{marasco_mstar}) and fit them with exponential profiles as a function of the disc luminosity ($L_{\rm d}$):
\begin{equation}
\Upsilon_{\rm d} = A\, e^{B\, \log(L_{\rm d}/L_\odot)} + C~.   
\end{equation}
Our best-fitting values for the 3.6\,$\mu$m data are $A=0.016, B = 0.303$, and $C=0.105$. In turn, for the 1.65\,$\mu$m data the best-fitting coefficients are $A=0.034, B = 0.277$, and $C=0.200$. The exponential functions (evaluated at a given $L_{\rm d}$) set the centre of our Gaussian priors. Based on the scatter around the best-fitting relations, we assign an uncertainty (i.e. the standard deviation of our Gaussian priors) of $0.1\,M_\odot/L_\odot$ and $0.15\,M_\odot/L_\odot$ for the 3.6\,$\mu$m and 1.65\,$\mu$m data, respectively. These uncertainties (represented with confidence bands in Fig.~\ref{fig:m2l_priors}) agree well with typically quoted values in the literature (e.g. \citealt{lelliBTFR,kirby08} and references therein). Finally, we truncate the Gaussian priors to avoid $\Upsilon_{\rm d} < 0.05$, which has no repercussions for our results but ensures that only physically motivated values are sampled during our Nested Sampling routine.
We emphasise that our fits are not associated with any particular star formation history framework. They are purely empirical descriptions that are useful as they approximate well the $L_{\rm d}-\Upsilon_{\rm d}$ relations apparent from the analysis by \cite{marasco_mstar}. We remind the reader that out of our 49 galaxies, 43 have photometry in the $3.6\,\mu \rm{m}$ band and six in $1.65\,\mu \rm{m}$.

\section{Comparison of mass models with previous works}
\label{sec:comparison}

In this Appendix, we compare the dark matter parameters found in this work for some of our galaxies against values previously reported in the literature. Specifically, we compare against 11 galaxies from \cite{read2017}, six from \cite{postinomissing}, six form \cite{li_massmodels}, and 26 from \cite{paper_massmodels}. Before delving into the comparison, we list the main differences with respect to these works\footnote{We chose the above works because they more closely resemble our analysis, but some of our galaxies have also been studied elsewhere (e.g. \citealt{begeman,deblok08,randriamampandry2014,oh2015,frank2016}).}. 

First, we highlight significant differences in the data. The differences between us and \cite{read2017} and \cite{paper_massmodels} are minor since we use the same rotation curves and stellar and gas surface density profiles. This is not the case for \cite{postinomissing} and \cite{li_massmodels}. Those works rely on the simple bulge-disc decomposition from \cite{sparc}, while we instead use the 2D bulge-disc decomposition modelling from \cite{salo2015}. Additionally, their rotation curves come from a collection of different studies with somewhat different methodologies \citep{sparc}, while our measurements come from accurate kinematic modelling using the same software and techniques. Our massive galaxies also include H$_2$ surface densities, unlike \cite{postinomissing} or \cite{li_massmodels}.

There are also important differences in the methodology. The most obvious one is the halo profiles. \cite{postinomissing} fitted NFW haloes, while \cite{read2017,li_massmodels,paper_massmodels}, and us used \textsc{coreNFW} profiles. However, \cite{read2017,li_massmodels}, and \cite{paper_massmodels} fixed $n$ and $\eta$ to the calibrations from \cite{coreNFW}, while we fixed $n=1$ and treated $\eta$ as a free parameter.
Second, different priors in halo mass were used. \cite{li_massmodels} imposed a prior centred in the abundance-matching SHMR from \cite{moster2010}, while we and the other three works used a flat prior.
Third, different priors in the concentration parameter were assumed. \cite{read2017} used a flat prior (resulting in some concentrations being very high, see below). \cite{postinomissing}, \cite{li_massmodels}, and \cite{paper_massmodels} used all a Gaussian prior centred in the $c_{200}-M_{200}$ relation from \cite{duttonmaccio2014}. Instead, we used the \citetalias{diemer2019} relation, which relies on more sophisticated techniques and is more appropriate for our halo mass range.
Fourth, there were differences in the stellar mass-to-light ratios. \cite{read2017} did not fit $\Upsilon_{\rm d}$, and instead used $M_\ast$ values from \cite{zhang} based on SED fitting. \cite{postinomissing}, \cite{li_massmodels}, and \cite{paper_massmodels} used a Gaussian prior across their full mass regime centred at $\Upsilon_{\rm d}^{3.6}=0.5$, and assumed $\Upsilon_{\rm b}^{3.6} \approx 1.4\,\Upsilon_{\rm d}^{3.6}$. We instead used the empirical priors defined in Appendix~\ref{app:m2l}.
Another difference is that only our work and \cite{paper_massmodels} considered the gas scale heights when deriving the mass models (see Sec.~\ref{sec:impact}).
Last, small differences could also arise from slightly different adopted distances and fitting methods (MCMC in the previous works vs. MC nested sampling in this paper).

 \begin{figure}[h]
     \centering
         \includegraphics[width=0.75\linewidth]{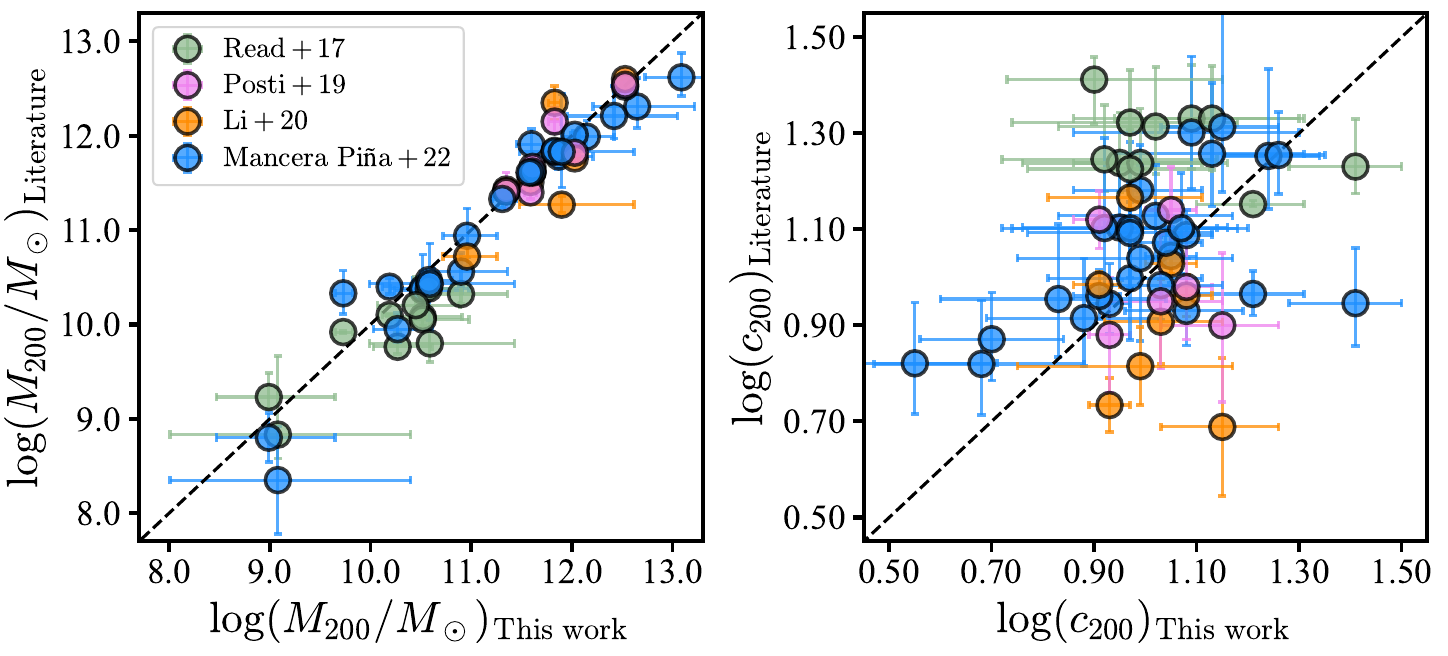}
         \caption{Comparison between halo masses and concentrations derived in this work and previous determinations.}
         \label{fig:comparison}
 \end{figure}

Fig.~\ref{fig:comparison} shows the comparison of halo masses and concentration parameters between this paper and the above studies. The halo masses are in reasonably good agreement. However, there is a slight but systematic offset of our halo masses being slightly higher than the literature values. The concentration parameters show more scatter, mainly driven by the high values from \cite{read2017}, but are usually consistent with our work within $3\,\sigma$. Most of the time, our concentrations are lower, somewhat expectedly given that the concentration-mass relation from \citetalias{diemer2019} has a lower normalisation than that of \cite{duttonmaccio2014}. Considering the numerous differences highlighted above, the agreement with literature values for the subset of galaxies in common is reasonable.

\section{The impact of potential bars in our mass models}
\label{app:bars}
As discussed in the main text, we have selected galaxies whose NIR photometry indicates only a weak presence of a bar-like structure, or no bar at all. Yet, while small, and even though we account for radial motions during our kinematic models, it is possible that the bars introduce non-circular motions that could affect our kinematic measurements (e.g. \citealt{kim2024_bar,liu2025_bar} and references therein) and therefore our mass models.

To test the potential impact of bars in our results, we have performed the following exercise\footnote{More sophisticated methods to account for bar effects (e.g. \citealt{randriamampandry2018,liu2025_bar}) exist, but their implementation is beyond our current scope.}. First, we selected our galaxies with $\log(M_\ast/M_\odot) \gtrsim 8.3$; observations indicate that below this mass threshold bars are relatively rare \citep{diazgarcia2016,erwin2018}. We do this regardless of whether or not the available bulge-disc decomposition indicates the presence of a bar (i.e. even for `pure discs'), testing an extreme scenario. Second, we estimate the extent of a putative bar ($r_{\rm bar}$) following \cite{erwin2019}, who provide NIR $r_{\rm bar}$ estimates as a function of $R_{\rm eff}$. Thirdly, we find which of our selected galaxies have rotation curve data at radii $R < r_{\rm bar}$ (we note that the maximum bar torque always occurs inside $r_{\rm bar}$, see \citealt{kim2012}), and we mask the data with $R < r_{\rm bar}$ and re-derive our mass models.
We find that the mass models derived with the masked data are in excellent agreement with our fiducial values, with the differences being usually smaller than $0.5\,\sigma$ and always smaller than $1.5\,\sigma$. Therefore, we conclude that thanks to our selection cuts and our accounting for potential radial motions in the kinematics, our mass models are robust against the presence of small bars.

\section{Could the baryon-deficient dwarfs result from observational or modelling errors?}
\label{app:vmax}

As discussed in Sec.~\ref{sec:scalinglaws}, our analysis revealed a population of baryon-deficient dwarfs (BDDs) with $20-60$ times less (more) stellar (dark matter halo) mass than expected from theoretical models (see Figs.~\ref{fig:shmr}-\ref{fig:bhmr}). In this Appendix, we investigate the possibility that the feature arises artificially by an overestimation of $M_{200}$. We note that an underestimation of $M_\ast$ by  $\sim1-1.5$ orders of magnitude is extremely improbable according to our dynamical modelling and in general with $\Upsilon$ values from SED fitting and SPMs (e.g. \citealt{schombert2022,marasco_mstar}). Instead, we consider three potential reasons for a halo mass overestimation (we note in advance that these factors do not explain the observations).

The first explanation could be that the $M_{200}$ values are overestimated due to observational errors. For instance, if our inclinations were underestimated, $V_{\rm circ}$ would be biased high, and so would $M_{200}$. However, the BDDs are DDO 190, LVHIS 017, LVHIS 072, LVHIS 080, and UGC 8508 (and note from Fig.~\ref{fig:shmr} that more galaxies are on the verge of surpassing our threshold), which are seen at high inclinations ($i\sim70^\circ$, except for DDO 190 with $i=41^\circ$) and therefore there is little room for a significant change lowering $V_{\rm circ}$. We also checked that their $M_{200}$ are robust to reasonable changes in the deprojection of the highly inclined surface density profiles (for which the correction is larger). Moreover, there are other galaxies with $i\sim70^\circ$ that are not baryon-deficient as well as systems of lower inclination close to our baryon-deficient threshold. Since by selection all the distances to our galaxies are well constrained, we do not find it likely that observational errors are the culprit. 

A second explanation we explore is that those galaxies may have suffered from strong adiabatic contraction (e.g. \citealt{blumenthal1986,dutton2016}) so that our fit interprets a central density excess as an overly massive halo. However, adiabatic contraction could elevate the concentration parameters but does not bias the recovery of halo masses \citep{coreEinasto,li_adiabatic}. Moreover, the gap galaxies do not have high $c_{200}$ values for their mass (if anything the opposite, see Fig.~\ref{fig:c200M200}), so it is unlikely that adiabatic contraction is playing a significant role. 

\begin{figure*}[!h]
    \centering
\includegraphics[width=1\linewidth]{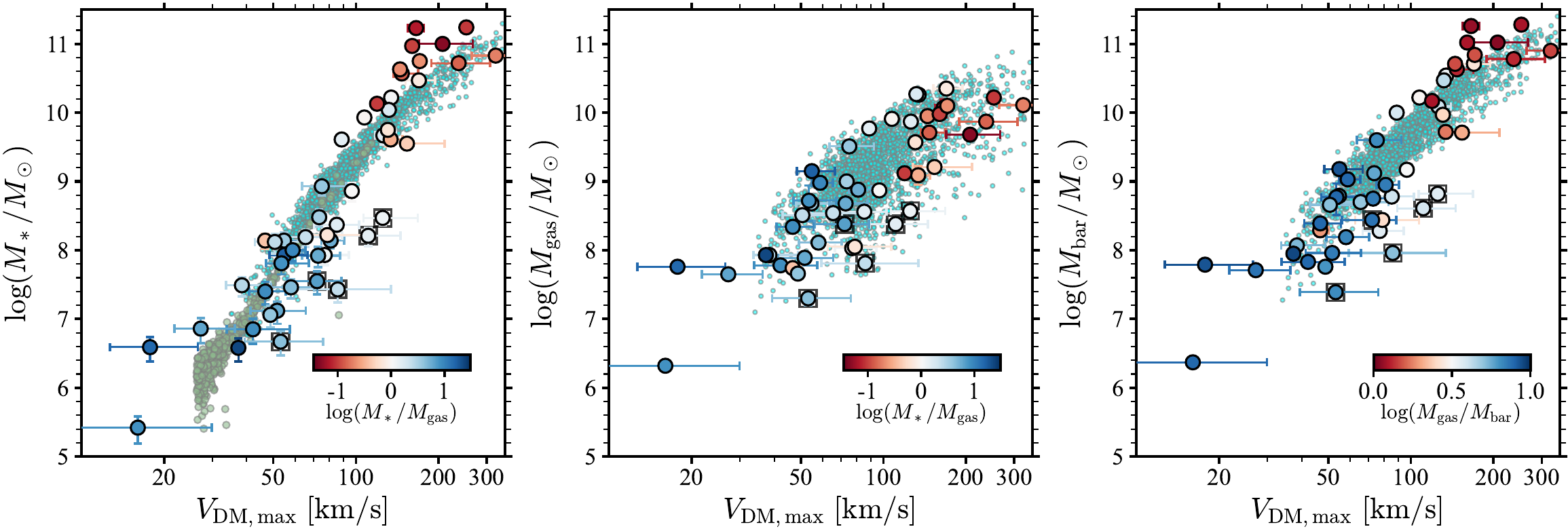}
    \caption{Stellar (\emph{left}), gas (\emph{middle}), and baryonic (\emph{right}) mass of our galaxies as a function of the maximum circular speed of their host dark matter haloes. Symbols and colours are as in Fig.~\ref{fig:shmr}. }
    \label{fig:vmax}
\end{figure*}

The third possibility could be that our $M_{200}$ values are biased by the large extrapolation between $R_{200}$ (between tens and a couple of hundred kiloparsecs) and the radial extent of our rotation curves (a few kiloparsecs for dwarfs, tens of kiloparsecs for massive galaxies). To test this, we use another proxy for $M_{200}$, namely the maximum circular speed of the halo $V_{\rm DM,max}$. This reduces the impact of any extrapolation (and mergers or environmental processes) since $V_{\rm DM,max}$ typically occurs at a radius within or close to the radial coverage of our rotation curves. To measure $V_{\rm DM,max}$, we sample the posterior distributions of our best-fitting mass models and extract the maximum circular speed of the halo. 
Fig.~\ref{fig:vmax} shows $M_\ast$, $M_{\rm gas}$, and $M_{\rm bar}$ as a function of $V_{\rm DM,max}$. Symbols and colours are as in Fig.~\ref{fig:shmr}. Reassuringly, the same observed features and diversity in $M_\ast$, $M_{\rm gas}$, and $M_{\rm bar}$ at fixed $M_{200}$ can be seen at fixed $V_{\rm DM,max}$. This indicates that the baryon fractions studied in Sec.~\ref{sec:scalinglaws} are robust, and our $M_{200}$ estimates are not significantly affected by the extrapolation between the extent of our rotation curves and $R_{200}$. 

From all of the above, we consider it highly unlikely that the baryon-deficient population is spurious. Moreover, additional and independent weak-lensing measurements appear to agree with our results (see Sec.~\ref{fig:shmr} and Fig.~\ref{fig:shmr_lensing}).

\clearpage

\section{Halo best-fitting parameters and baryonic budget}
Table~\ref{tab:mm} lists the best-fitting dark matter halo parameters obtained through our mass modelling (Sec.~\ref{sec:massmodels}). Table~\ref{tab:mbar} provides the information on the baryonic content of our galaxy sample (Sec.~\ref{sec:scalinglaws}).

\begin{table*}[!h]
    \caption{Best-fitting mass model parameters}
    \label{tab:mm}
    \resizebox{1\textwidth}{!}{
    \centering
    \begin{tabular}{lccccccccccccccc}
    \hline \noalign{\vskip 1.1pt} 
       ID  & \multicolumn{3}{c}{$\log(M_{200}/M_\odot)$} & \multicolumn{3}{c}{$\log(c_{200})$} & \multicolumn{3}{c}{$\log(\eta)$} & \multicolumn{3}{c}{$\log(\Upsilon_{\rm d}/[M_\odot/L_\odot])$}  & \multicolumn{3}{c}{$\log(\Upsilon_{\rm b}/[M_\odot/L_\odot])$} \\ 
       \noalign{\vskip 1.1pt} 
       & 50th pctl & $\sigma_-$ & $\sigma_+$ & 50th pctl & $\sigma_-$ & $\sigma_+$ & 50th pctl & $\sigma_-$ & $\sigma_+$  & 50th pctl & $\sigma_-$ & $\sigma_+$ & 50th pctl & $\sigma_-$ & $\sigma_+$  \\ \noalign{\vskip 1.1pt}
       \hline \noalign{\vskip 1.1pt} 
CVn I dW & 8.99 & 0.52 & 0.66 & 1.13 & 0.19 & 0.18 & 0.13 & 0.51 & 0.31 & -0.60 & 0.21 & 0.14 & nan & nan & nan \\  \noalign{\vskip 1.1pt}
DDO 126 & 10.27 & 0.24 & 0.34 & 1.02 & 0.19 & 0.15 & 0.23 & 0.17 & 0.18 & -0.56 & 0.19 & 0.13 & nan & nan & nan \\  \noalign{\vskip 1.1pt}
DDO 133 & 10.45 & 0.38 & 0.44 & 0.90 & 0.17 & 0.25 & -0.48 & 0.29 & 0.31 & -0.49 & 0.16 & 0.12 & nan & nan & nan \\  \noalign{\vskip 1.1pt}
DDO 154 & 10.19 & 0.07 & 0.11 & 1.21 & 0.11 & 0.10 & 0.44 & 0.10 & 0.09 & -0.55 & 0.18 & 0.13 & nan & nan & nan \\  \noalign{\vskip 1.1pt}
DDO 168 & 10.90 & 0.27 & 0.46 & 0.97 & 0.20 & 0.16 & 0.18 & 0.14 & 0.15 & -0.52 & 0.17 & 0.12 & nan & nan & nan \\  \noalign{\vskip 1.1pt}
DDO 181 & 10.20 & 0.39 & 0.50 & 0.93 & 0.17 & 0.16 & 0.46 & 0.14 & 0.08 & -0.62 & 0.21 & 0.15 & nan & nan & nan \\  \noalign{\vskip 1.1pt}
DDO 183 & 9.59 & 0.38 & 0.48 & 0.99 & 0.18 & 0.17 & 0.44 & 0.18 & 0.10 & -0.64 & 0.21 & 0.15 & nan & nan & nan \\  \noalign{\vskip 1.1pt}
DDO 190 & 11.22 & 0.66 & 0.82 & 0.88 & 0.23 & 0.22 & -0.19 & 0.48 & 0.28 & -0.54 & 0.18 & 0.13 & nan & nan & nan \\  \noalign{\vskip 1.1pt}
DDO 210 & 9.08 & 1.07 & 1.32 & 1.09 & 0.23 & 0.21 & 0.27 & 0.46 & 0.22 & -0.67 & 0.23 & 0.16 & nan & nan & nan \\  \noalign{\vskip 1.1pt}
DDO 52 & 10.52 & 0.28 & 0.39 & 0.95 & 0.19 & 0.21 & -0.11 & 0.28 & 0.24 & -0.51 & 0.16 & 0.12 & nan & nan & nan \\  \noalign{\vskip 1.1pt}
DDO 87 & 10.52 & 0.32 & 0.46 & 0.92 & 0.20 & 0.22 & -0.29 & 0.32 & 0.26 & -0.52 & 0.16 & 0.12 & nan & nan & nan \\  \noalign{\vskip 1.1pt}
IC 2574 & 10.96 & 0.24 & 0.30 & 0.97 & 0.16 & 0.14 & 0.31 & 0.21 & 0.17 & -0.41 & 0.13 & 0.10 & nan & nan & nan \\  \noalign{\vskip 1.1pt}
LVHIS 009 & 10.51 & 0.23 & 0.31 & 0.98 & 0.18 & 0.17 & 0.28 & 0.23 & 0.18 & -0.50 & 0.15 & 0.12 & nan & nan & nan \\  \noalign{\vskip 1.1pt}
LVHIS 011 & 11.10 & 0.15 & 0.21 & 0.85 & 0.14 & 0.10 & 0.49 & 0.10 & 0.06 & -0.40 & 0.10 & 0.08 & nan & nan & nan \\  \noalign{\vskip 1.1pt}
LVHIS 012 & 10.89 & 0.09 & 0.15 & 0.97 & 0.16 & 0.14 & 0.26 & 0.13 & 0.14 & -0.46 & 0.15 & 0.11 & nan & nan & nan \\  \noalign{\vskip 1.1pt}
LVHIS 017 & 11.55 & 0.37 & 0.49 & 0.81 & 0.15 & 0.13 & -0.68 & 0.21 & 0.25 & -0.49 & 0.15 & 0.11 & nan & nan & nan \\  \noalign{\vskip 1.1pt}
LVHIS 019 & 10.73 & 0.37 & 0.47 & 0.85 & 0.18 & 0.19 & 0.02 & 0.62 & 0.38 & -0.31 & 0.16 & 0.12 & nan & nan & nan \\  \noalign{\vskip 1.1pt}
LVHIS 020 & 10.81 & 0.32 & 0.42 & 0.88 & 0.18 & 0.16 & 0.45 & 0.15 & 0.09 & -0.27 & 0.14 & 0.11 & nan & nan & nan \\  \noalign{\vskip 1.1pt}
LVHIS 026 & 10.98 & 0.26 & 0.37 & 0.95 & 0.13 & 0.10 & -0.77 & 0.16 & 0.18 & -0.29 & 0.15 & 0.11 & nan & nan & nan \\  \noalign{\vskip 1.1pt}
LVHIS 055 & 10.04 & 0.15 & 0.20 & 1.31 & 0.13 & 0.12 & 0.22 & 0.36 & 0.18 & -0.36 & 0.12 & 0.09 & nan & nan & nan \\  \noalign{\vskip 1.1pt}
LVHIS 060 & 10.59 & 0.40 & 0.47 & 0.79 & 0.15 & 0.14 & -0.73 & 0.18 & 0.26 & -0.38 & 0.19 & 0.13 & nan & nan & nan \\  \noalign{\vskip 1.1pt}
LVHIS 072 & 11.66 & 0.32 & 0.55 & 0.89 & 0.21 & 0.17 & 0.35 & 0.11 & 0.12 & -0.25 & 0.11 & 0.09 & nan & nan & nan \\  \noalign{\vskip 1.1pt}
LVHIS 077 & 11.13 & 0.10 & 0.13 & 1.13 & 0.09 & 0.09 & 0.13 & 0.08 & 0.08 & -0.55 & 0.15 & 0.11 & nan & nan & nan \\  \noalign{\vskip 1.1pt}
LVHIS 078 & 11.04 & 0.43 & 0.64 & 0.96 & 0.20 & 0.16 & -0.87 & 0.09 & 0.16 & -0.45 & 0.12 & 0.09 & nan & nan & nan \\  \noalign{\vskip 1.1pt}
LVHIS 080 & 11.14 & 0.49 & 0.46 & 0.67 & 0.15 & 0.21 & -0.50 & 0.32 & 0.52 & -0.58 & 0.19 & 0.14 & nan & nan & nan \\  \noalign{\vskip 1.1pt}
NGC 0253 & 12.65 & 0.44 & 0.57 & 0.68 & 0.21 & 0.20 & 0.40 & 0.70 & 0.13 & -0.41 & 0.06 & 0.05 & -0.26 & 0.06 & 0.05 \\  \noalign{\vskip 1.1pt}
NGC 0925 & 11.63 & 0.15 & 0.23 & 0.90 & 0.15 & 0.14 & 0.18 & 0.12 & 0.13 & -0.52 & 0.13 & 0.10 & nan & nan & nan \\  \noalign{\vskip 1.1pt}
NGC 1313 & 11.60 & 0.15 & 0.21 & 1.08 & 0.12 & 0.11 & 0.18 & 0.13 & 0.10 & -0.35 & 0.09 & 0.07 & -0.16 & 0.12 & 0.10 \\  \noalign{\vskip 1.1pt}
NGC 2403 & 11.59 & 0.06 & 0.06 & 1.05 & 0.05 & 0.05 & -0.72 & 0.21 & 0.63 & -0.34 & 0.09 & 0.08 & nan & nan & nan \\  \noalign{\vskip 1.1pt}
NGC 2541 & 10.96 & 0.05 & 0.06 & 1.23 & 0.08 & 0.09 & 0.01 & 0.07 & 0.07 & -0.29 & 0.07 & 0.06 & nan & nan & nan \\  \noalign{\vskip 1.1pt}
NGC 2841 & 12.53 & 0.04 & 0.04 & 0.93 & 0.04 & 0.04 & -0.31 & 0.46 & 0.30 & -0.07 & 0.03 & 0.03 & 0.09 & 0.05 & 0.05 \\  \noalign{\vskip 1.1pt}
NGC 3198 & 11.61 & 0.02 & 0.03 & 1.08 & 0.05 & 0.05 & -0.01 & 0.09 & 0.08 & -0.34 & 0.07 & 0.06 & nan & nan & nan \\  \noalign{\vskip 1.1pt}
NGC 3351 & 11.87 & 0.20 & 0.34 & 0.88 & 0.19 & 0.15 & -0.48 & 0.35 & 0.35 & -0.21 & 0.05 & 0.05 & -0.02 & 0.04 & 0.04 \\  \noalign{\vskip 1.1pt}
NGC 3621 & 11.61 & 0.07 & 0.07 & 1.04 & 0.06 & 0.07 & 0.20 & 0.19 & 0.14 & -0.50 & 0.08 & 0.06 & nan & nan & nan \\  \noalign{\vskip 1.1pt}
NGC 3992 & 11.83 & 0.06 & 0.06 & 1.15 & 0.12 & 0.11 & -0.47 & 0.37 & 0.27 & -0.12 & 0.06 & 0.05 & 0.08 & 0.11 & 0.09 \\  \noalign{\vskip 1.1pt}
NGC 4535 & 13.09 & 0.36 & 0.40 & 0.55 & 0.15 & 0.16 & -0.77 & 0.15 & 0.20 & -0.28 & 0.05 & 0.04 & -0.04 & 0.07 & 0.05 \\  \noalign{\vskip 1.1pt}
NGC 4536 & 12.15 & 0.23 & 0.27 & 0.70 & 0.14 & 0.14 & -0.49 & 0.36 & 0.45 & -0.52 & 0.09 & 0.05 & -0.45 & 0.04 & 0.04 \\  \noalign{\vskip 1.1pt}
NGC 4559 & 11.35 & 0.09 & 0.10 & 1.03 & 0.11 & 0.12 & 0.23 & 0.50 & 0.18 & -0.36 & 0.09 & 0.07 & -0.19 & 0.13 & 0.10 \\  \noalign{\vskip 1.1pt}
NGC 4651 & 11.58 & 0.08 & 0.09 & 1.24 & 0.09 & 0.10 & 0.17 & 0.90 & 0.17 & -0.22 & 0.05 & 0.03 & 0.01 & 0.09 & 0.06 \\  \noalign{\vskip 1.1pt}
NGC 4725 & 11.83 & 0.08 & 0.10 & 1.07 & 0.14 & 0.13 & -0.27 & 0.48 & 0.33 & -0.19 & 0.06 & 0.05 & -0.03 & 0.10 & 0.08 \\  \noalign{\vskip 1.1pt}
NGC 4736 & 11.31 & 0.10 & 0.14 & 1.26 & 0.11 & 0.09 & -0.02 & 0.59 & 0.20 & -0.57 & 0.05 & 0.04 & -0.54 & 0.03 & 0.03 \\  \noalign{\vskip 1.1pt}
NGC 5005 & 12.42 & 0.46 & 0.63 & 0.83 & 0.23 & 0.21 & -0.52 & 0.33 & 0.37 & -0.20 & 0.05 & 0.04 & -0.11 & 0.03 & 0.03 \\  \noalign{\vskip 1.1pt}
NGC 5055 & 12.03 & 0.04 & 0.04 & 0.91 & 0.05 & 0.04 & 0.20 & 0.09 & 0.08 & -0.36 & 0.03 & 0.02 & nan & nan & nan \\  \noalign{\vskip 1.1pt}
NGC 7793 & 11.90 & 0.42 & 0.72 & 0.99 & 0.24 & 0.18 & -0.05 & 0.19 & 0.14 & -0.33 & 0.09 & 0.07 & nan & nan & nan \\  \noalign{\vskip 1.1pt}
NGC 2366 & 10.58 & 0.16 & 0.23 & 0.99 & 0.13 & 0.12 & -0.07 & 0.10 & 0.11 & -0.51 & 0.16 & 0.12 & nan & nan & nan \\  \noalign{\vskip 1.1pt}
NGC 4190 & 9.89 & 0.30 & 0.47 & 1.20 & 0.19 & 0.15 & -0.22 & 0.36 & 0.23 & -0.50 & 0.15 & 0.11 & nan & nan & nan \\  \noalign{\vskip 1.1pt}
UGC 1501 & 11.28 & 0.39 & 0.52 & 0.83 & 0.18 & 0.16 & 0.45 & 0.13 & 0.09 & -0.55 & 0.15 & 0.11 & nan & nan & nan \\  \noalign{\vskip 1.1pt}
UGC 8508 & 10.59 & 0.60 & 0.84 & 0.97 & 0.23 & 0.21 & -0.14 & 0.33 & 0.26 & -0.59 & 0.20 & 0.14 & nan & nan & nan \\  \noalign{\vskip 1.1pt}
WLM & 9.73 & 0.08 & 0.09 & 1.41 & 0.13 & 0.09 & 0.40 & 0.15 & 0.11 & -0.58 & 0.20 & 0.14 & nan & nan & nan \\  \noalign{\vskip 1.1pt}
 \hline
    \end{tabular}} 
    \tablefoot{Halo mass, concentration, $\eta$ parameter, and disc and bulge mass-to-light ratios (with nan values for galaxies without bulges). The first column of each parameter represents the 50th percentile (median) of its corresponding posterior distribution. The lower and upper uncertainties correspond to the absolute difference between the distribution's median and the 16th and 84th percentiles, respectively. The corner plots of the posterior distributions of all our mass models are available at \href{https://www.dropbox.com/scl/fo/oc1tai7t4f4vp6f7iivls/AJOFSqUDfltJ9x3DNsSQSzE?rlkey=9iocjnkbzn4zfydy2p6effito&st=j2mcmbfi&dl=0}{this link}.}
\end{table*}

\begin{landscape}

\begin{table*}[h]
    \caption{Baryonic budget of our dwarf galaxy sample.}\vspace{-0.2cm}
    \label{tab:mbar}
    \resizebox{1.28\textwidth}{!}{
    \centering
    \begin{tabular}{lccccccccccccccccccccc}
    \hline \noalign{\vskip 1.1pt} 
       ID  & \multicolumn{3}{c}{$\log(M_\ast/M_\odot)$} & \multicolumn{3}{c}{$\log(M_{\rm gas}/M_\odot)$} & \multicolumn{3}{c}{$\log(M_{\rm bar}/M_\odot)$} & \multicolumn{3}{c}{$f_{\rm gas}$} & \multicolumn{3}{c}{$\log(\tilde{f}_{\rm \ast})$} & \multicolumn{3}{c}{$\log(\tilde{f}_{\rm gas})$} & \multicolumn{3}{c}{$\log(\tilde{f}_{\rm bar})$} \\ 
       \noalign{\vskip 1.1pt} 
       & 50th pctl & $\sigma_-$ & $\sigma_+$ & 50th pctl & $\sigma_-$ & $\sigma_+$ & 50th pctl & $\sigma_-$ & $\sigma_+$ & 50th pctl & $\sigma_-$ & $\sigma_+$ & 50th pctl & $\sigma_-$ & $\sigma_+$ & 50th pctl & $\sigma_-$ & $\sigma_+$ & 50th pctl & $\sigma_-$ & $\sigma_+$ \\ \noalign{\vskip 1.1pt}
       \hline \noalign{\vskip 1.1pt} 
CVn I dW A  &   6.59   &  0.21  &  0.15   &  7.76   &  0.06   &  0.06    &  7.79    & 0.06   &  0.06   &   0.94   &  0.03   &  0.02   &   -1.65   &  0.67   &  0.58   &   -0.44  &   0.66   &  0.54   &   -0.41  &   0.66  &   0.54          \\  \noalign{\vskip 1.pt}      
DDO 52      &   7.93   &  0.17  &  0.12   &  9.15   &  0.06   &  0.06    &  9.18    & 0.05   &  0.06   &   0.94   &  0.02   &  0.02   &   -1.81   &  0.42   &  0.31   &   -0.57  &   0.39   &  0.28   &   -0.54  &   0.39  &   0.28          \\  \noalign{\vskip 1.pt}      
DDO 87      &   7.81   &  0.16  &  0.12   &  8.72   &  0.30   &  0.22    &  8.77    & 0.26   &  0.20   &   0.89   &  0.10   &  0.05   &   -1.94   &  0.49   &  0.34   &   -1.05  &   0.55   &  0.41   &   -0.99  &   0.52  &   0.39          \\  \noalign{\vskip 1.pt}      
DDO 126     &   7.40   &  0.19  &  0.14   &  8.34   &  0.08   &  0.07    &  8.39    & 0.07   &  0.07   &   0.90   &  0.04   &  0.04   &   -2.12   &  0.38   &  0.31   &   -1.15  &   0.34   &  0.26   &   -1.10  &   0.34  &   0.26          \\  \noalign{\vskip 1.pt}      
DDO 133     &   8.12   &  0.17  &  0.12   &  8.51   &  0.06   &  0.06    &  8.66    & 0.06   &  0.06   &   0.71   &  0.07   &  0.07   &   -1.56   &  0.49   &  0.41   &   -1.15  &   0.45   &  0.39   &   -1.00  &   0.46  &   0.38      \\  \noalign{\vskip 1.pt}      
DDO 154     &   7.06   &  0.18  &  0.13   &  7.66   &  0.05   &  0.04    &  7.76    & 0.05   &  0.04   &   0.80   &  0.06   &  0.06   &   -2.36   &  0.21   &  0.15   &   -1.73  &   0.12   &  0.09   &   -1.64  &   0.12  &   0.09          \\  \noalign{\vskip 1.pt}
DDO 168     &   7.92   &  0.17  &  0.13   &  8.68   &  0.07   &  0.07    &  8.75    & 0.06   &  0.06   &   0.85   &  0.05   &  0.05   &   -2.22   &  0.47   &  0.32   &   -1.43  &   0.46   &  0.28   &   -1.36  &   0.46  &   0.28          \\  \noalign{\vskip 1.pt}      
DDO 181     &   6.85   &  0.21  &  0.15   &  7.78   &  0.05   &  0.05    &  7.83    & 0.05   &  0.05   &   0.90   &  0.04   &  0.04   &   -2.59   &  0.51   &  0.43   &   -1.63  &   0.50   &  0.40   &   -1.58  &   0.50  &   0.40          \\  \noalign{\vskip 1.pt}      
DDO 183     &   6.86   &  0.21  &  0.15   &  7.65   &  0.06   &  0.05    &  7.71    & 0.05   &  0.05   &   0.86   &  0.05   &  0.05   &   -1.97   &  0.53   &  0.43   &   -1.15  &   0.49   &  0.37   &   -1.08  &   0.49  &   0.37          \\  \noalign{\vskip 1.pt}      
DDO 190     &   7.43   &  0.19  &  0.13   &  7.81   &  0.05   &  0.05    &  7.96    & 0.06   &  0.06   &   0.71   &  0.07   &  0.08   &   -3.03   &  0.84   &  0.71   &   -2.62  &   0.82   &  0.67   &   -2.46  &   0.83  &   0.68          \\  \noalign{\vskip 1.pt}      
DDO 210     &   5.42   &  0.23  &  0.16   &  6.32   &  0.06   &  0.06    &  6.37    & 0.06   &  0.05   &   0.89   &  0.05   &  0.04   &   -2.92   &  1.31   &  1.11   &   -1.98  &   1.32   &  1.08   &   -1.92  &   1.31  &   1.08          \\  \noalign{\vskip 1.pt}      
IC 2574     &   8.93   &  0.13  &  0.10   &  9.51   &  0.05   &  0.04    &  9.60    & 0.04   &  0.04   &   0.81   &  0.04   &  0.04   &   -1.31   &  0.31   &  0.26   &   -0.66  &   0.30   &  0.24   &   -0.57  &   0.30  &   0.23          \\  \noalign{\vskip 1.pt}      
LVHIS 009   &   8.14   &  0.16  &  0.12   &  8.68   &  0.06   &  0.06    &  8.79    & 0.06   &  0.05   &   0.78   &  0.06   &  0.06   &   -1.60   &  0.33   &  0.28   &   -1.03  &   0.32   &  0.24   &   -0.92  &   0.32  &   0.24          \\  \noalign{\vskip 1.pt}      
LVHIS 011   &   8.14   &  0.11  &  0.09   &  8.88   &  0.05   &  0.04    &  8.95    & 0.04   &  0.04   &   0.85   &  0.03   &  0.03   &   -2.18   &  0.21   &  0.17   &   -1.42  &   0.22   &  0.16   &   -1.35  &   0.21  &   0.16          \\  \noalign{\vskip 1.pt}      
LVHIS 012   &   8.48   &  0.14  &  0.11   &  9.00   &  0.05   &  0.04    &  9.12    & 0.05   &  0.04   &   0.77   &  0.05   &  0.06   &   -1.64   &  0.19   &  0.15   &   -1.10  &   0.15   &  0.10   &   -0.98  &   0.15  &   0.10          \\  \noalign{\vskip 1.pt}      
LVHIS 017   &   8.21   &  0.15  &  0.12   &  8.38   &  0.05   &  0.05    &  8.61    & 0.06   &  0.06   &   0.60   &  0.07   &  0.08   &   -2.58   &  0.50   &  0.37   &   -2.38  &   0.51   &  0.37   &   -2.16  &   0.50  &   0.36          \\  \noalign{\vskip 1.pt}      
LVHIS 019   &   7.46   &  0.16  &  0.12   &  8.11   &  0.06   &  0.06    &  8.19    & 0.06   &  0.05   &   0.82   &  0.05   &  0.05   &   -2.50   &  0.51   &  0.43   &   -1.83  &   0.47   &  0.38   &   -1.74  &   0.47  &   0.39          \\  \noalign{\vskip 1.pt}      
LVHIS 020   &   8.19   &  0.14  &  0.11   &  8.54   &  0.06   &  0.06    &  8.70    & 0.06   &  0.05   &   0.70   &  0.06   &  0.07   &   -1.85   &  0.41   &  0.34   &   -1.47  &   0.41   &  0.32   &   -1.31  &   0.41  &   0.32          \\  \noalign{\vskip 1.pt}      
LVHIS 026   &   7.93   &  0.15  &  0.11   &  8.03   &  0.06   &  0.06    &  8.28    & 0.07   &  0.06   &   0.56   &  0.07   &  0.09   &   -2.28   &  0.38   &  0.27   &   -2.16  &   0.37   &  0.26   &   -1.91  &   0.36  &   0.26          \\  \noalign{\vskip 1.pt}      
LVHIS 055   &   8.14   &  0.12  &  0.10   &  7.74   &  0.06   &  0.06    &  8.29    & 0.09   &  0.07   &   0.29   &  0.05   &  0.07   &   -1.13   &  0.21   &  0.18   &   -1.51  &   0.20   &  0.16   &   -0.97  &   0.20  &   0.16          \\  \noalign{\vskip 1.pt}      
LVHIS 060   &   7.12   &  0.19  &  0.13   &  7.89   &  0.06   &  0.06    &  7.96    & 0.06   &  0.05   &   0.85   &  0.05   &  0.05   &   -2.70   &  0.51   &  0.44   &   -1.90  &   0.47   &  0.41   &   -1.84  &   0.47  &   0.41          \\  \noalign{\vskip 1.pt}      
LVHIS 072   &   8.47   &  0.11  &  0.09   &  8.57   &  0.06   &  0.06    &  8.82    & 0.06   &  0.05   &   0.56   &  0.06   &  0.07   &   -2.41   &  0.55   &  0.33   &   -2.30  &   0.54   &  0.32   &   -2.04  &   0.55  &   0.32          \\  \noalign{\vskip 1.pt}      
LVHIS 077   &   8.86   &  0.15  &  0.12   &  8.87   &  0.06   &  0.06    &  9.17    & 0.08   &  0.07   &   0.51   &  0.08   &  0.09   &   -1.50   &  0.18   &  0.15   &   -1.47  &   0.14   &  0.12   &   -1.18  &   0.14  &   0.12          \\  \noalign{\vskip 1.pt}      
LVHIS 078   &   8.22   &  0.12  &  0.09   &  8.05   &  0.06   &  0.06    &  8.44    & 0.07   &  0.06   &   0.40   &  0.06   &  0.07   &   -2.03   &  0.60   &  0.38   &   -2.19  &   0.66   &  0.43   &   -1.80  &   0.62  &   0.39          \\  \noalign{\vskip 1.pt}      
LVHIS 080   &   7.55   &  0.19  &  0.14   &  8.38   &  0.06   &  0.06    &  8.44    & 0.05   &  0.05   &   0.87   &  0.05   &  0.04   &   -2.83   &  0.50   &  0.55   &   -1.97  &   0.46   &  0.49   &   -1.91  &   0.46  &   0.50     \\  \noalign{\vskip 1.pt}      
NGC 0253    &   10.72  &  0.05  &  0.03   &  9.87   &  0.05   &  0.05    &  10.78   & 0.04   &  0.03   &   0.13   &  0.01   &  0.02   &   -1.12   &  0.59   &  0.43   &   -1.98  &   0.57   &  0.45   &   -1.07  &   0.59  &   0.44          \\  \noalign{\vskip 1.pt}      
NGC 0925    &   9.67   &  0.13  &  0.10   &  9.87   &  0.05   &  0.05    &  10.09   & 0.06   &  0.05   &   0.62   &  0.06   &  0.07   &   -1.19   &  0.25   &  0.18   &   -0.97  &   0.23   &  0.16   &   -0.75  &   0.23  &   0.15          \\  \noalign{\vskip 1.pt}      
NGC 1313    &   9.61   &  0.09  &  0.07   &  9.09   &  0.14   &  0.13    &  9.72    & 0.08   &  0.06   &   0.24   &  0.06   &  0.07   &   -1.21   &  0.22   &  0.16   &   -1.74  &   0.24   &  0.21   &   -1.09  &   0.21  &   0.15          \\  \noalign{\vskip 1.pt}      
NGC 2366    &   8.00   &  0.16  &  0.13   &  8.98   &  0.09   &  0.08    &  9.03    & 0.08   &  0.07   &   0.91   &  0.04   &  0.03   &   -1.80   &  0.31   &  0.21   &   -0.81  &   0.24   &  0.18   &   -0.77  &   0.24  &   0.18          \\  \noalign{\vskip 1.pt}      
NGC 2403    &   9.75   &  0.09  &  0.08   &  9.57   &  0.06   &  0.05    &  9.97    & 0.06   &  0.06   &   0.40   &  0.05   &  0.06   &   -1.06   &  0.06   &  0.07   &   -1.23  &   0.08   &  0.08   &   -0.83  &   0.05  &   0.05          \\  \noalign{\vskip 1.pt}      
NGC 2541    &   9.61   &  0.07  &  0.06   &  9.77   &  0.05   &  0.05    &  10.00   & 0.04   &  0.04   &   0.59   &  0.04   &  0.05   &   -0.55   &  0.08   &  0.08   &   -0.40  &   0.07   &  0.07   &   -0.16  &   0.06  &   0.06          \\  \noalign{\vskip 1.pt}      
NGC 2841    &   11.24  &  0.02  &  0.03   &  10.22  &  0.06   &  0.05    &  11.28   & 0.02   &  0.02   &   0.09   &  0.01   &  0.01   &   -0.49   &  0.03   &  0.03   &   -1.52  &   0.07   &  0.07   &   -0.45  &   0.03  &   0.03          \\  \noalign{\vskip 1.pt}      
NGC 3198    &   10.22  &  0.07  &  0.06   &  10.26  &  0.06   &  0.06    &  10.54   & 0.05   &  0.04   &   0.52   &  0.05   &  0.05   &   -0.59   &  0.07   &  0.06   &   -0.56  &   0.07   &  0.06   &   -0.27  &   0.04  &   0.04          \\  \noalign{\vskip 1.pt}      
NGC 3351    &   10.57  &  0.04  &  0.03   &  9.71   &  0.05   &  0.04    &  10.63   & 0.04   &  0.03   &   0.12   &  0.01   &  0.02   &   -0.49   &  0.33   &  0.16   &   -1.36  &   0.34   &  0.20   &   -0.44  &   0.33  &   0.16          \\  \noalign{\vskip 1.pt}      
NGC 3621    &   10.04  &  0.08  &  0.06   &  10.27  &  0.07   &  0.07    &  10.47   & 0.05   &  0.05   &   0.63   &  0.05   &  0.05   &   -0.78   &  0.11   &  0.11   &   -0.54  &   0.10   &  0.10   &   -0.35  &   0.08  &   0.09          \\  \noalign{\vskip 1.pt}      
NGC 3992    &   11.23  &  0.06  &  0.05   &  10.07  &  0.14   &  0.12    &  11.26   & 0.06   &  0.05   &   0.06   &  0.02   &  0.02   &   0.20    &  0.09   &  0.09   &   -0.96  &   0.15   &  0.14   &   0.23   &  0.09   &  0.09          \\  \noalign{\vskip 1.pt}      
NGC 4190    &   7.49   &  0.16  &  0.12   &  7.93   &  0.06   &  0.06    &  8.07    & 0.06   &  0.05   &   0.74   &  0.06   &  0.07   &   -1.64   &  0.47   &  0.33   &   -1.16  &   0.47   &  0.30   &   -1.03  &   0.47  &   0.30          \\  \noalign{\vskip 1.pt}      
NGC 4535    &   10.83  &  0.05  &  0.04   &  10.11  &  0.04   &  0.04    &  10.90   & 0.04   &  0.04   &   0.16   &  0.02   &  0.02   &   -1.48   &  0.40   &  0.33   &   -2.20  &   0.41   &  0.36   &   -1.40  &   0.40  &   0.34          \\  \noalign{\vskip 1.pt}      
NGC 4536    &   10.47  &  0.06  &  0.05   &  10.35  &  0.05   &  0.05    &  10.71   & 0.04   &  0.04   &   0.43   &  0.04   &  0.04   &   -0.90   &  0.25   &  0.20   &   -1.01  &   0.28   &  0.23   &   -0.65  &   0.26  &   0.21          \\  \noalign{\vskip 1.pt}      
NGC 4559    &   9.93   &  0.09  &  0.07   &  9.91   &  0.06   &  0.05    &  10.22   & 0.05   &  0.05   &   0.49   &  0.05   &  0.06   &   -0.63   &  0.14   &  0.12   &   -0.65  &   0.11   &  0.11   &   -0.34  &   0.12  &   0.11          \\  \noalign{\vskip 1.pt}      
NGC 4651    &   10.63  &  0.05  &  0.04   &  9.95   &  0.07   &  0.07    &  10.71   & 0.04   &  0.04   &   0.17   &  0.03   &  0.03   &   -0.17   &  0.10   &  0.10   &   -0.84  &   0.11   &  0.11   &   -0.08  &   0.10  &   0.10          \\  \noalign{\vskip 1.pt}
NGC 4725    &   10.97  &  0.06  &  0.05   &  9.98   &  0.06   &  0.05    &  11.02   & 0.06   &  0.05   &   0.09   &  0.01   &  0.02   &   -0.08   &  0.09   &  0.10   &   -1.05  &   0.11   &  0.10   &   -0.03  &   0.09  &   0.10          \\  \noalign{\vskip 1.pt}      
NGC 4736    &   10.13  &  0.04  &  0.03   &  9.12   &  0.07   &  0.06    &  10.17   & 0.03   &  0.03   &   0.09   &  0.01   &  0.01   &   -0.39   &  0.12   &  0.10   &   -1.40  &   0.16   &  0.13   &   -0.35  &   0.12  &   0.10          \\  \noalign{\vskip 1.pt}      
NGC 5005    &   11.00  &  0.04  &  0.04   &  9.68   &  0.08   &  0.08    &  11.02   & 0.04   &  0.03   &   0.05   &  0.01   &  0.01   &   -0.63   &  0.61   &  0.45   &   -1.96  &   0.62   &  0.46   &   -0.61  &   0.61  &   0.45          \\  \noalign{\vskip 1.pt}      
NGC 5055    &   10.75  &  0.03  &  0.03   &  10.10  &   0.05  &   0.04   &   10.84  &  0.03  &   0.03  &    0.19  &   0.02  &   0.02  &    -0.49  &   0.04  &   0.04  &    -1.13 &    0.06  &   0.05  &    -0.40 &    0.04 &    0.04   \\  \noalign{\vskip 1.pt}      
NGC 7793    &   9.55   &  0.09  &  0.07   &  9.21   &  0.05   &  0.05    &  9.71    & 0.06   &  0.05   &   0.32   &  0.04   &  0.05   &   -1.57   &  0.72   &  0.42   &   -1.89  &   0.72   &  0.42   &   -1.40  &   0.72  &   0.42          \\  \noalign{\vskip 1.pt}      
UGC 1501    &   8.37   &  0.16  &  0.11   &  8.56   &  0.06   &  0.06    &  8.78    & 0.06   &  0.06   &   0.61   &  0.07   &  0.09   &   -2.14   &  0.52   &  0.39   &   -1.92  &   0.53   &  0.39   &   -1.71  &   0.53  &   0.38          \\  \noalign{\vskip 1.pt}  
UGC 8508    &   6.67   &  0.20  &  0.15   &  7.30   &  0.10   &  0.09    &  7.39    & 0.08   &  0.08   &   0.81   &  0.07   &  0.07   &   -3.14   &  0.83   &  0.61   &   -2.50  &   0.82   &  0.60   &   -2.40  &   0.82  &   0.60          \\  \noalign{\vskip 1.pt}      
WLM         &   6.58   &  0.20  &  0.14   &  7.93   &  0.06   &  0.05    &  7.95    & 0.06   &  0.05   &   0.96   &  0.02   &  0.02   &   -2.36   &  0.21   &  0.16   &   -1.01  &   0.10   &  0.10   &   -0.99  &   0.10  &   0.10          \\  \noalign{\vskip 1.pt}      
 \hline
    \end{tabular}} 
    \tablefoot{The first column of each parameter represents the 50th percentile (median) of its corresponding posterior distribution. The lower and upper uncertainties correspond to the absolute difference between the distribution's median and the 16th and 84th percentiles, respectively. We note that $f_{\rm gas}=M_{\rm gas}/M_{\rm bar}$, and $\tilde{f}_{i} = M_i/(0.16\,M_{200})$.}
\end{table*}
\end{landscape}

\end{appendix}

\end{document}